\newcommand{\pa}{\partial}
\newcommand{\be}{\begin{equation}}
\newcommand{\ee}{\end{equation}}
\newcommand{\bea}{\begin{eqnarray}}
\newcommand{\eea}{\end{eqnarray}}
\newcommand{\nn}{\nonumber}
\newcommand{\G}{\Gamma}
\newcommand{\A}{{\cal A}}
\newcommand{\B}{{\cal B}}
\newcommand{\C}{{\cal C}}
\renewcommand{\a}{\alpha}
\renewcommand{\d}{\delta}
\newcommand{\D}{\Delta}
\newcommand{\w}{\vec{w}}
\newcommand{\rmd}{{\rm d}}

\newcommand{\bt}[1]{{\bar t}}
\newcommand{\ts}{\textstyle}
\renewcommand{\t}{\tau}
\newcommand{\ea}{\!\!\! = \!\!\!}
\newcommand{\lb}{{\mbox{\Large (}}}
\newcommand{\rb}{{\mbox{\Large )}}}
\newcommand{\oD}{{\overline D}}
\newcommand{\half}{{\ts \frac{1}{2}}}
\newcommand{\pr}{\partial}
 
\documentclass[11pt]{article}
\usepackage{epsfig}
\usepackage{amssymb}
\usepackage{curves}
\input FEYNMAN

\csname @addtoreset\endcsname{equation}{section}
\begin{document}

\begin{titlepage}
{\hbox to\hsize{\hfill {AEI--2002--096}}} {\hbox to\hsize{${~}$\hfill 
{DAMTP-02-148}}}
{\hbox to\hsize{${~}$ \hfill {LAPTH-953/02}}} \vskip 0.2pt \hfill {\tt 
hep-th/0212116}

\begin{center}
\vglue .3in {\Large\bf Correlation Functions and Massive Kaluza-Klein 
Modes\\ in the AdS/CFT Correspondence}

\vskip 1cm

{\large {$\rm{G.~ Arutyunov}^{(a)}$}}, {\large{$\rm{F.A. ~Dolan}^{(b)}$, 
{\large{$\rm{H. ~Osborn}^{(b)}$}}, $\rm{E.~Sokatchev^{(c)}}$}}
\\[.3in]
\small

$^{\rm(a)}$ {\it Max-Planck-Institut f\"ur Gravitationsphysik,
Albert-Einstein-Institut, \\
Am M\"uhlenberg 1, D-14476 Golm, Germany}
\\
[.03in]
$^{\rm(b)}${\it Department of Applied Mathematics and Theoretical Physics, \\
Silver Street, Cambridge, CB3 9EW, England}
\\
[.03in] $^{\rm(c)}${\it Laboratoire d'Annecy-le-Vieux de Physique 
Th\'{e}orique\footnote[1]{UMR 5108 associ{\'e}e {\`a}
 l'Universit{\'e} de Savoie}
LAPTH, B.P. 110, \\ F-74941 Annecy-le-Vieux et l'Universit\'{e} de Savoie}
\\[.3in]
\normalsize

{\bf ABSTRACT}\\[.0015in]
\end{center}
We study four-point correlation functions of $\half$-BPS operators in
${\cal N}=4$ SYM which are dual to massive KK modes in AdS${}_5$ supergravity. 
On the field theory side, the procedure of inserting the SYM action yields 
partial non-renormalisation of the four-point amplitude for such operators. 
In particular, if the BPS operators have dimensions equal to three or four, 
the corresponding four-point amplitude is determined by one or two independent 
functions of the two conformal cross-ratios, respectively. This restriction 
on the amplitude does not merely follow from the superconformal Ward identities, 
it also encodes dynamical information related to the structure of the gauge 
theory Lagrangian. \\
The dimension 3 BPS operator is the AdS/CFT dual of the first 
non-trivial massive Kaluza-Klein mode of the compactified type IIB 
supergravity, whose interactions go beyond the level of the 
five-dimensional gauged ${\cal N}=8$ supergravity. We show that the 
corresponding effective Lagrangian has a surprisingly simple 
sigma-model-type form with at most two derivatives. We then compute the 
supergravity-induced four-point amplitude for the dimension 3 operators. 
Remarkably, this amplitude splits into a ``free" and an ``interacting" 
parts in exact agreement with the structure predicted by the insertion 
procedure. The underlying OPE fulfills the requirements of 
superconformal symmetry and unitarity.

\vskip 7pt ${~~~}$ \newline

\end{titlepage}

\section{Introduction}
In the past few years the holographic relation between supergravity 
(string) theories on AdS backgrounds and certain conformal field 
theories living on the corresponding AdS boundaries has been studied and 
tested by various means. The most typical example is the ${\cal N}=4$ 
super-Yang-Mills (SYM) theory whose conjectured holographic dual is the 
type IIB supergravity (string) theory on $AdS_5 \times S^5$. Our current 
understanding of this basic example is primarily based on superconformal 
kinematics, as the isometry group of the supergravity theory coincides 
with the superconformal symmetry group of its dual. In this respect it 
is highly desirable to learn how to separate the actual dynamical 
statements from those due to superconformal symmetry and to subsequently 
put the former to the test.

The string spectrum on the $AdS_5 \times S^5$ background is presently 
unavailable. Thus, our current studies are confined to $AdS_5$ 
supergravity that is dual to the limit of the gauge theory where the 't 
Hooft coupling $\lambda=g^2_{YM}N$ is infinite and the rank $N$ of the 
gauge group SU($N$) is large. Compactifying type IIB supergravity on the 
five-dimensional sphere results in an infinite tower of (generically 
massive) Kaluza-Klein (KK) modes. According to the superconformal 
kinematics, their field-theory duals are the so-called $\half$-BPS 
operators. They form short superconformal multiplets whose lowest-weight 
states are annihilated by half of the Poincar\'e supercharges. The 
conformal dimensions and more generally, the two- and three-point 
correlation functions of the $\half$-BPS operators are protected from 
quantum corrections, but the four-point functions are not. The OPE 
spectrum of two $\half$-BPS operators is rich, coupling-dependent and 
generically contains unprotected (long) supermultiplets. Thus, the 
four-point correlators of $\half$-BPS operators encode some genuinely 
dynamical information and hence are interesting objects to study both in 
field theory and in the supergravity approximation.

The superconformal kinematics puts constraints on the four-point 
amplitude in the form of superconformal Ward identities. Their solution 
admits a functional freedom providing a window for the non-trivial 
dynamics. Further restrictions on the correlation functions can be 
obtained by a {\it dynamical mechanism}. It consists in generating the 
quantum corrections to the amplitude by insertion of the SYM action (the 
insertion procedure) and has the effect of reducing the functional 
freedom in the amplitude (partial non-renormalisation). This procedure 
is purely field-theoretic and it relies on the existence of a Lagrangian 
description of the theory. However, its prediction, i.e. the particular 
form of the amplitude compared to the general solution of the 
superconformal Ward identities, can be confronted with concrete 
supergravity-induced correlators computed in $AdS_5$ supergravity. This 
is one of the most probing tests of the AdS/CFT duality available today.

Our concrete knowledge of the supergravity-induced four-point amplitudes 
has up to now been exhausted by a single example, the correlator of 
$\half$-BPS operators of the lowest allowed dimension 2. The corresponding 
supermultiplet is rather special, as it contains the conserved 
R symmetry current and the stress tensor of the ${\cal N}=4$ theory. It 
is dual to the graviton multiplet of the gauged ${\cal N}=8$ 
five-dimensional supergravity comprising the {\it massless KK modes} of 
the type IIB supergravity compactification.

Clearly, to start bringing out the flavor of the more involved 
ten-dimensional physics one has to go beyond the massless sector of the 
theory and obtain new examples of supergravity-induced four-point 
correlators involving {\it BPS operators of higher dimension}. In this 
paper we make a first step in this direction by studying the general 
form of the four-point amplitude for $\half$-BPS operators of dimension 3 
and then comparing it to an explicit supergravity computation.

As mentioned earlier, this time we cannot restrict ourselves to the 
gauged ${\cal N}=8$ supergravity but should rather start from the full 
ten-dimensional theory. The $\half$-BPS operators of dimension $k$ are 
AdS/CFT dual to the KK modes $s_k$ transforming in the irrep $[0,k,0]$ 
of SO(6). To compute the corresponding supergravity-induced four-point 
amplitude one first has to find the action for the fields $s_k$ up to 
quartic order. This is a hard problem primarily due to the absence of a 
suitable Lagrangian formulation of the type IIB supergravity (the 
well-known self-duality problem). One way to solve it is to expand the 
covariant equations of motion of type IIB supergravity around the 
background solution and find the quadratic and cubic corrections to the 
free AdS equations of motion. Using the freedom of perturbative field 
redefinitions, one is able to recast these equations into a Lagrangian 
form. This approach has lead to an effective action in $AdS_5$ space 
which allows the computation of four-point correlators for $\half$-BPS 
operators of {\it arbitrary dimension} (see Section \ref{amplitude} for 
references and a brief review). The effective action has the following 
remarkable properties:
\begin{itemize}
\item the quartic couplings contain terms with four and two derivatives only;

\item it admits a consistent truncation to the massless graviton multiplet and
the corresponding action coincides with that of the gauged ${\cal N}=8$ 
five-dimensional supergravity on $AdS_5$;

\item the quartic couplings corresponding to the so-called extremal or subextremal correlators vanish; 
\item the four-point correlation functions in the boundary CFT derived from this action obey the predictions of the field-theoretic insertion procedure.
\end{itemize}

The latter property is probably the most non-trivial one. It has so far 
been verified for $\half$-BPS operators of dimension 2. One of the principal 
aims of the present paper is to carry out a similar test in the case of 
dimension 3. Again we find a remarkable agreement. One of the surprising 
features of the supergravity-induced amplitude in both cases is the 
splitting into a ``free" and an ``interacting" parts, exactly following 
the field theory pattern where the free amplitude is supposed to receive quantum corrections. This is indeed unexpected, since in supergravity there is no analog of the coupling constant which makes the splitting natural in 
field theory. All this is not only a non-trivial check of the effective 
supergravity action but it also supports the field-theoretic arguments 
for the partially non-renormalised form of the amplitude. Last but not 
least, it provides strong evidence for the AdS/CFT duality.

It should also be pointed out that after specifying the explicit 
representation content of the fields involved, the resulting quartic 
effective action turns out to be rather simple. The quartic couplings with 
four-derivative vertices actually vanish, at least in the case we 
consider here. In other words, the action is of the sigma-model type. 
Therefore it would be highly desirable to develop a systematic 
superspace procedure for the construction of the supergravity effective 
action. This could eventually unravel its hidden symmetry structure and 
simplicity.

Another interesting problem is to understand the relationship between 
four-point amplitudes of $\half$-BPS operators of different dimensions. 
Indeed, adding at least one KK mode to the massless graviton multiplet 
causes the whole infinite tower of massive KK modes to emerge. In other 
words, all the higher KK modes are tightly bound together in a unique 
interacting Lagrangian. It seems plausible that this would have 
implications for the correlators derived from such a Lagrangian.

Finally, one can reverse the logic of passing from supergravity to 
gauge theory and ask the question to what extent are the quartic 
(contact) terms in the effective action (or even the whole action) fixed 
by requiring the corresponding four-point amplitude to satisfy the 
restrictions imposed by the insertion procedure. We hope to come back to 
this question in the future.

The paper is organised as follows. In Section 2 we discuss the general 
form of the four-point amplitude of $\half$-BPS operators based on the 
conformal, $R$ and crossing symmetries. In Section 3 we provide 
field-theoretic arguments leading to the partially non-renormalised form 
of the amplitude. We show that for the ``quantum part'' of the 
four-point correlator of dimension 3 operators the insertion formula 
predicts a single function of the conformal cross-ratios instead of 
three such functions. In Section 4 we compute the corresponding 
supergravity-induced four-point amplitude and in Section 5 we verify the 
CFT predictions from Section 3. Finally, in Section 6 we perform a 
partial OPE analysis of the amplitude and obtain the large $N$
corrections to the scaling dimensions of the long multiplets. The computational details are gathered in several appendices.

  \setcounter{equation}0
\section{General four-point amplitudes}
\label{General}

According to the AdS/CFT duality conjecture \cite{M,GKP,W}, the KK modes of the 
$AdS_5\times S^5$ compactification of type-IIB supergravity are mapped 
into $\half$-BPS multiplets of the ${\cal N}=4$ SYM theory whose lowest 
components are scalar fields ${\cal O}^I$ of conformal dimension $k \geq 
2$ in the irrep $[0,k,0]$ of the $R$ symmetry group SO(6) $\sim$ SU(4). 
For $k=2$ this is the so-called stress-tensor multiplet corresponding to 
the massless AdS supergravity multiplet. For $k\geq 3$ we are dealing 
with non-trivial (massive) KK modes. In this section we discuss the 
general structure of the correlator of four identical operators ${\cal 
O}^I$, predicted on the basis of conformal, $R$ and crossing symmetry. In 
particular, we determine the number of independent conformal invariant 
functions which specify such correlators for $k=2,3,4$ and we outline a 
method to do this for general $k$.

The composite operators ${\cal O}^I$ with a suitably normalised 
two-point function are of the form 
\bea \label{ibasis} {\cal O}^I= 
C_{i_1\cdots i_k}^I \mbox{Tr}(\phi^{i_1} \cdots \phi^{i_k}) \, . \eea 
Here $\phi^i$, $i=1,\ldots, 6$ are the ${\cal N}=4$ SYM scalars and 
$C^I_{i_1\cdots i_k}$ are traceless symmetric tensors obeying the 
normalisation condition $C^I_{i_1\cdots i_k}C^J_{i_1\cdots 
i_k}=\d^{IJ}$, which describe the irreducible representations $[0,k,0]$.

The $R$ symmetry content of the OPE of two identical operators ${\cal 
O}^{I_1}$ and ${\cal O}^{I_2}$ is determined by the corresponding tensor 
product decomposition
\begin{equation}\label{decgen}
[0,k,0] \times [0,k,0] = \sum_{p = 0}^k \sum_{q=0}^{k-p} \ 
[q, 2k-2q-2p,q] \ .
\end{equation}
The total number of terms in the right-hand side of eq.(\ref{decgen}) 
is $\half (k+1)(k+2)$. According to the general arguments in 
\cite{AES}-\cite{HH2}, all the OPE channels with $q=0,1$ contain only 
protected (BPS or semishort) operators.

The decomposition (\ref{decgen}) provides the basis for an OPE expansion 
of the four-point amplitude $\langle {\cal O}(1){\cal O}(2){\cal 
O}(3){\cal O}(4) \rangle$. However, in field theory there exists a 
different basis which naturally arises by connecting the four points 
with free propagators (Wick contractions). The free propagator for the 
elementary ${\cal N}=4$ SYM scalars is 
\begin{equation}\label{frpr} 
\langle \phi^i(x_1)\phi^j(x_2) \rangle = \frac{\delta^{ij}}{x^2_{12}} \;. 
\end{equation}
A convenient way of keeping track of the SO(6) indices is to project them 
with ``harmonic variables", i.e. a complex vector $u_i$ satisfying the 
conditions 
\be  
u_iu_i = 0\;, \qquad u_i\bar{u}_i = 1\;. 
\label{defhar}
\end{equation}

This vector provides a harmonic description of the coset space 
SO(6)/SO(2)$\times$SO(4). Indeed, the vector $u_i$ subject to the 
constraints (\ref{defhar}) and modulo U(1) $\sim$ SO(2) phase 
transformations contains exactly  the eight real coordinates of the 
manifold. With its help we can project the operator ${\cal O}^I$ onto 
the highest-weight state of the representation $[0,k,0]$: 
\begin{equation}\label{prO} 
{\cal O}^{(k)} = u_{i_1} \cdots u_{i_k} \mbox{Tr}(\phi^{i_1} \cdots 
\phi^{i_k}) \, . \end{equation}
Here the Dynkin label $k$ is identified with the 
U(1) charge of the projection (\ref{prO}) (assuming that the vector 
$u_i$ carries U(1) charge $+1$). Further, using two copies of the 
harmonic variables, one for each point, we can project the propagator 
(\ref{frpr}): 
\begin{equation}\label{prpr} \langle \phi(1)\phi(2) \rangle = 
\frac{u^i_{1} u^i_{2}}{x^2_{12}} \equiv \frac{(12)}{x^2_{12}} = 
\frac{(21)}{x^2_{21}}\;. \end{equation}

Now we can start constructing four-point functions by connecting pairs 
of points by the propagators (\ref{prpr}). The simplest case corresponds 
to $k=1$, i.e., to only one propagator leaving or entering each point. 
In this way we find the three basic propagator structures depicted in 
Figure 1: \vskip 1cm
\begin{figure}[ht]
\begin{center}
\input{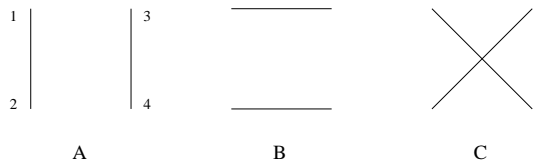}
\end{center}
\caption{Basic contractions}
\end{figure}
The corresponding expressions involving four sets of space-time and 
harmonic variables are: \be A = \frac{(12)(34)}{x^2_{12}\, x^2_{34}}\;, 
\quad B = \frac{(14)(23)}{x^2_{14}\, x^2_{23}}\;, \quad C = 
\frac{(13)(24)}{x^2_{13}\, x^2_{24}}\;. \end{equation}

Symbolically, the propagator basis for a four-point amplitude with 
arbitrary $k$ can be obtained by the following ``binomial expansion"
\begin{equation}\label{symb}
(A+B+C)^k \ \rightarrow \sum A^mB^nC^l\ .
\end{equation}
Here the sum goes over the integers $(m,n,l)$ such that $m\geq n\geq 
l\geq 0$ and $m+n+l =k$. For a given set $(m,n,l)$ \footnote{The labels 
$(m,n,l)$ should not be confused with the Dynkin labels $[m,n,l]$.} the 
permutations of $A,B,C$ correspond to all possible graphs in the 
equivalence class obtained by crossing symmetry. It is then clear that 
the number of such permutations is equal to six if $m\neq n \neq l$, or 
to three if any two of the labels coincide but are different from the 
third, and to one if $m=n=l$. For example, in the case $k=2$ we obtain 
six graphs combined into two crossing-equivalence classes $(2,0,0)$ and 
$(1,1,0)$ and depicted in Figure 2.

\begin{figure}[ht]
\begin{center}
\input{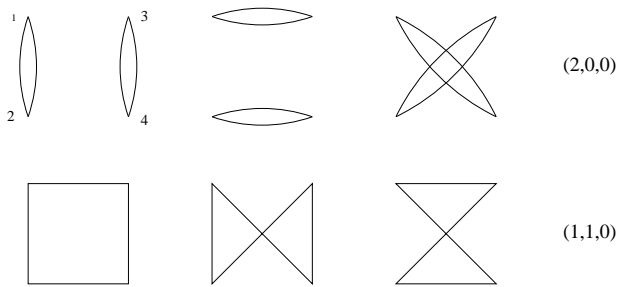}
\end{center}\caption{The case $k=2$}
\end{figure}

The expressions of, e.g., the first graphs in each line in Figure 2 
read: 
\bea
 AA & \ea & \frac{(12)^2(34)^2}{x^4_{12}\, x^4_{34}} 
=  1_{i_1} 1_{j_1}2_{i_2}2_{j_2}3_{i_3}3_{j_3}4_{i_4}4_{j_4}\ 
\frac{\delta^{\{i_1}_{i_2} \delta^{j_1 \}}_{ j_2} 
\delta^{\{i_3}_{i_4} \delta^{j_3 \}}_{ j_4}}{x^4_{12}\, x^4_{34}}  \;, \nonumber \\
 AB &\ea & \frac{(12)(23)(34)(41)}{x^2_{12}\, x^2_{23}\, x^2_{34}\, x^2_{41}}
 = 1_{i_1} 1_{j_1}2_{i_2}2_{j_2}3_{i_3}3_{j_3}4_{i_4}4_{j_4}\ 
\frac{\delta^{\{i_1}_{\{i_2} \delta^{j_1 \}}_{ \{i_4} \delta^{\{i_3}_{j_4\}} 
\delta^{j_3 \}}_{ j_4\}}} {x^2_{12}\, x^2_{23}\, x^2_{34}\, x^2_{41}} \;, 
\label{AA}
\eea 
where $n_i$ is a shorthand for the harmonic $u^i_{n}$ at point 
$n$. Note that the traceless symmetrisation denoted by $\{\}$ is 
automatic, given the fact that the harmonic variables commute and 
satisfy the defining conditions (\ref{defhar}). Since the harmonics at 
each point are independent variables, we can remove them and thus obtain 
the explicit tensor structure made out of Kronecker deltas.

Each of the six propagator structures in Figure 2 can be multiplied by 
an arbitrary function of the conformal invariant cross-ratios
$$
s=\frac{x^2_{12}\, x^2_{34}}{x^2_{13}\, x^2_{24}}\,, \qquad 
t=\frac{x^2_{14}\, x^2_{23}}{x^2_{13}\, x^2_{24}}\,.
$$ In this way we obtain the most general four-point amplitude for operators 
${\cal O}^{(2)}$ with the required SO(6) and conformal transformation properties:
\bea \hskip -1.5cm
&& \langle {\cal O}^{(2)}(x_1){\cal O}^{(2)}(x_2){\cal O}^{(2)}(x_3)
{\cal O}^{(2)}(x_4)\rangle \nonumber\\
\hskip -1.5cm && \quad {}= a_1(s,t)\frac{(12)^2(34)^2}{x^4_{12}\, x^4_{34}} +
a_2(s,t)\frac{(13)^2(24)^2}{x^4_{13}\, x^4_{24}} + 
a_3(s,t)\frac{(14)^2(23)^2}{x^4_{14}\, x^4_{23}}\nonumber\\  
\hskip -1.5cm &&\qquad  {} + b_1(s,t) 
\frac{(13)(14)(23)(24)}{x^2_{13}\, x^2_{14}\, x^2_{23}\, x^2_{24}}
+ b_2(s,t)\frac{(12)(14)(23)(34)}{x^2_{12}\, x^2_{14}\, x^2_{23}\, x^2_{34}}
+ b_3(s,t) \frac{(12)(13)(24)(34)}
{x^2_{12}\, x^2_{13}\, x^2_{24}\, x^2_{34}}\;. \label{befcro2}
\eea 
We still need to impose the full crossing symmetry of this 
correlator. This is very easy to do, since the crossing properties of 
the structures in Figure 2 are obvious. Thus, the three coefficients 
$a_i$ in (\ref{befcro2}) correspond to the crossing class $(2,0,0)$ and 
transform into each other: \bea
a_1(s,t)&\ea&a_3(t,s)=a_1(s/t,1/t) \nonumber\\
a_2(s,t)&\ea&a_2(t,s)=a_3(s/t,1/t) \label{cr1}\, ; \eea the same applies 
to the coefficients $b_i$ from the class $(1,1,0)$: \bea
b_1(s,t)&\ea&b_3(t,s)=b_1(s/t,1/t) \nonumber\\
b_2(s,t)&\ea&b_2(t,s)=b_3(s/t,1/t) \label{cr2}\, . \eea 
Thus, the crossing 
invariant correlator for ${\cal O}^{(2)}$ is in general determined by 
{\it two independent functions}, one of the $a_i$'s and one of the 
$b_i$'s.

Having explained the case $k=2$ in detail, we can immediately generalize 
to the main case of interest in this paper, $k=3$ (or to any higher 
value of $k$). The decomposition of the tensor product (\ref{decgen}) 
reads \bea \nonumber
[0,3,0]_{50}\times [0,3,0]_{50}&\ea&[0,0,0]_1+[0,2,0]_{20}+[0,4,0]_{105}\\
\label{irreps}
&&{}+ [0,6,0]_{336}+[2,0,2]_{84}+[2,2,2]_{729}\\
\nonumber &&{}+ [1,0,1]_{15}+[1,2,1]_{175}+[3,0,3]_{300}+[1,4,1]_{735}\ . 
\eea 
The subscripts indicate the dimension of the corresponding irrep. 
The irreps in the first two lines of eq.(\ref{irreps}) are symmetric 
and those in the third line are antisymmetric in the indices $I_1,I_2$. 
Thus, the corresponding OPE will have ten different SO(6) channels.
 Note that among them only three may contain unprotected operators: 
$[0,0,0]$, $[0,2,0]$ and $[1,0,1]$.

In the propagator basis the ten structures are organised in three 
crossing-equivalence classes, $(3,0,0)$, $(2,1,0)$ and $(1,1,1)$, 
depicted in Figure 3.
\begin{figure}[ht]
\begin{center}
\input{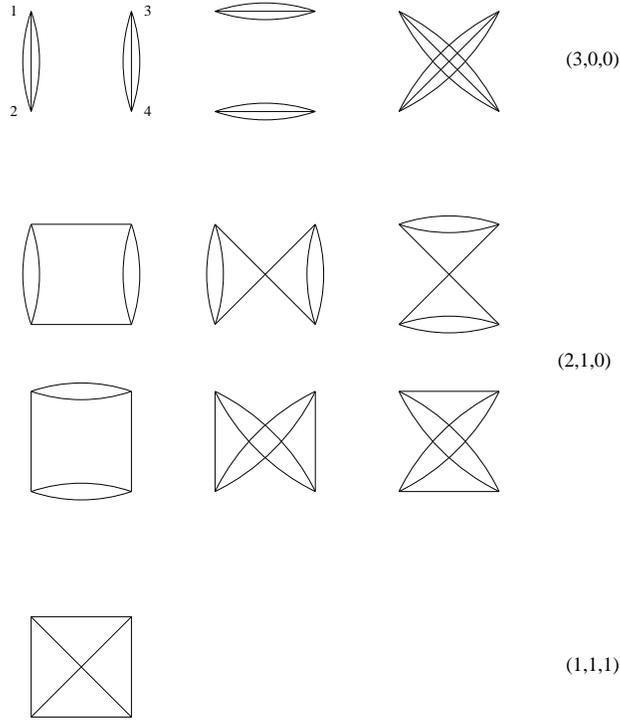}
\end{center}\caption{The case $k=3$}
\end{figure}
The most general SO(6) and conformally covariant four-point function for 
the operators ${\cal O}^{(3)}$ is of the form 
\bea \nonumber && \langle 
{\cal O}^{(3)}(x_1){\cal O}^{(3)}(x_2){\cal O}^{(3)}(x_3){\cal O}^{(3)}(x_4) 
\rangle \\
&& \quad {} = 
a_1(s,t)\frac{(12)^3(34)^3}{x_{12}^6\, x_{34}^6}+ 
a_2(s,t)\frac{(13)^3(24)^3}{x_{13}^6\, x_{24}^6}+
a_3(s,t)\frac{(14)^3(23)^3}{x_{14}^6\, x_{23}^6} \nonumber \\
\nonumber &&\qquad {}+ 
b_1(s,t)\frac{(12)^2(34)^2(13)(24)}{x_{12}^4\, x_{34}^4\, x_{13}^2\, x_{24}^2}
+b_2(s,t)\frac{(12)^2(34)^2(14)(23)}{x_{12}^4\, x_{34}^4\, x_{14}^2\, x_{23}^2} 
\nonumber\\
&& \qquad {}+
b_3(s,t)\frac{(14)^2(23)^2(13)(24)}{x_{14}^4\, x_{23}^4\, x_{13}^2\, x_{24}^2} + 
b_4(s,t)\frac{(14)^2(23)^2(12)(34)}{x_{14}^4\, x_{23}^4\, x_{12}^2\, x_{34}^2} 
\nonumber\\ \qquad && \qquad {}+ 
b_5(s,t)\frac{(13)^2(24)^2(12)(34)}{x_{13}^4\, x_{24}^4\, x_{12}^2\, x_{34}^2} 
+b_6(s,t)\frac{(13)^2(24)^2(14)(23)}{x_{13}^4\, x_{24}^4\, x_{14}^2\, x_{23}^2} 
\nonumber \\ && \qquad {}+ 
c(s,t)\frac{(12)(13)(14)(23)(24)(34)}{x_{12}^2\, x_{13}^2\, x_{14}^2\, x_{23}^2\, 
x_{24}^2x_{34}^2}\, . \label{g4pt}\eea 
Crossing symmetry imposes relations among the 
functions $a_i(s,t)$, $b_i(s,t)$. For the coefficients $a_i$ they are 
the same as in (\ref{cr1}), for the coefficients $b_i$ they are: \bea
b_2(s,t)&\ea &b_1(s/t,1/t), ~~~~b_3(s,t)=b_1(t,s),~~~~b_4(s,t)=b_1(t/s,1/s)\nn \\
b_5(s,t)&\ea &b_1(1/s,t/s),~~~~b_6(s,t)=b_1(1/t,s/t) \, ; 
\label{cr2'}\eea 
finally, the coefficient $c(s,t)$ must obey the symmetry conditions 
\bea c(s,t)=c(t,s)=c(s/t,1/t) \, . \label{cr3}\eea

Thus, after taking crossing symmetry into account we have {\it three 
independent coefficient functions}, for instance, $a_1(s,t)$, $b_1(s,t)$ 
and $c(s,t)$, one for each equivalence class in Figure 3. They can be 
split into a free field theory contribution and a quantum correction 
part. In free field theory the coefficient functions for the canonically 
normalised operators are given by ($b_i$ and $c$ have been calculated in 
the large $N$ limit) \bea \label{ff} a_i=1\, ,~~~~~~~b_i=\frac{9}{N^2}\, 
,~~~~~~c=\frac{18}{N^2}\, . \eea
 As will be shown in the next section, the dynamics of the ${\cal N}=4$ SYM
theory encoded in the insertion formula implies further non-trivial 
algebraic relations among the ``quantum parts'' of these coefficient 
functions leaving a {\it single} function of the conformal cross-ratios, 
e.g. $a_2(s,t)$, as the only undetermined functional freedom.

The method developed in this section and explicitly applied to $k=2,3$ 
can easily be generalised to any value of $k$.\footnote{A different 
counting method as well as a general formula for the number of independent 
functions has been given in \cite{Heslop:2002hp}.} As explained earlier, 
the  number of crossing-equivalence classes and thus of independent 
functions  in the amplitude corresponds to the possible splittings
$(m,n,l)$ of 
$k$. We mention that for $k=4$ there are four such splittings: 
$(4,0,0)$, $(3,1,0)$, $(2,2,0)$ and $(2,1,1)$ with multiplicities 3, 6, 
3 and 3, respectively.

Concluding this section we point out that the discussion of AdS 
supergravity in Section \ref{amplitude} requires a different set of 
harmonic variables, this time on the coset space SO(6)/SO(5) $\sim$ 
$S^5$. In contrast to the four-dimensional complex manifold 
SO(6)/SO(2)$\times$SO(4) considered in this section, $S^5$ is a real 
five-dimensional manifold. Thus, we are forced to describe the same 
representations of SO(6) $\sim$ SU(4) in terms of the complex harmonics 
$u_i$ in the context of CFT and in terms of the real ones on $S^5$ in 
the context of AdS supergravity. The transition from one description to 
the other is not direct, we need to exhibit the index structure of the 
representations. To this end we can use the tensor $C^I$ defined in 
(\ref{ibasis}). Take, for instance, the case $k=2$ and consider the 
four-point block $AA$ (\ref{AA}). Removing the harmonics and then 
converting the pairs of vector indices $ij$ into indices $I$ of the 
irrep {\bf 20}, we obtain \be
 (AA)^{I_1I_2I_3I_4} = C^{I_1}_{i_1j_1} C^{I_2}_{i_2j_2} C^{I_3}_{i_3j_3} 
C^{I_4}_{i_4j_4} \ \frac{\delta^{\{i_1}_{i_2} \delta^{j_1 \}}_{ j_2} 
\delta^{\{i_3}_{i_4} \delta^{j_3 \}}_{ j_4}}{x^4_{12}\, x^4_{34}}  \equiv  
\frac{\delta^{12} \delta^{34}}{x^4_{12}\, x^4_{34}}\;. \label{simple}
\end{equation}
In what follows we will systematically use a compact notation for 
the tensor structures where the indices $I_1,I_2,I_3,I_4$ will be 
replaced by $1234$. In (\ref{simple}) the tensor is just the identity 
$\delta^{12} \delta^{34}$. For the block $AB$ we find \be
 (AB)^{I_1I_2I_3I_4} = C^{I_1}_{i_1j_1} C^{I_2}_{i_2j_2} C^{I_3}_{i_3j_3} 
C^{I_4}_{i_4j_4} \ \ \frac{\delta^{\{i_1}_{\{i_2} \delta^{j_1 \}}_{ \{i_4} 
\delta^{\{i_3}_{j_4\}} \delta^{j_3 \}}_{ j_2\}}} 
{x^2_{12}\, x^2_{23}\, x^2_{34}\, x^2_{41}}   \equiv  
\frac{T^{1234}}{x^2_{12}\, x^2_{23}\, x^2_{34}\,  x^2_{41}}\;. 
\end{equation} 
For $k=3$ we need three types of tensors: 
\bea
\delta^{12} \delta^{34} &\ea & C_{ijk}^{I_1}C_{ijk}^{I_2}
C_{mln}^{I_3}C_{mln}^{I_4}  \, , \nonumber\\
C^{1234} &\ea & C_{ijk}^{I_1}C_{ij s}^{I_2}C_{ml k}^{I_3}C_{ml s}^{I_4} 
\, , \nonumber \\
S^{1234} &\ea & C_{ijk}^{I_1}C_{i m n}^{I_2}C_{j m s}^{I_3}C_{k n s}^{I_4} 
\, , \label{defC}
\eea 
and their permutations. The tensor $S^{1234}$ is totally symmetric, 
while $C^{1234}$ obeys the following symmetry relations
$$
C^{1234}=C^{2143}=C^{3412}=C^{4321}\;.
$$
In this notation eq.(\ref{g4pt}) becomes 
\bea \nonumber && \langle {\cal 
O}^{1}(x_1){\cal O}^{2}(x_2){\cal O}^{3}(x_3){\cal O}^{4}(x_4) \rangle \\
&& \nonumber \quad{} =  
a_1(s,t)\frac{\delta^{12}\delta^{34}}{x_{12}^6\, x_{34}^6}+ 
a_2(s,t)\frac{\delta^{13}\delta^{24}}{x_{13}^6\, x_{24}^6}+
a_3(s,t)\frac{\delta^{14}\delta^{23}}{x_{14}^6\, x_{23}^6}\\
\nonumber &&\qquad {}+ b_1(s,t)\frac{C^{1234}}
{x_{12}^4\, x_{34}^4\, x_{13}^2\, x_{24}^2} 
+b_2(s,t)\frac{C^{1243}}{x_{12}^4\, x_{34}^4\, x_{14}^2\, x_{23}^2} 
+b_3(s,t)\frac{C^{3214}}{x_{23}^4\, x_{14}^4\, x_{13}^2\, x_{24}^2}
\\
&& \nonumber\qquad  {}+ b_4(s,t)\frac{C^{3241}}
{x_{23}^4\, x_{14}^4\, x_{34}^2\, x_{12}^2} 
+b_5(s,t)\frac{C^{4231}}{x_{24}^4\, x_{13}^4\, x_{12}^2\, x_{34}^2}
+b_6(s,t)\frac{C^{4213}}{x_{24}^4\, x_{13}^4\, x_{14}^2\, x_{23}^2} \\
&& \qquad {}+ 
c(s,t)\frac{S^{1234}}
{x_{12}^2\, x_{13}^2\, x_{14}^2\, x_{23}^2\, x_{24}^2\, x_{34}^2}\, 
. \label{g4pt'}\eea

\setcounter{equation}0
\section{The insertion formula and partial non-renormalisation}
\label{prediction}

Further, dynamical information about the four-point correlators can be 
obtained by using a well-known quantum field theory procedure, the 
insertion of the action as an extra fifth point.\footnote{In the context 
of ${\cal N}=4$ SYM theory this procedure was first considered in 
\cite{I}.} Let us consider the correlator
\begin{equation}\label{4ptgen}
  \langle{\cal O}^{(k_1)}{\cal O}^{(k_2)}{\cal O}^{(k_3)}{\cal
O}^{(k_4)} \rangle
\end{equation}
of four, a priori different, $\half$-BPS operators ${\cal O}^{(k_i)}$ in the 
SO(6) irreps $[0,k_i,0]$. A particular case is obtained by setting 
$k=2$, and this is the so-called stress-tensor multiplet. Its $\theta$ 
expansion contains the on-shell ${\cal N}=4$ SYM Lagrangian at the level 
$(\theta)^4$. Now, the derivative of the correlator (\ref{4ptgen}) with 
respect to the YM coupling $g^2$ can be represented in the form
\begin{equation}\label{Intr}
 \frac{\pa}{\pa g^2} \langle{\cal O}^{(k_1)}{\cal O}^{(k_2)}{\cal O}^{(k_3)}{\cal
O}^{(k_4)} \rangle \propto \int {\rm d}^4x_0 {\rm d}^4\theta_0 \
 \langle{\cal O}^{(2)}(0){\cal O}^{(k_1)}{\cal O}^{(k_2)}{\cal O}^{(k_3)}{\cal
O}^{(k_4)} \rangle \;.
\end{equation}
The integration over the insertion point 0 with a specially chosen 
superspace measure corresponds to a ``superaction" in the terminology of 
Ref. \cite{HST}. In principle, the smallest invariant subspace of ${\cal 
N}=4$ superspace, suitable for describing $\half$-BPS multiplets, involves 
$8 =16/2$ $\theta$'s, while the integration in (\ref{Intr}) goes over 
only four $\theta$'s. The point however is that the operator ${\cal 
O}^{(2)}$ is not just $\half$-BPS short but it is even ``ultrashort" in the 
terminology of \cite{Andrianopoli:1999vr}. This means that its $\theta$ 
expansion effectively terminates at four $\theta$'s, all the 
higher-order terms being total space-time derivatives of the lower 
terms. This explains why the integration in (\ref{Intr}) is 
supersymmetric. Note also that the measure $d^4\theta_0$ in (\ref{Intr}) 
involves only chiral $\theta$'s (left-handed, or right-handed in the 
conjugate form).

Using the ${\cal N}=4$ version \cite{N4HSS} of harmonic superspace 
\cite{GIOS}, one can show \cite{ehw,hssw,Howe:2001je,Heslop:2002hp} that the 
five-point correlator 
$\langle{\cal O}^{(2)}(0){\cal O}^{(k_1)}{\cal O}^{(k_2)}{\cal 
O}^{(k_3)}{\cal O}^{(k_4)} \rangle$ gives rise to a nilpotent 
superconformal covariant. Nilpotent covariants for $\half$-BPS operators do 
not exist if the number of points $n\leq 4$. Indeed, the $\half$-BPS 
condition tells us that at each point such a covariant should depend on 
(a particularly chosen, or harmonic-projected) half of the $\theta$'s. 
Since this object must be covariant under two full supersymmetries 
(Poincar\'e and special conformal), for $n\leq 4$ there exist no 
invariant combinations of the $\theta$'s and hence no way to form 
nilpotent covariants. However, starting with $n=5$ this becomes 
possible. On the other hand, since the $\theta$ measure in (\ref{Intr}) 
is chiral, the $\theta$ expansion of the five-point correlator must 
start with four left-handed $\theta$'s. So, it must be precisely of the 
nilpotent type.

Although it is in principle known how to construct such nilpotent 
covariants \cite{ehw}, the explicit expression in ${\cal N}=4$ harmonic 
superspace is rather complicated and is currently not available. 
Instead, it is much easier to carry out the insertion procedure in 
${\cal N}=2$ harmonic superspace. The idea is to project the ${\cal 
N}=4$ composite operators ${\cal O}^{(k_i)}$ on ${\cal N}=2$. In 
particular, one finds ${\cal N}=2$ projections which involve only 
hypermultiplets. All the coefficient functions of the initial ${\cal 
N}=4$ correlator can be read off from a few such hypermultiplet 
projections. The advantage of this ${\cal N}=2$ approach is that the 
explicit form of the corresponding five-point nilpotent covariants is 
much simpler. A further, and even more important feature of the 
${\cal N}=2$ formalism, is the possibility to formulate both ingredients 
of the ${\cal N}=4$ theory (${\cal N}=2$ SYM and hypermultiplets) 
{\it off shell} and to develop a straightforward quantisation scheme 
\cite{Galperin:bj}. Thus, formal manipulations like the insertion of the 
action (\ref{Intr}) become well justified.

The simplest case $k_i=2$, i.e., the four-point correlator of 
stress-tensor multiplets, has been treated in detail in Refs. 
\cite{hssw,4pt',EPSS} in the ${\cal N}=2$ framework (see also Appendix 
\ref{-A} for a short summary). The important result is that this 
correlator is determined by a single function of the conformal 
cross-ratios, and not by two functions, as predicted by the general 
analysis in Section \ref{general}. Conversely, this ${\cal N}=2$ result 
can be translated back in terms of the ${\cal N} = 4$ insertion 
procedure. In order to be consistent with the ${\cal N}=2$ analysis, the 
relevant term in the corresponding five-point nilpotent covariant in 
(\ref{Intr}) must have the following general form (after setting 
$\theta_1=\ldots=\theta_4=0$):
\begin{equation}\label{facn1}
\langle{\cal O}^{(2)}(0){\cal O}^{(2)}{\cal O}^{(2)}{\cal O}^{(2)}{\cal
O}^{(2)} \rangle = R^{2222}\; (\theta_0)^4\; F(x_0,\ldots,x_4)
\end{equation}
with
\begin{eqnarray}
\hskip -1cm R^{2222}&\ea & \frac{(12)^2\, (34)^2}{x^4_{12}\, x^4_{34}}\ s +
 \frac{(13)^2(24)^2}{x^4_{13}\, x^4_{24}} +
  \frac{(14)^2(23)^2}{x^4_{14}\, x^4_{23}}\ t  +
  \frac{(13)(14)(23)(24)}{x^2_{13}\, x^2_{14}\, x^2_{23}\, x^2_{24}}\ (s-t-1)
\nonumber\\
 &&{}+ \frac{(12)(14)(23)(34)}{x^2_{12}\, x^2_{14}\, x^2_{23}\, x^2_{34}}\
(1-s-t)
 + \frac{(12)(13)(24)(34)}{x^2_{12}\, x^2_{13}\, x^2_{24}\, x^2_{34}}\
  (t-s-1)   \nn \\
&\ea & \frac{1}{x^2_{13}\, x^2_{24}} \left[\frac{(12)^2(34)^2}{x^2_{12}\,
x^2_{34}} + \frac{(13)(14)(23)(24)}
{x^2_{13}\, x^2_{14}\, x^2_{23}\, x^2_{24}}\ (x^2_{12}\, x^2_{34} 
- x^2_{14}\, x^2_{23} - x^2_{13}\, x^2_{24}) + 
\mbox{cycle}   \right] \label{R} .  
\end{eqnarray}
Note that the second form of the prefactor $R^{2222}$ is manifestly 
cyclic (i.e., crossing) symmetric, apart from the factor 
$1/x^2_{13}x^2_{24}$. This results in the following properties of 
$R^{2222}$ under point permutations:
\begin{eqnarray}
1\leftrightarrow 2: && R^{2222} \ \rightarrow \ \frac{1}{t}\; R^{2222}
\, , \nonumber\\ 
1\leftrightarrow 3: && R^{2222} \ \rightarrow \ R^{2222}\;. 
\label{crossR}
\end{eqnarray}

It is important to realize that the prefactor $R^{2222}$ does not depend 
on the coordinates, in particular, on the harmonics at the insertion 
point, so it is an SO(6) singlet at this point. At the same time, 
$R^{2222}$ has absorbed the entire harmonic dependence (i.e., the SO(6) 
irreps) at points 1 to 4, while the harmonic dependence at point 0 is 
contained in the chiral Grassmann factor $(\theta_0)^4$. Consequently, 
the only part of the amplitude (\ref{facn1}) which is not fixed by 
superconformal invariance, the function $F(x_0,\ldots,x_4)$, is not 
allowed to depend on any of the harmonics (in an analytic, i.e., 
polynomial fashion). Thus, this coefficient function is an SO(6) 
singlet.

Substituting (\ref{facn1}) in (\ref{Intr}) and integrating over 
$\theta_0$ (the Grassmann superaction measure in (\ref{Intr}) exactly 
matches the nilpotent factor $(\theta_0)^4$ in (\ref{facn1})) and over 
$x_0$ at the insertion point, we obtain the allowed form of the quantum 
part of the four-point correlator:
\begin{equation}\label{pred}
  \frac{\pa}{\pa g^2} \langle{\cal O}^{(k_1)}{\cal O}^{(k_2)}{\cal O}^{(k_3)}{\cal
O}^{(k_4)} \rangle = R^{2222}\; {\cal F}(s,t)\;.
\end{equation}
Here the four-point conformal invariant
\begin{equation}\label{F}
  {\cal F}(s,t) = \int {\rm d}^4x_0\; F(x_0,\ldots,x_4)\;
\end{equation}
satisfies the crossing symmetry conditions
\begin{equation}\label{crs}
  {\cal F}(s,t) = {\cal F}(t,s) = 1/t\; {\cal F}(s/t,1/t)\;,
\end{equation}
as it easily follows from (\ref{crossR}).

The prediction of conformal supersymmetry (kinematics) combined with the 
insertion formula (dynamics) consists in fixing the relative 
coefficients of all six terms in (\ref{R}), leaving undetermined the 
single conformal invariant ${\cal F}(s,t)$. This should be compared to 
the two arbitrary functions allowed by the general arguments in Section 
\ref{general} for $k=2$, see (\ref{befcro2}), (\ref{cr1}) and 
(\ref{cr2}). The single function ${\cal F}(s,t)$ encodes the dynamical 
information about the quantum part of the correlator. This phenomenon 
was revealed in Ref. \cite{EPSS} and was called ``partial 
non-renormalisation". The predicted form has been confirmed by a number 
of explicit calculations: perturbative at order $g^2$ 
\cite{EHSSW}-\cite{BKRS2} and $g^4$ \cite{ESS},\cite{BKRS4}, instanton 
\cite{BGKR} and AdS supergravity \cite{AF3},\cite{EPSS}.

The generalisation of the ${\cal N}=4$ $\rightarrow$ ${\cal N}=2$ 
projection procedure to arbitrary $k_i\geq 2$ becomes cumbersome, due to 
the increasing number of terms in the tensor product $[0,k_i,0] \times 
[0,k_j,0]$ and to the large amount of linear algebra needed to 
reconstruct the ${\cal N}=4$ amplitude from its ${\cal N}=2$ 
projections.\footnote{For sufficiently low values of $k$ this is still 
doable, as we show in Appendix \ref{-A} for $k=3$.} Here we choose an 
indirect way which efficiently exploits the knowledge from the simplest 
case $k_i=2$. We assume that the five-point nilpotent covariant in 
(\ref{Intr}) can always be factorised into a ``kernel" with $k_i=2$ and 
an extra factor carrying the rest of the SO(6) quantum numbers 
$k'_i=k_i-2$ at each point:
\begin{equation}\label{facn2}
  \langle{\cal O}^{(2)}(0){\cal O}^{(k_1)}{\cal O}^{(k_2)}{\cal O}^{(k_3)}{\cal
O}^{(k_4)} \rangle = R^{2222}\; (\theta_0)^4\; 
F^{k'_1k'_2k'_3k'_4}(x_0,\ldots,x_4; u_1,\ldots,u_4)\;.
\end{equation}
The main difference from (\ref{facn1}) is that now the factor 
$F^{k'_1k'_2k'_3k'_4}$ depends on the harmonic variables 
$u_1,\ldots,u_4$, i.e., it is not an SO(6) singlet at points 1 to 4. The 
Grassmann factor $(\theta_0)^4$ still carries the full harmonic 
dependence at the insertion point. Thus, the functional freedom in 
(\ref{facn2}) and, after integration over the insertion point, in the 
quantum correction (\ref{Intr}) is determined by the SO(6) structure of 
the factor $F^{k'_1k'_2k'_3k'_4}$.

Take for example arbitrary $k_1$, $k_2$ and $k_3$, but keep $k_4=2$. The 
factor $F^{k'_1k'_2k'_3\; 0}$ still contains a unique harmonic 
structure:
\begin{equation}\label{3322}
  F^{k'_1k'_2k'_3\; 0} = 
\left(\frac{(12)}{x^{2}_{12}}\right)^{\!\frac{1}{2}(k_1+k_2-k_3-2)}
\left(\frac{(13)}{x^{2}_{13}}\right)^{\!\frac{1}{2}(k_1+k_3-k_2-2)}
\left(\frac{(23)}{x^{2}_{23}}\right)^{\!\frac{1}{2}(k_2+k_3-k_1-2)}\; 
\!\!\!\! f(x_0,\ldots,x_4)\;,
\end{equation}
where $f$ is an arbitrary SO(6) singlet function. After integration over 
$\theta_0,x_0$ we again find that the amplitude is determined by one 
function, $\phi(s,t) = \int \rmd^4x_0\; f(x_0,\ldots,x_4)$, just like in 
(\ref{F}).

Next we move to the main case of interest in this paper, $k_i=3\ 
\rightarrow \ k'_i=1$. According to Section \ref{general}, the factor 
$F^{1111}$ contains three terms:
\begin{equation}\label{3333}
  F^{1111} =  \frac{(12)(34)}{x^2_{12}\, x^2_{34}}\; \alpha(x_0,\ldots,x_4) +
  \frac{(14)(23)}{x^2_{14}\, x^2_{23}}\; \beta(x_0,\ldots,x_4) +
  \frac{(13)(24)}{x^2_{13}\, x^2_{24}}\; \gamma(x_0,\ldots,x_4) \;.
\end{equation}
After integration over the insertion point $x_0$ the coefficients in 
(\ref{3333}) give rise to four-point conformal invariants, $\alpha(s,t), 
\ \beta(s,t), \ \gamma(s,t)$. Inserting all this back into (\ref{facn2}) 
and then into (\ref{Intr}) and using the explicit form (\ref{R}) of 
$R^{2222}$, we obtain the following {\it factorised form} of the quantum 
part of the four-point correlator: 
\bea\label{qu3} 
&& \frac{\pa}{\pa g^2} 
\langle{\cal O}^{(3)}{\cal O}^{(3)}{\cal O}^{(3)}{\cal O}^{(3)} \rangle 
= R^{2222} F^{1111} \nonumber\\
&& \quad {}= 
\left[\frac{(12)^2\, (34)^2}{x^4_{12}\, x^4_{34}}\ s +
 \frac{(13)^2\,(24)^2} {x^4_{13}\, x^4_{24}} +
  \frac{(14)^2\, (23)^2} {x^4_{14}\, x^4_{23}}\ t  +
  \frac{(13)(14)(23)(24)}{x^2_{13}\, x^2_{14}\, x^2_{23}\, x^2_{24}}\ (s-t-1)
   \right. \nonumber\\
&& \quad \qquad \left.{} + \frac{(12)(14)(23)(34)}
{x^2_{12}\, x^2_{14}\, x^2_{23}\, x^2_{34}}\ 
(1-s-t) + \frac{(12)(13)(24)(34)}{x^2_{12}\, x^2_{13}\, x^2_{24}\, x^2_{34}}\
  (t-s-1)   \right]  \nonumber\\
&&\qquad \times \left[\frac{(12)(34)}{x^2_{12}x^2_{34}}\; \alpha +
  \frac{(14)(23)}{x^2_{14}x^2_{23}}\; \beta +
  \frac{(13)(24)}{x^2_{13}x^2_{24}}\; \gamma  \right] \;.
\eea 
Doing the multiplication and bringing the result to the standard 
form (\ref{g4pt}), we may express the ten coefficients $a,b,c$ in terms of 
only three functions:
\begin{equation}\label{ai}
  a_1 = s\alpha(s,t)\,, \qquad a_2 = \gamma(s,t)\,, \qquad a_3= t\beta(s,t)\,;
\end{equation}
\begin{eqnarray}
  && b_1 = s\gamma + (t-s-1)\alpha \, , \nonumber\\
  && b_2 = s\beta +(1-s-t)\alpha \, ,\nonumber\\
  && b_3 = (s-t-1)\beta + t\gamma \, ,\nonumber\\
  && b_4 = t\alpha + (1-s-t)\beta \, ,\nonumber\\
  && b_5 = \alpha + (t-s-1)\gamma \, ,\nonumber\\
  && b_6 = \beta + (s-t-1) \gamma\,; \label{bi}
\end{eqnarray}
\begin{equation}\label{c}
  c=(s-t-1)\alpha + (t-s-1)\beta + (1-s-t)\gamma \,.
\end{equation}

We still have to impose the crossing symmetry condition on the 
correlator (\ref{facn2}) with respect to points $1,\ldots,4$. Taking 
into account the property (\ref{crossR}) of the prefactor $R^{2222}$, it 
is easy to derive that the coefficients $\alpha(s,t), \ \beta(s,t), \ 
\gamma(s,t)$ are not independent:
\begin{equation}\label{relabc}
  \alpha(s,t) = 1/s\; \gamma(t/s,1/s)\,, \qquad \beta(s,t) =
  {1}/{t}\;\gamma(s/t,1/t)\,, \qquad \gamma(s,t) = \gamma(t,s)\,.
\end{equation}
Thus, all the ten coefficients in the quantum part of the amplitude 
(\ref{g4pt}) are expressed in terms of a {\it single 
function}, for instance,  $\a(s,t)$ obeying the symmetry relation 
$\a(s/t,1/t)=t\a(s,t)$. This is the content of the partial 
non-renormalisation theorem for the correlator of four $\half$-BPS operators 
of weight 3.

Clearly, the same analysis can be carried out for correlators with 
arbitrary weights $k_i$. The effect of the insertion procedure is to 
reduce the weight at each point by two units, $k_i \ \rightarrow \ 
k'_i=k_i-2$. The resulting four-point object $F^{k'_1k'_2k'_3k'_4}$ 
depends on as many functions as predicted by its SO(6) structure (and 
eventually by crossing symmetry). For example, if $k_i=4$ initially the 
correlator can depend on four functions (see the end of Section 
\ref{general}). However, in this case $k'_i=2$ and we know from Section 
\ref{general} that the fully crossing symmetric factor $F^{2222}$ only 
involves two functions. The conclusion is that the quantum part of the 
weight four correlator can only involve two independent functions.

Concluding this section, we mention that a somewhat different approach 
presented in the recent paper \cite{Heslop:2002hp} leads to the same 
number of independent functions in an amplitude of weights $k_1=\ldots = 
k_4 = k$. Instead of applying directly the insertion procedure to the 
four-point correlators, the authors rely on their earlier result 
\cite{HH2} on the non-renormalisation of three-point functions of all kinds 
of protected (BPS or semishort) operators. Combining this with double-OPE 
arguments of the type developed in Ref. \cite{DO}, they are able to 
predict the same general form of the amplitude. It should however be 
stressed that the non-renormalisation theorem of Ref. \cite{HH2} is also 
based on the insertion procedure, but this time at the level of 
three-point rather than four-point functions. Another important remark 
is that this non-renormalisation theorem {\it cannot be directly tested 
in supergravity} simply because the supergravity spectrum has only fields 
dual to $\half$-BPS short operators.

\setcounter{equation}0
\section{Supergravity-induced four-point amplitude}
\label{amplitude} In this section we provide a novel example of a 
conformal four-point amplitude induced by type IIB supergravity on an 
$AdS_5\times S^5$ background. So far the knowledge of the 
strongly-coupled four-point functions \cite{DHFMMR,AF3} has been limited 
to those of the stress-tensor multiplet dual to the field content of 
gauged ${\cal N}=8$ supergravity. We thus provide for the first time a 
four-point amplitude for operators corresponding to higher KK states and 
verify its compatibility with the field-theoretic predictions.

The computation of the supergravity-induced correlation functions 
proceeds in the standard way. One first evaluates the on-shell gravity 
action $S$ with Dirichlet boundary conditions on the fields; then, 
varying the generating functional $Z=e^{-S}$ with respect to the 
boundary data one obtains correlation functions of the boundary CFT. In 
particular, the derivation of the four-point amplitude requires the 
knowledge of the effective supergravity action on $AdS_5$ up to fourth 
order in the fields. The quadratic and cubic supergravity terms needed 
to compute the four-point function of {\it arbitrary $\half$-BPS operators} 
were found in \cite{LMRS}-\cite{Lee}. Their contribution to the extrema 
of the supergravity action is interpreted as the AdS exchange graphs. 
The most difficult part of this program is however to obtain the quartic 
effective action corresponding to contact interactions. This formidable 
problem was solved in \cite{AF2}, where it was found in particular that 
the quartic Lagrangian contains derivative interactions (up to four 
derivatives).

The $\half$-BPS operator ${\cal O}^I$ in the irrep $[0,3,0]$ (or {\bf 50}) 
is dual to the gravity scalar $s^I$ with AdS mass $m^2=-3$. The relevant 
part of the effective five-dimensional action from \cite{AF2} is given 
by 
\bea S=\frac{N^2}{8\pi^2}\int {\rm d}^5 z \sqrt{g_a} \left( {\cal L}_2
+{\cal L}_3+{\cal L}_4 \right) \, , \eea 
i.e. it is a sum of quadratic, cubic 
and quartic terms. Here $g_a$ is the determinant of the Euclidean AdS 
metric $\rmd s^2=\frac{1}{z_0^2}(\rmd z_0^2+ \rmd x_a \rmd x_a)$, $a=1,2,3,4$.

To distinguish supergravity scalars and vectors belonging to different 
SO(6) representations we introduce the subscript $k$, i.e. $s_k^I$ 
(irrep $[0,k,0]$), $A_{\mu,k}^I$ (irrep $[1,k-1,1]$, $k$ odd). It is 
related to the conformal dimension of the corresponding CFT operator as 
$\Delta=k$ for scalars and $\Delta=k+2$ for vectors. Then the Lagrangian 
involves the scalar fields $s^I_{ 2}$, $s^I_{ 3}$ and $s^I_{ 4}$, as 
well as two vectors $A^I_{{\mu},{ 1}}$, $A^I_{{\mu},{ 3}}$, the graviton 
$\phi_{\mu\nu}$ and a massive symmetric tensor $\varphi_{\mu\nu}^I$ 
transforming in the irrep $[0,2,0]$ of SO(6). Its quadratic part is 
normalised as follows\footnote{To simplify the action we have performed 
suitable rescalings of the fields.} 
\bea \nonumber {\cal 
L}_2&\ea &\frac{1}{4}\left(\nabla_\mu s_{ 2}^1\nabla^\mu s_{ 2}^1 -4 s_{ 
2}^1s_{ 2}^1 \right) +\frac{1}{4}\left(\nabla_\mu s_{ 3}^1\nabla^\mu s_{ 
3}^1 -3s_{ 3}^1s_{ 3}^1 \right)
+\frac{3}{2}\nabla_\mu  s_{ 4}^1\nabla^\mu s_{ 4}^1 \\
\nonumber
&&{}+ \frac{1}{2}(F^1_{\mu\nu,{ 1}})^2+\frac{1}{2}(F^1_{\mu\nu,{ 3}})^2+
8(A^1_{\mu,{ 3}})^2\\
\nonumber 
&&{}+ \frac{1}{4}\nabla_{\rho}\phi_{\mu\nu}\nabla^{\rho}\phi^{\mu\nu} 
-\frac{1}{2}\nabla_{\mu}\phi_{\mu\rho}\nabla^{\nu}\phi_{\nu\rho} 
+\frac{1}{2}\nabla_{\mu}\phi_{\rho}^{\rho}\nabla_{\nu}\phi^{\mu\nu}
-\frac{1}{4}\nabla_{\rho}\phi_{\mu}^{\mu}\nabla^{\rho}\phi^{\nu}_{\nu}\\
\nonumber &&{}- \frac{1}{2}\phi_{\mu\nu}\phi^{\mu\nu}
+\frac{1}{2}(\phi_{\nu}^{\nu})^2\\
\nonumber &&{}+ 6\Big[ 
\frac{1}{4}\nabla_{\rho}\varphi_{\mu\nu}^1\nabla^{\rho}\varphi^{\mu\nu~1} 
-\frac{1}{2}\nabla_{\mu}\varphi_{\mu\rho}^1\nabla^{\nu}\varphi_{\nu\rho}^1 
+\frac{1}{2}\nabla_{\mu}\varphi_{\rho}^{\rho~1}\nabla_{\nu}\varphi^{\mu\nu~1}
-\frac{1}{4}\nabla_{\rho}\varphi_{\mu}^{\mu~1}\nabla^{\rho}
\varphi^{\nu~1}_{\nu}\\
&&\quad {}+ \frac{5}{2}\varphi_{\mu\nu}^1\varphi^{\mu\nu~1} 
-\frac{7}{2}(\varphi_{\nu}^{\nu~1})^2 \Big] \, , \eea 
(summation over the repeated superscript indices is implied).

To write down the cubic Lagrangian we introduce the so-called C-tensors. 
They are the Clebsh-Gordon coefficients for tensor products of different 
SO(6) irreps and describe the cubic interactions of various supergravity 
fields. For all necessary definitions and the properties of the 
C-tensors we refer the reader to Appendix \ref{A}, where in particular 
the summation formulae which we term as ``C-algebra'' are established. 
Explicitly, the cubic Lagrangian is 
\bea \nonumber {\cal 
L}_3&\ea&-3\langle C^1C^2C^3_{[0,2,0]}\rangle s^1_{ 3}s^2_{ 3}s^3_{ 2} -
18\langle C^1C^2C^3_{[0,4,0]}\rangle s^1_{ 3}s^2_{ 3}s^3_{ 4} \\
&&{}- \frac{1}{4}\Big( \nabla^\mu s^1_{ 3} \nabla^\nu s^1_{ 
3}\;\phi_{\mu\nu}-\frac{1}{2}\left( \nabla^\mu s^1_3 \nabla_\mu s^1_3 
-3s^1_3s^1_3 \Big )\phi_\nu^\nu
\right) \nn \\
\nonumber &&{}- \frac{3}{4} \langle C^1C^2C^3_{[0,2,0]}\rangle \Big ( 
\nabla^\mu s^1_{ 3} \nabla^\nu s^2_{ 3} 
\;\varphi_{\mu\nu}^3-\frac{1}{2}\left( \nabla^\mu s^1_3 \nabla_\mu s^2_3 
-9s^1_3s^2_3 \right)\varphi_\nu^{\nu~3}
\Big )\\
&&{}- \frac{3}{2}\langle C^1C^2C^3_{[1,0,1]}\rangle s^1_{ 
3}\nabla^{\mu} s^2_{ 3} A^3_{\mu,{ 1}} -3\langle 
C^1C^2C^3_{[1,2,1]}\rangle s^1_{ 3}\nabla^{\mu} s^2_{ 3} A^3_{\mu,{ 3}} 
\, . \label{cc} \eea

{}From this Lagrangian we deduce that the interactions of the scalar 
field $s_{ 3}$ are mediated by two neighboring multiplets: one of them 
is the massless graviton multiplet $(s^I_{ 2},A_{\mu,{ 
1}}^I,\phi_{\mu\nu})$ which includes the $m^2=-4$ scalar, the massless 
vector and the graviton and the massive multiplet $(s^I_{ 4},A_{\mu,{ 
3}}^I,\varphi_{\mu\nu}^I)$ whose lowest component is the massless scalar 
$s_{ 4}$.

Since the leading (i.e. most singular) and the subleading terms in the 
OPE are determined by the three-point functions derived from the cubic 
Lagrangian, the cubic couplings (\ref{cc}) are responsible for these 
contributions to the supergravity-induced OPE. Thus, in the leading and 
the subleading terms in the OPE we expect the following irreps to 
appear: leading $[0,0,0]$, $[1,0,1]$ and [0,2,0] corresponding to the 
stress tensor, the conserved current and a $\half$-BPS operator with 
conformal dimension $\Delta=2$; subleading $[0,4,0]$, $[1,2,1]$ and 
$[0,2,0]$ corresponding to a $\half$-BPS operator with $\Delta=4$, and 
higher vector and tensor currents, respectively. In principle, one could 
also expect the appearance of a cubic interaction with a $\half$-BPS 
multiplet whose lowest component is a scalar of $\Delta=6$. However, the 
corresponding coupling is then extremal and therefore it must vanish 
\cite{D'Hoker, AF4}. The representations which do not appear among the 
cubic couplings correspond, in free-theory terms, to the contribution of 
the double-trace operators.


\begin{picture}(30000,15000)
\drawline\scalar[\W\REG](15000,5000)[4] \global\advance\pmidx by -2500 
\global\advance\pmidy by 500 
\put(\pmidx,\pmidy){$s_2^I,A_{\mu}^I,\phi_{\mu\nu}$}
\drawline\fermion[\SW\REG](\particlebackx,\particlebacky)[4000] 
\global\advance\pbackx by -1200 \put(\pbackx,\pbacky){$s_3^{I_2}$}
\drawline\fermion[\NW\REG](\scalarbackx,\scalarbacky)[4000] 
\global\advance\pbackx by -1200 \put(\pbackx,\pbacky){$s_3^{I_1}$}
\drawline\fermion[\SE\REG](\scalarfrontx,\scalarfronty)[4000] 
\global\advance\pbackx by 300 \put(\pbackx,\pbacky){$s_3^{I_3}$}
\drawline\fermion[\NE\REG](\scalarfrontx,\scalarfronty)[4000] 
\global\advance\pbackx by 300 \put(\pbackx,\pbacky){$s_3^{I_3}$}
\drawline\scalar[\W\REG](35000,5000)[4] \global\advance\pmidx by -2500 
\global\advance\pmidy by 500 
\put(\pmidx,\pmidy){$s_4^I,A_{\mu}^I,\varphi_{\mu\nu}^I$}
\drawline\fermion[\SW\REG](\particlebackx,\particlebacky)[4000] 
\global\advance\pbackx by -1200 \put(\pbackx,\pbacky){$s_3^{I_2}$}
\drawline\fermion[\NW\REG](\scalarbackx,\scalarbacky)[4000] 
\global\advance\pbackx by -1200 \put(\pbackx,\pbacky){$s_3^{I_1}$}
\drawline\fermion[\SE\REG](\scalarfrontx,\scalarfronty)[4000] 
\global\advance\pbackx by 300 \put(\pbackx,\pbacky){$s_3^{I_3}$}
\drawline\fermion[\NE\REG](\scalarfrontx,\scalarfronty)[4000] 
\global\advance\pbackx by 300 \put(\pbackx,\pbacky){$s_3^{I_3}$}
\end{picture}
\begin{center}
\begin{tabular}{l}
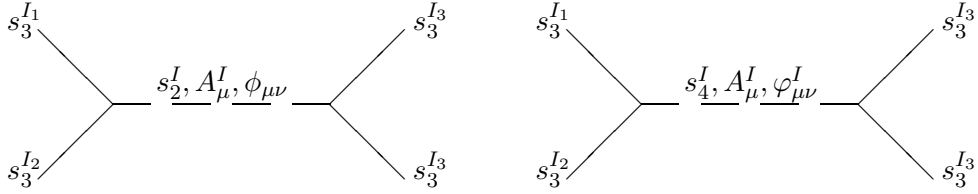
 Figure 4: Exchange graphs contributing to the four-point function of 
operators ${\cal O}^I$.  \\
The first graph represents the exchange of fields from the massless 
graviton multiplet:\\ the scalar $s_2$ in the {\bf 20}, the vector 
$A_{\mu}$ in the {\bf 15} and the graviton $\phi_{\mu\nu}$.
The other \\ graph involves a massive multiplet whose lowest scalar component 
$s_4$ is in the {\bf 105}. \\
These graphs are responsible for the leading and the subleading
terms in  the OPE.
\end{tabular}
\end{center}

\vskip 0.5cm

One of the most involved parts of our computation is to extract the 
relevant contact interactions from the general quartic action of Ref. 
\cite{AF2}. This action contains the quartic terms with two and four 
derivatives as well as terms without derivatives and for generic fields 
$s^I$ looks pretty complicated (see Appendix A of Ref. \cite{AF2}). 
However, by heavy usage of the C-algebra, the symmetry properties of the 
derivative vertices and integration by parts (see Appendix \ref{B} for 
details) we reduced it to the remarkably simple expression \bea 
\label{cont} {\cal 
L}_{4}=\left(\frac{3}{2}S^{1234}+\frac{39}{32}C^{1234}-\frac{3}{16}\d^{12}\d^{34} 
\right)s^1s^2s^3s^4 -\frac{3}{32}C^{1234}s^1\nabla_\mu s^2 \nabla^\mu 
s^3 s^4 \, . \eea It is worthwhile noting that the quartic terms with 
four derivatives completely disappear, i.e. {\it the final action is of 
the sigma model type}. This suggests the interesting possibility that 
the quartic four-derivative Lagrangian of Ref. \cite{AF2} is 
intrinsically zero, although this is not seen for generic fields, i.e. 
without specifying the explicit representation content. If true, this 
might imply that the extension of the the five-dimensional gauged ${\cal 
N}=8$ supergravity by including the massive KK modes of the type IIB 
supergravity compactification is described by some sigma model.

Having established the relevant supergravity Lagrangian we can now 
proceed to the evaluation of the AdS exchange graphs. In comparison to 
other calculations existing in the literature \cite{DHFMMR,AF3,AS}, a 
novelty is the appearance of a massive symmetric tensor. The 
corresponding exchange graph can be computed by generalizing the method 
of Ref. \cite{dHFR} (see Appendix \ref{D}). Omitting the details of the 
tedious calculations, here we present the final result for the 
supergravity-induced four-point function of the canonically normalised 
$\half$-BPS operators of dimension 3: 
\bea &&\langle {\cal O}^{1}(x_1)...{\cal O}^{4}(x_4)\rangle =
\frac{\d^{12}\d^{34}}{x_{12}^6\, x_{34}^6} \nn \\
&&~~~~+\frac{72}{\pi^2N^2}\Big[S^{1234}A^{S}_{1234}+ 
\d^{12}\d^{34}A^{\d}_{1234}+C^{1234}A^C_{1234}+C^{1243}A^C_{1243}\Big]+t+u  \, .
\eea 
Here we explicitly exhibit only the expression in the $s$-channel 
as the $t$-channel is obtained by replacing $1\leftrightarrow 4$, and 
the $u$-channel by $1\leftrightarrow 3$. The terms without $1/N^2$ in 
front represent the contribution of the disconnected AdS graphs. The 
coefficients $A_{1234}$ are expressed in terms of the $D$-functions (see 
Appendix \ref{C}) as follows \bea \label{s4pt} 
A^{S}_{1234}&\ea & 8\left(\frac{D_{3322}}{x_{34}^2}
-\frac{2}{3} D_{3333}\right) \, ,\nn \\
\nonumber A^{\d}_{1234} &\ea & \frac{3}{4} 
(x_{13}^2x_{24}^2+x_{14}^2x_{23}^2-x_{12}^2x_{34}^2)\left(\frac{D_{4433}}{x_{34}^2} 
+\frac{D_{4422}}{x_{34}^4}\right) -\frac{1}{2}\frac{D_{3311}}{x_{34}^4}
+\frac{5}{3}D_{3333}  \, ,\\ \nonumber 
A^{C}_{1234}&\ea& \frac{3}{4}
(x_{13}^2x_{24}^2+x_{14}^2x_{23}^2-x_{12}^2x_{34}^2) 
\frac{D_{4433}}{x_{34}^2} 
+\frac{D_{3311}}{x_{34}^4}-\frac{D_{3322}}{x_{34}^2} 
+2\frac{D_{3232}}{x_{24}^2}-3x_{14}^2D_{4334}\\
\nonumber &&-\frac{2}{x_{24}^2}\lb D_{3243}x_{34}^2-D_{3342}x_{23}^2 \rb
-\frac{1}{2x_{34}^2}\lb D_{3423}x_{24}^2-D_{3432}x_{23}^2 \rb  \\ &&
{} -\frac{3}{2x_{34}^4}\lb D_{3412}x_{24}^2-D_{3421}x_{23}^2 \rb \, . \eea

\setcounter{equation}0
\section{Verifying the CFT predictions}
\label{check} As argued in Section \ref{prediction}, the insertion 
formula, being a dynamical constraint on the theory, predicts that the 
general four-point amplitude (\ref{g4pt}) depends on one undetermined 
coefficient function $a_2(s,t)$, while the nine others are expressed in 
terms of it. Here we compare the supergravity-induced four-point 
amplitude obtained in the previous section to this field-theoretic 
prediction and indeed find an almost miraculous agreement.

Initially we express all $D$-functions entering (\ref{s4pt}) in terms of
corresponding functions $\oD$ of the invariants $s,t$ (for a detailed
definition see Appendix \ref{C}) so that the functions $c(s,t)$, $a_1(s,t)$ 
(the $1/N^2$-part) and $b_1(s,t)$ can be represented as follows
\bea \nonumber 
c(s,t)&\ea &\frac{2^4\cdot 9}{N^2}\, st
\Big[ -\frac{3}{2}{\oD}_{3333} + {\oD}_{2332} 
+ {\oD}_{3232} + {\oD}_{3322} \Big ] \, , \\  
a_1(s,t)&\ea &\frac{9}{2N^2}\, s^3\Big[ 
(1+t-s) \left( {\oD}_{4433} + {\oD}_{4422}\right) +
5 {\oD}_{3333} - {\oD}_{3311} \Big] \, ,  \nn \\
b_1(s,t)&\ea &\frac{9}{2N^2}\, s^2 \Big[ (1-s+t) {\oD}_{4433} -
4{\oD}_{4334}+t {\oD}_{3432} - {\oD}_{3423} \nn \\
&&{}+  4t {\oD}_{3342}-4 {\oD}_{3243}+
2t {\oD}_{3421}- 2{\oD}_{3412} + 8{\oD}_{3232}-
4{\oD}_{3322}+2 {\oD}_{3311} \Big ] \, .  \label{abcgr} \eea 
Our strategy is then similar to that in Refs. \cite{EPSS,AS}. 
For integer $\D_i$ and $\sum_i \D_i$ even we may express 
$\oD_{\D_1 \D_2 \D_3 \D_4}(s,t)$ in terms of  $s$- and 
$t$-derivatives of the standard four-dimensional one-loop box integral 
$\Phi(s,t)=\oD_{1111}(s,t)$. This representation is convenient since the 
derivatives $\pa_s\Phi(s,t)$ and $\pa_t\Phi(s,t)$ are expressed in a simple 
manner in terms of the function $\Phi(s,t)$ itself (see Appendix \ref{C}).
Using the notation $\overline{{\bf D}}_{\D_1\D_2\D_3\D_4}$ to denote
the appropriate differential operator we have
\bea
\oD_{\D_1\D_2\D_3\D_4}(s,t) =\overline{\bf D}_{\D_1\D_2\D_3\D_4}\Phi(s,t) \, .
\label{dPhi}
\eea
Substituting the explicit expressions 
for the $\overline{{\bf D}}$-operators from Appendix \ref{C} and successively 
using the identities (\ref{identity}), we express all the coefficient 
functions via $\Phi(s,t)$. Upon substitution of our findings into 
eqs.(\ref{bi}) and (\ref{c}), and use of the symmetry properties 
\bea 
\Phi(s,t)=\Phi(t,s) \, ,~~~~\Phi(s/t,1/t)=t\Phi(s,t)\, , 
~~~~\Phi(t/s,1/s)=s\Phi(s,t)\,, \eea we finally get 
\bea \label{amaz}
&&b_1-s\gamma-(t-s-1)\alpha=\frac{9}{N^2} \, , \nn \\
&&c-(s-t-1)\alpha -(t-s-1)\beta-(1-s-t)\gamma = \frac{18}{N^2} 
\, . \eea 
The r.h.s. of these formulae literally coincide with the free 
field theory values of the coefficients $b_1$ and $c$ in eq.(\ref{ff})! 
We thus observe the surprising property of the supergravity-induced 
four-point amplitude to split into a ``free'' and a ``quantum'' parts, 
precisely reproducing the form predicted by SCFT. Moreover, the 
numerical coefficients in the ``free" part of this amplitude are exactly 
those given by the free Wick contractions. What makes this so unexpected 
is that contrary to the SYM theory, the supergravity action has no 
coupling constant and therefore no natural separation of the free part 
from the interacting part. Yet, we observe the same splitting in the 
four-point amplitudes derived from this action. Thus, supergravity retains 
memory of the Lagrangian formulation of the gauge theory even with an 
infinite value of the 't Hooft coupling.
In the absence of an obvious explanation why the supergravity-induced 
amplitude has this particular form, we consider this fact as a strong 
evidence in favor of the AdS/CFT duality conjecture. At the same time, 
our result is a very non-trivial check of the effective action derived 
in \cite{AF2}.

Finally, we note that in fact the results for the coefficient functions
given in (\ref{abcgr})  can be drastically simplified by using  various 
identities for the $\oD$-functions. In Appendix \ref{C} we present an 
independent proof of eqs.(\ref{amaz})  based on the ${\overline D}$-algebra. 
In the process a surprisingly  simple expression for $a_1(s,t)$ emerges: 
\bea a_1(s,t) = - {9\over N^2} 
\lb s^2 {\overline D}_{3335}(s,t) + s {\overline D}_{2235}(s,t)\rb \, . 
\eea 
According to our analysis in Section \ref{prediction} this 
function can be taken as the unique function describing 
the supergravity-induced four-point amplitude of dimension 3 BPS 
operators. In the next Section we will use this representation to shed 
some light on the properties of the corresponding OPE.

\section{Operator Product Expansion}
\label{general} The operator product expansion of the stress-energy 
tensor multiplets in ${\cal N}=4$ theory and the $AdS_5$ supergravity is 
by now a well-developed subject \cite{DHMMR}-\cite{DO1},\cite{DO}. Here 
we will exhibit some general properties of the OPE underlying the 
supergravity four-point amplitude for our new example of dimension 3 
operators.

\subsection{OPE and anomalous dimensions of some long multiplets}
For the purposes of analyzing the operator product expansion we may 
expand the four-point function in the form \be \nonumber \langle {\cal 
O}^{1}(x_1){\cal O}^{2}(x_2){\cal O}^{3}(x_3){\cal O}^{4}(x_4) \rangle = 
\frac{1}{x_{12}^6\, x_{34}^6} \, \sum_{\cal J} A_{\cal J}(s,t) 
P^{1234}_{\cal J}\, , \end{equation}
and using the explicit formulae for the 
projectors $P^{1234}_{\cal J}$ from Appendix B we may write 
\bea
A_{[0,0,0]} & \ea {}& 50 a_1 + s^3 a_2 + {s^3\over t^3} a_3 + {25\over 3} \Big (
s b_1 + {s\over t} b_2 \Big ) + {5\over 2 } \Big ( s^2 b_5
+ {s^2\over t^2} b_4 \Big ) \nn \\
&&{} + {1\over 6}  \Big ( {s^3\over t} b_6
+ {s^3\over t^2} b_3 \Big ) + {5\over 6} \, {s^2\over t} c \, , \nn  \\
A_{[1,0,1]} & \ea  {}&  s^3 a_2 - {s^3\over t^3} a_3 + {35\over 9} \Big (
s b_1 - {s\over t} b_2 \Big ) + {35\over 18 } \Big ( s^2 b_5 -
{s^2\over t^2} b_4 \Big ) + {1\over 18}  \Big ( {s^3\over t} b_6
- {s^3\over t^2} b_3 \Big ) \, ,  \nn \\
A_{[0,2,0]} & \ea {}& s^3 a_2 + {s^3\over t^3} a_3 + {35\over 12} \Big (
s b_1 + {s\over t} b_2 \Big ) + {7 \over 4 } \Big ( s^2 b_5
+ {s^2\over t^2} b_4 \Big ) + {1\over 12}  \Big ( {s^3\over t} b_6
+ {s^3\over t^2} b_3 \Big ) + {7\over 24} \, {s^2\over t} c \, , \nn \\
A_{[2,0,2]} & \ea  {}& s^3 a_2 + {s^3\over t^3} a_3  + {10 \over 9} \Big ( s^2 b_5
+ {s^2\over t^2} b_4 \Big ) - {1\over 9}  \Big ( {s^3\over t} b_6
+ {s^3\over t^2} b_3 \Big ) - {5\over 9} \, {s^2\over t} c \, , \nn \\
A_{[0,4,0]} & \ea  {}& s^3 a_2 + {s^3\over t^3} a_3  + {7 \over 9} \Big ( s^2 b_5
+ {s^2\over t^2} b_4 \Big ) + {2\over 9}  \Big ( {s^3\over t} b_6
+ {s^3\over t^2} b_3 \Big ) + {7\over 9} \, {s^2\over t} c \, , \nn \\
A_{[1,2,1]} & \ea  {}& s^3 a_2 - {s^3\over t^3} a_3  + s^2 b_5
- {s^2\over t^2} b_4  \, , \nn \\
A_{[3,0,3]} & \ea  {}& s^3 a_2 - {s^3\over t^3} a_3  - {1\over 3}
\Big ( {s^3\over t} b_6 - {s^3\over t^2} b_3 \Big ) \, , \nn \\
A_{[0,6,0]} & \ea  {}& s^3 a_2 + {s^3\over t^3} a_3 +
{s^3\over t} b_6 + {s^3\over t^2} b_3 \, , \nn \\
A_{[2,2,2]} & \ea  {}& s^3 a_2 + {s^3\over t^3} a_3  - {1\over 9}
\Big ( {s^3\over t} b_6 + {s^3\over t^2} b_3 \Big ) \, , \nn \\
A_{[1,4,1]} & \ea  {}& s^3 a_2 - {s^3\over t^3} a_3  + {1\over 3}
\Big ( {s^3\over t} b_6 - {s^3\over t^2} b_3 \Big ) \, . 
\eea
Using the expressions for $a_i , b_i$ and $c$ in terms of $\alpha,\beta,\gamma$
this may be simplified. For the purposes here it is convenient to define
\be
\A =  {s\over t}\, \alpha  \, , \qquad
\B  = {s^2 \over t^2} ( t \gamma - \beta ) \, , \qquad
\C =  {s^2 \over t^2} ( t \gamma  + \beta  ) \, , 
\end{equation}
and also to separate the results into two parts
\be
A_{\cal J} = A^{\rm free}_{\cal J} + A^{\rm int.}_{\cal J} \, ,
\end{equation}
where the first term corresponds to the result in free field theory
\bea
A^{\rm free}_{[0,6,0]} & \ea {}& s^3 + {s^3\over t^3} + 
{9\over N^2} \Big ( {s^3 \over t} + {s^3 \over t^2} \Big ) \, , \nn \\
A^{\rm free}_{[1,4,1]} & \ea {}&  s^3 - {s^3\over t^3} 
+ {3\over N^2} \Big ( {s^3 \over t} - {s^3 \over t^2} \Big ) \, , \nn \\
A^{\rm free}_{[2,2,2]} & \ea {}&  s^3 + {s^3\over t^3} - 
{1\over N^2} \Big ( {s^3 \over t} + {s^3 \over t^2} \Big ) \, , \nn \\
A^{\rm free}_{[3,0,3]} & \ea {}&  s^3 - {s^3\over t^3} - 
{3\over N^2} \Big ( {s^3 \over t} - {s^3 \over t^2} \Big ) \, , \nn \\
A^{\rm free}_{[0,4,0]} & \ea {}&  s^3 + {s^3\over t^3} + 
{2\over N^2} \Big ( {s^3 \over t} + {s^3 \over t^2} \Big ) + 
{7\over N^2} s^2 \Big ( 1 + {1\over t} \Big )^2 \, ,\nn \\
A^{\rm free}_{[1,2,1]} & \ea {}&  s^3 - {s^3\over t^3} + 
{9\over N^2} \Big ( {s^2 \over t} - {s^2 \over t^2} \Big  )  \, , \nn \\
A^{\rm free}_{[2,0,2]} & \ea {}&  s^3 + {s^3\over t^3} 
- {1\over N^2} \Big ( {s^3 \over t}
+ {s^3 \over t^2} \Big ) + {10\over N^2} s^2 \Big ( 1 - {1\over t} +
{1\over t^2} \Big ) \, , \nn \\
A^{\rm free}_{[0,2,0]} & \ea {}&  s^3 + {s^3\over t^3} + 
{3\over 4N^2} \Big ( {s^3 \over t}
+ {s^3 \over t^2} \Big ) + {63\over 4N^2} s^2 \Big ( 1 + {1\over 3t} +
{1\over t^2} \Big ) + {105\over 4N^2} s \Big ( 1 + {1\over t} \Big ) \, , \nn \\
A^{\rm free}_{[1,0,1]} & \ea {}&  s^3 - {s^3\over t^3} + {1\over 2N^2} s 
\Big ( 1 - {1\over t}\Big )
\bigg ( 70 + 35 s \Big ( 1 + {1\over t}\Big ) + {s^2 \over t} \bigg ) \, , \nn \\
\hskip-0.2cm
A^{\rm free}_{[0,0,0]} & \ea {}&  50 + s^3 + {s^3\over t^3} + {3\over 2N^2} 
\Big ( {s^3 \over t}
+ {s^3 \over t^2} \Big ) + {45\over 2N^2} s^2 \Big ( 1 + {2\over 3t} +
{1\over t^2} \Big ) + {75\over N^2} s \Big ( 1 + {1\over t} \Big ) \, . 
\label{freep}
\eea
The remaining parts contain the essential dynamics
\bea
A^{\rm int.}_{[0,6,0]} & \ea {}& s^2  \C \, , \nn \\
A^{\rm int.}_{[1,4,1]} & \ea {}&  
- {\ts {2 \over 3}} (1-t)s\, \C  + {\ts{1 \over 3}} s^2 \B \, , \nn \\
A^{\rm int.}_{[2,2,2]} & \ea {}&  
{\ts{1\over 9}}\lb 5(1+t)-s \rb s\, \C
- {\ts{5 \over 9}}(1-t)s \, \B \, , \nn \\
A^{\rm int.}_{[3,0,3]} & \ea {}&  
- {\ts{1 \over 3}} (1-t)s\, \C
+ {\ts{1 \over 3}}\lb 3(1+t) - s \rb s \, \B \, , \nn \\
A^{\rm int.}_{[0,4,0]} & \ea {}&  
{\ts{1 \over 18}}\lb 14(1-t)^2 - 7s(1+t) +4s^2 \rb \C
- {\ts{7 \over 18}}(1-t)s \, \B + {\ts{7 \over 9}}s^2 \A \, , \nn \\
A^{\rm int.}_{[1,2,1]} & \ea {}&  
- {\ts{1 \over 2}} (1-t^2) \C
+ {\ts{1 \over 2}} (1-t)^2 \B - (1-t)s \, \A\, , \nn \\
A^{\rm int.}_{[2,0,2]} & \ea {}&  
{\ts{1 \over 18}}\lb 5(1-t)^2 + 5 s(1+t) - 2s^2 \rb \C
- {\ts{5 \over 18}}\lb 3(1+t) - s \rb (1-t) \B  \nn \\
&& {} + {\ts{ 5 \over 9}}\lb 3(1+t) - s \rb s \, \A \, , \nn \\
A^{\rm int.}_{[0,2,0]} & \ea {}&  
{\ts{1 \over 48}}\lb 35 (1 + t)^2 -14 (1-t)^2 - 27 s(1+t) +4 s^2 \rb
\C \nn \\
&&{} - {\ts{1 \over 48 }}\lb 35 (1+t) - 13 s \rb (1-t) \B
+ {\ts{7 \over 24}}\lb 10(1-t)^2 - 5s(1+t) + s^2 \rb \A \, , \nn \\
A^{\rm int.}_{[1,0,1]} & \ea {}&  
{} - {\ts{1 \over 36}} \lb 35(1+t) - 16s \rb (1-t) \C
+ {\ts{1 \over 36}} \lb 35(1+t)^2 - 20s(1+t) + 2s^2 \rb \B \nn \\
&&{} - {\ts{ 35\over 18}}\lb 2(1+t) -s \rb (1-t) \A\, , \nn \\
A^{\rm int.}_{[0,0,0]} & \ea {}&  
{\ts{1 \over 12}}\lb 25(1 + t)^2  - 5(1-t)^2 - 15 s(1+t) + 2 s^2 \rb \C
- {\ts{5 \over 12}}\lb 2 (1+t) -  s \rb (1-t) \B  \nn \\
&&{}+ {\ts{5 \over 6}}\lb 15(1+t)^2 - 5(1-t)^2 - 8 s(1+t) + s^2 \rb
\A \, .
\label{Aint}\eea

For the operator product expansion we expand $\A, \B$ and $\C$ in the form
\be
\sum_{\Delta,\ell} a_{\Delta,\ell} \,
s^{{1\over 2}(\Delta - \ell)} G^{(\ell)}_{\Delta+4} (s,t) 
\end{equation}
which represents them each as a sum of  contributions of  operators 
of scale dimension $\Delta$
and spin $\ell$, belonging to a $({1\over 2}\ell,{1\over 2}\ell)$ spin
representation. Explicit results for $G^{(\ell)}_{\Delta} (s,t)$ are
known but here we note the important recurrence relations from \cite{DO},
\bea
- {\ts {1\over 2}} (1-t)\, G^{(\ell)}_{\Delta} (s,t)
& \ea &
G^{(\ell+1)}_{\Delta-1} (s,t) + {\ts{1\over 4}}s G^{(\ell-1)}_{\Delta-1} (s,t)
+ {\ts{1\over 4}}f(\Delta+\ell) \,s G^{(\ell+1)}_{\Delta+1} (s,t) \nn \\
&&{}+ {\ts{1\over 16}}f(\Delta-\ell-2)\, s^2 G^{(\ell-1)}_{\Delta+1}(s,t)\, ,\nn \\
{\ts {1\over 2}} (1+t)\, G^{(\ell)}_{\Delta} (s,t)
& \ea & 
G^{(\ell)}_{\Delta-2} (s,t) + f(\Delta+\ell) \,s G^{(\ell+2)}_{\Delta} (s,t)
+{\ts{1\over 4}}s G^{(\ell)}_{\Delta} (s,t)  \nn \\
&&{} + {\ts{1\over 16}}f(\Delta-\ell-2) \,s^2 G^{(\ell-2)}_{\Delta} (s,t) \nn \\
&&{} + {\ts{1\over 4}}f(\Delta+\ell)f(\Delta-\ell-2)\, 
s^2 G^{(\ell)}_{\Delta+2}(s,t)\, ,
\label{recur}
\eea
for $f(\lambda)=\frac{1}{4}\lambda^2/(\lambda^2-1)$.
If we consider the contribution of a single term
in the expansion of $\A, \B$ then using this in the above generates
associated contributions in $A_{\cal J}$ for all operators expected for
a long multiplet whose lowest dimension operator have scale dimension 
$\Delta$ and spin $\ell$ and which belong to the $[0,0,0], \ [1,0,1]$
representations respectively. In each case the expected representations
with the appropriate $\Delta$ and spin $\ell$ arise, for  the
$[0,0,0]$ and $[1,0,1]$  cases the scale dimension varies
from $\Delta$ to $\Delta+8$ and for scale dimension $\Delta+4$ the spin
varies from $\ell-4$ to $\ell+4$ as expected. For $\C$ we would find
that the set of operators contributing to the $[0,0,0]$ representation
is not in accord with that expected for an operator based on the $[0,2,0]$
representation. This problem is easily cured by setting 
\be
\A = \A' - {\ts {1\over 6}} \, \C \, ,
\end{equation}
which removes the $(1+t)^2$ term in the coefficient of $\C$ in $A_{[0,0,0]}$
in (\ref{Aint}) which is the cause of this problem.  This change then leads to
\bea
A^{\rm int.}_{[0,6,0]} & \ea {}&  s^2  \C \, , \nn \\
A^{\rm int.}_{[1,4,1]} & \ea {}&   -
{\ts {2 \over 3}} (1-t)s\, \C  + {\ts{1 \over 3}} s^2 \B \, , \nn \\
A^{\rm int.}_{[2,2,2]} & \ea{}& 
{\ts{1\over 9}}\lb 5(1+t)-s \rb s\, \C
- {\ts{5 \over 9}}(1-t)s \, \B \, , \nn \\
A^{\rm int.}_{[3,0,3]} & \ea{}&  - {\ts{1 \over 3}} (1-t)s\, \C
+ {\ts{1 \over 3}}\lb 3(1+t) - s \rb s \, \B \, , \nn \\
A^{\rm int.}_{[0,4,0]} & \ea{}& 
{\ts{1 \over 54}}\lb 42(1-t)^2 - 21s(1+t) +5s^2 \rb \C
- {\ts{7 \over 18}}(1-t)s \, \B + {\ts{7 \over 9}}s^2 \A' \, , \nn \\
A^{\rm int.}_{[1,2,1]} & \ea{}& - {\ts{1 \over 6}} \lb 3(1+t) 
- s \rb (1-t)\C + {\ts{1 \over 2}} (1-t)^2 \B - (1-t)s \, \A'\, , \nn \\
A^{\rm int.}_{[2,0,2]} & \ea{}& 
{\ts{1 \over 54}}\lb 15(1-t)^2 - s^2 \rb \C
- {\ts{5 \over 18}}\lb 3(1+t) - s \rb (1-t) \B  \nn \\
&& {} + {\ts{ 5 \over 9}}\lb 3(1+t) - s \rb s \, \A' \, , \nn \\
A^{\rm int.}_{[0,2,0]} & \ea{}&  {\ts{1 \over 144}}
\lb 105 (1 + t)^2 - 28 (1-t)^2- 46 s(1+t)+ 5 s^2 \rb \C \nn \\
&&{} - {\ts{1 \over 48 }}\lb 35 (1+t) - 13 s \rb (1-t) \B
 + {\ts{7 \over 24}}\lb 10(1-t)^2 - 5s(1+t) + s^2 \rb \A' \, , \nn \\
A^{\rm int.}_{[1,0,1]} & \ea{}& - {\ts{1 \over 108}} \lb 35(1+t) - 
13 s \rb (1-t) \C
+ {\ts{1 \over 36}} \lb 35(1+t)^2 - 20s(1+t) + 2s^2 \rb \B \nn \\
&&{} - {\ts{ 35\over 18}}\lb 2(1+t) -s \rb (1-t) \A'\, , \nn \\
A^{\rm int.}_{[0,0,0]} & \ea{}& 
{\ts{1 \over 36}}\lb 10 (1-t)^2 - 5 s(1+t) +  s^2 \rb \C
- {\ts{5 \over 12}}\lb 2 (1+t) -  s \rb (1-t) \B  \nn \\
&& {} + {\ts{5\over 6}}\lb 15(1+t)^2 - 5(1-t)^2- 8 s(1+t) + s^2 \rb \A' \, . 
\label{Aint2}\eea
The operators arising from $\C$ now correspond exactly to those present in a 
long multiplet with the lowest dimension operator belonging to a $[0,2,0]$ 
representation. It is convenient to write (\ref{Aint2}) succinctly as a linear
function of $\A',\B,\C$,
\be
A^{\rm int.}_{\cal J} = F_{\cal J}(\A',\B,\C) \, .
\label{defFJ}
\ee

The detailed form for $\A,\B,\C$ to leading order in $1/N^2$ is
\bea
\A(s,t)  & \ea{}& - {9\over N^2} \, s^3 \lb  {\overline D}_{3533}(s,t) +
{\overline D}_{3522}(s,t) \rb \, , \nn \\
\B(s,t)  & \ea{} & {9\over N^2} \, s^3 \lb {\overline D}_{2532}(s,t) -
{\overline D}_{2523}(s,t) \rb \, , \nn \\
\C(s,t)  & \ea{}& - {9\over N^2}\, \Big ( s^3 \lb 2 \,{\overline D}_{3533}(s,t)
-  {\overline D}_{3351}(s,t) \rb  + {s^2 \over t^2} \Big )  \, ,
\eea
where we used the results in appendix D to express them in a form with
the maximal overall power of $s$.

To analyse the operator product expansion for long multiplets we focus
on $A_{\cal J}$ for ${\cal J}= [0,6,0],[1,4,1],[0,4,0]$. For these ${\cal J}$
it is convenient to define $\A_0,\B_0,\C_0$ such that
$A^{\rm free}_{\cal J} = F_{\cal J}(\A_0,\B_0,\C_0)$ to leading order in $1/N^2$.

We first consider $A_{[0,6,0]}$. The free part determines $\C_0$ which has
the expansion
\be
\C_0(s,t) = s + {s \over t^3} =  \sum_{{\t=1,2,\dots\atop \ell =0,2,\dots}}
c_{\t,\ell}\,  s^{\t} G^{(\ell)}_{2\t+\ell+4}(s,t) \, , 
\label{defC0}
\ee
for
\bea
c_{\t,\ell}  & \ea & 2^{\ell-2}\, {(\ell+\t+1)!\, (\ell+\t+2)!\, \t!\, (\t+1)!\over
(2\ell+2\t+1)!\, (2\t)!}\,  \t(\ell+1)(\ell+2\t+2) \, . \label{A06}
\eea
We may also obtain, by using the results of Appendix \ref{D},
\bea 
&& \C(s,t) \sim - {1\over N^2} \ln s \sum_{{\t=3,4,\dots\atop \ell =0,2,\dots}}
{\hat c}_{\t,\ell}\,  s^{\t} G^{(\ell)}_{2\t+\ell+4}(s,t)\, ,  \nn  \\
{\hat c}_{\t,\ell} & \ea & 2^{\ell-2}\, 
{(\ell+\t)!\, (\ell+\t+1)!\, \t!\, (\t+3)!\over (2\ell+2\t+1)!\, (2\t)!}\nn \\ 
&&{}\times \t(\t-1)(\t-2)\lb  (2\ell+3\t+4)(\ell+\t+1) - \ell \t \rb \, . 
\label{A06I} \eea
For $\t=1,2$ there are no contributions involving $\ln s$ and the corresponding
operators in the $[0,6,0]$ representation belong to protected semi-short
multiplets. For $\t=3,4,\dots $, $\ell=0,2,\dots$ the $\ln s$ contributions generate 
anomalous dimensions for long multiplets whose lowest dimension operator belongs to 
a $[0,2,0]$ with scale dimensions $\Delta_{\t,\ell} = 2\t + \ell + \eta_{\t,\ell}$.
Assuming there is just a single long multiplet for each $\t,\ell$ the 
leading large $N$ contribution to $\eta_{\t,\ell}$ is given by
$-2 {\hat c}_{\t,\ell} / c_{\t,\ell} N^2$ giving
\be
\eta_{\t,\ell} = - {2 \over N^2} \, {(\t+3)(\t+2)(\t-1)(\t-2) \over
(\ell+\t+1)(\ell+\t+2)(\ell+2\t+2)(\ell+1)}\, 
\lb  (2\ell+3\t+4)(\ell+\t+1) - \ell \t \rb \, .
\end{equation}

The next step is to consider $A_{[1,4,1]}$ where the expansion
of $\B$ gives results for anomalous dimensions for long multiplets 
with the lowest dimension operator belonging to a $[1,0,1]$ representation. From
(\ref{freep}) and (\ref{Aint2}) the corresponding free contribution to leading
order at large $N$, after removing the parts corresponding to the $[0,2,0]$ long
supermultiplets, is given by defining $\B_0$ analogously to $\C_0$,
\be
\B_0(s,t) = 3 \Big ( s - {s \over t^3} \Big ) + 2(1-t)
\Big (1 + \frac{1}{t^3} \Big ) = \sum_{{\t=0,1,\dots\atop
\ell =1,3,\dots}} b_{\t,\ell}\,  s^{\t} G^{(\ell)}_{2\t+\ell+4}(s,t) \, , 
\label{defB0}
\ee
for 
\bea 
b_{\t,\ell}  & \ea & 2^{\ell-2}\, {(\ell+\t)!\, (\ell+\t+1)!\, (\t!)^2 \over
(2\ell+2\t+1)!\, (2\t)!}\,  (\t-1)(\t+2)(\ell+1)(\ell+2\t+2)(\ell+\t)(\ell+\t+3) \, .
\nn \\ 
\label{free11}
\eea
For the expansion of the $\ln s$ terms in $\B$ we use from (\ref{logexp})
\bea \hskip - 0.9cm 
&& \B(s,t) \sim {9\over N^2}\, \ln s \ s^3 \! \sum_{m,n=0}^\infty 
{(m+2)!\, (m+n+3)!\, (m+n+5)! \over m!\, n!\,(2m+n+7)!}\, s^m (1-t)^{n+1}
\nn \\
&& \qquad \qquad {}= - {1\over N^2} \ln s \sum_{{\t=3,4,\dots\atop \ell =1,3,\dots}}
{\hat b}_{\t,\ell}\,  s^{\t} G^{(\ell)}_{2\t+\ell+4}(s,t)\, ,  \nn \\
\hskip - 0.9cm {\hat b}_{\t,\ell} & \ea & 2^{\ell-1}\, 
{(\ell+\t)!\, (\ell+\t+1)!\, \t!\, (\t+3)!\over (2\ell+2\t+1)!\, (2\t)!}\,  \t(\t-1)(\t-2)
(\ell+1)(\ell+2\t+2) \, .
\eea
We may then read off the anomalous dimensions for $\t=3,4\dots$, $\ell=1,3,\dots$,
\be
\eta_{\t,\ell} = - {4 \over N^2} \, {(\t+3)(\t+1)\t(\t-2) \over
(\ell+\t)(\ell+\t+3)} \, 
 \, .
\ee
{}From (\ref{free11}) $b_{0,\ell} <0$. However this case should not be taken
in isolation as for the corresponding twist 4 operators it is necessary to analyse
the contribution of  semi-short multiplets which we undertake later.

The final step is a similar analysis of $A_{[0,4,0]}$ where the expansion
of $\A'$ yields anomalous dimensions for long multiplets
with the lowest dimension operator belonging to the $[0,0,0]$
representation.  After removing the free contributions of  the $[0,2,0]$
and $[1,0,1]$ long multiplets then, from
(\ref{freep}) and (\ref{Aint2}), the corresponding free contribution to
leading order at large $N$ is given by
\be
\A_0(s,t) = {{1\over 2}}\Big(1+t+{\ts{7\over 3}}s\Big)
\Big (1 + \frac{1}{t^3} \Big ) + {{3 \over 2}}(1-t)\Big(1 - {1 \over t^3}\Big) 
= \sum_{{\t=0,1,\dots\atop
\ell =0,2,\dots}} a_{\t,\ell}\,  s^{\t} G^{(\ell)}_{2\t+\ell+4}(s,t) \, ,
\label{defA0} 
\ee
where
\bea
a_{\t,\ell}  & \ea &  2^{\ell-3}\, {(\ell+\t)!\, (\ell+\t+1)!\, (\t!)^2
\over 3(2\ell+2\t+1)!\, (2\t)!} \nn \\
&&{}\times (\t-2)(\t+3)(\ell+1)(\ell+2\t+2)(\ell+\t-1)(\ell+\t+4) \, .
\eea
In a similar fashion as before we may find that for $\A'$ we have
\be
\A'(s,t) \sim - {1\over N^2} \ln s \sum_{{\t=3,4,\dots\atop \ell
=0,2,\dots}}
{\hat a}_{\t,\ell}\,  s^{\t} G^{(\ell)}_{2\t+\ell+4}(s,t)\, ,
\label{A0I}
\ee
for 
\bea
{\hat a}_{\t,\ell} & \ea & 2^{\ell-2}\,
{(\ell+\t)!\, (\ell+\t+1)!\, \t!\, (\t+3)!\over 3 (2\ell+2\t+1)!\,
(2\t)!}\nn \\  
&&{}\times \t(\t-1)(\t-2) \lb  (\ell+\t-1)(\ell+\t+4) + 5(\t-2)(\t+3)\rb \, . 
\eea
Taking these results into consideration then the  anomalous dimensions
for $\t=3,4\dots$, $\ell=0,2,\dots$, if there was a single long multiplet
for each $\t,\ell$, would be given by,
\be
\eta_{\t,\ell} = - {4 \over N^2} \, {(\t+2)(\t+1)\t(\t-1) \over
(\ell+1)(\ell+2\t+2)}\Big( 1+{5(\t-2)(\t+3) \over
(\ell+\t-1)(\ell+\t+4)}\Big)\, .
\end{equation}
{}From a similar analysis of the operator product expansion in \cite{DO}
for the four point function of $[0,2,0]$ $\half$-BPS operators a result
for $\eta_{\t,\ell}$ was obtained without the final term in parentheses. 
For $\t=2$ the result in \cite{DO} appears to be valid, and agrees with a
similar calculation in \cite{HMR}, but for $\t \ge 3$ the
difference has to be a reflection of more than one long multiplet for each
$\t,\ell$ being present so that it is then necessary to consider operator mixing
effects.

\subsection{Semi-short multiplets}

Besides the contributions of long multiplets which may have arbitrary scale
dimensions greater than the unitarity bound there are also contributions from
semi-short multiplets whose dimensions are protected. According to \cite{ES}
the relevant semi-short multiplets which may contribute to the operator product
expansion of two $[0,3,0]$ $\half$-BPS multiplets correspond to the case where the 
lowest dimension operator belongs to the representations $[0,0,0], \ [0,2,0], \ [2,0,2]$ 
and $[0,4,0]$, for even $\ell$, and $[1,0,1], \ [1,2,1]$ for odd $\ell$.
Denoting the multiplets by ${\cal C}_{{\cal J}\ell}$ these contain operators
with twist $\Delta-\ell$ according to the following table (obtained from 
\cite{DO2}),
\medskip

\begin{tabular}{|c|c|c|c|}\hline
$\C_{[0,0,0]\ell}$ & [0,0,0]& [0,2,0] & [1,0,1] \\ \hline
2 & $\ell,\ell+2,\ell+4$ & $\ell+2$ & $\ell+1,\ell+3$ \\ \hline
\end{tabular}

\begin{tabular}{|c|c|c|c|c|c|c|}\hline
$\C_{[1,0,1]\ell}$ & [0,0,0]& [0,2,0] & [0,4,0]& [2,0,2]& [1,0,1]&[1,2,1]
\\ \hline
4 &$\ell{+1},\ell{+3}$&$\ell{+1},\ell{+3}$&&$\ell+{1},\ell{+3}$&
$\ell,\ell{+2},\ell{+4}$&$\ell{+2}$
\\ \hline
6 &$\ell{-1},\ell{+1},\ell{+3}$&$\ell{-1},\ell{+1},\ell{+3}$&$\ell{+1}$
&$\ell{+1}$&$\ell,\ell{+2}$&$\ell,\ell{+2}$ \\ \hline
8&$\ell{-1},\ell{+1}$&$\ell{-1},\ell{+1}$&&$\ell{-1},\ell{+1}$&
$\ell{-2},\ell,\ell{+2}$&$\ell$ \\ \hline
10 &$\ell{-3},\ell{-1},\ell{+1}$&$\ell{-1}$&&&$\ell{-2},\ell$&
\\ \hline
\end{tabular}

\begin{tabular}{|c|c|c|c|c|c|c|}\hline
$\C_{[0,2,0]\ell}$ & [0,0,0]& [0,2,0] & [0,4,0]& [2,0,2]& [1,0,1]&[1,2,1]
\\ \hline
4 &$\ell{+2}$&$\ell,\ell{+2},\ell{+4}$&$\ell{+2}$&$\ell{+2}$&
$\ell{+1},\ell{+3}$&$\ell{+1},\ell{+3}$
\\ \hline
6 &$\ell,\ell{+2}$&$\ell,\ell{+2}$&
&$\ell,\ell{+2}$&$\ell{-1},\ell{+1},\ell{+3}$&$\ell{+1}$ \\ \hline
8&$\ell{-2},\ell,\ell{+2}$&$\ell$&&&
$\ell{-1},\ell{+1}$& \\ \hline
\end{tabular}

\hskip -1.2cm
\begin{tabular}{|c|c|c|c|c|c|c|c|c|c|}\hline
$\C_{[2,0,2]\ell}$ & [0,0,0]& [0,2,0] & [0,4,0]& [2,0,2]&[2,2,2]&[1,0,1]&[1,2,1]
&[3,0,3]&[1,4,1]
\\ \hline 6 &&$\ell{+2}$&&${\ell,\ell{+2}\atop\ell{+4}}$&
$\ell{+2}$&$\ell{+1},\ell{+3}$&
$\ell{+1},\ell{+3}$&$\ell{+1},\ell{+3}$& \\ \hline
8 &$\ell,\ell{+2}$&$\ell,\ell{+2}$&$\ell,\ell{+2}$
&$\ell,\ell{+2}$&$\ell,\ell{+2}$&${\ell{-1},\ell{+1}\atop\ell{+3}}$&
${\ell{-1},\ell{+1}\atop\ell{+3}}$&$\ell{+1}$&$\ell{+1}$
\\ \hline
10&$\ell$&${\ell{-2},\ell\atop\ell{+2}}$&$\ell$&${\ell{-2},\ell\atop\ell{+2}}$&
$\ell$&$\ell{-1},\ell{+1}$&$\ell{-1},\ell{+1}$&$\ell{-1},\ell{+1}$&
\\ \hline
12&$\ell{-2},\ell$&$\ell{-2},\ell$&&$\ell{-2},\ell$&&
${\ell{-3},\ell{-1}\atop\ell{+1}}$&$\ell{-1}$&&
\\ \hline
\end{tabular}

\hskip -1.2cm
\begin{tabular}{|c|c|c|c|c|c|c|c|c|c|c|}\hline
$\C_{[0,4,0]\ell}$ & [0,0,0]& [0,2,0] & [0,4,0]&[0,6,0]&[2,0,2]&[2,2,2]&[1,0,1]&[1,2,1]&[3,0,3]&[1,4,1]
\\ \hline 6
&&$\ell{+2}$&${\ell,\ell{+2}\atop\ell{+4}}$&$\ell{+2}$&&$\ell{+2}$&
&$\ell{+1},\ell{+3}$&&$\ell{+1},\ell{+3}$ \\ \hline
8 &&$\ell,\ell{+2}$&$\ell,\ell{+2}$
&&$\ell,\ell{+2}$&$\ell,\ell{+2}$&$\ell{+1}$&${\ell{-1},\ell{+1}\atop\ell{+3}}$&
$\ell{+1}$&$\ell{+1}$\\ \hline
10&$\ell$&${\ell{-2},\ell\atop\ell{+2}}$&$\ell$&&$\ell$&
&$\ell{-1},\ell{+1}$&$\ell{-1},\ell{+1}$&&
\\ \hline
\end{tabular}

\hskip -1.8cm
\begin{tabular}{|c|c|c|c|c|c|c|c|c|c|c|}\hline
$\C_{[1,2,1]\ell}$&[0,0,0]&[0,2,0]&[0,4,0]&[0,6,0]&[2,0,2]&[2,2,2]&[1,0,1]&[1,2,1]&
[3,0,3]&[1,4,1]
\\ \hline 6
&&$\ell{+1},\ell{+3}$&$\ell{+1},\ell{+3}$&&$\ell{+1},\ell{+3}$&$\ell{+1},\ell{+3}$&
$\ell{+2}$&${\ell,\ell{+2}\atop\ell{+4}}$&$\ell{+2}$&$\ell{+2}$ \\ \hline
8 &$\ell{+1}$&${\ell{-1},\ell{+1}\atop\ell{+3}}$&${\ell{-1},\ell{+1}\atop\ell{+3}}$
&$\ell{+1}$&${\ell{-1},\ell{+1}\atop\ell{+3}}$&$\ell{+1}$&$\ell,\ell{+2}$&
$\ell,\ell{+2}$&$\ell,\ell{+2}$&$\ell{+2}$
\\ \hline
10&$\ell{-1},\ell{+1}$&${\ell{-1},\ell{+1}}$&$\ell{-1},\ell{+1}$&&
$\ell{-1},\ell{+1}$&$\ell{+1}$&${\ell{-2},\ell\atop\ell{+2}}$&
${\ell{-2},\ell\atop\ell{+2}}$&$\ell$&$\ell$
\\ \hline
12&$\ell{-1}$&$\ell{-1},\ell{+1}$&$\ell{-1}$&&$\ell{-1}$&&$\ell{-2},\ell$&
$\ell{-2},\ell$&&
\\ \hline
\end{tabular}

\vskip 0.5cm
\begin{center}
\begin{tabular}{l}
Table 1: Spins of operators for given twist belonging to semi-short multiplets\\
contributing to the operator product expansion of two $[0,3,0]$ $\half$-BPS multiplets.
\end{tabular}
\end{center}
\vskip 0.5cm

It is important to note that
\be
\C_{[0,0,0]\ell} + \C_{[1,0,1]\ell-1} \, , \qquad 
\C_{[1,0,1]\ell} + \C_{[2,0,2]\ell-1}  \, , \qquad 
\C_{[0,2,0]\ell} + \C_{[1,2,1]\ell-1} \, , 
\ee
combine to form complete long multiplets with the lowest scale dimension
compatible with unitarity. These combinations may then be absorbed into
contributions represented by $\A',\, \B,\, \C$ respectively, which may gain
anomalous scale dimensions. Nevertheless the free field contributions given in
(\ref{freep}) cannot all be described by a choice for
$\A',\, \B, \, \C$ and so correspond to long multiplets. We show here, for 
simplicity to zeroth order in $1/N$, how the additional terms necessary to
accommodate (\ref{freep}) have restricted twists and are compatible with the
expected contributions in Table 1 so that for suitable $\t,\ell$,
\be
A^{\rm short}_{{\cal J}}(s,t) = \sum_{\t,\ell} d^{{\cal J}}_{\t,\ell}  \,
s^{\tau+2}G^{(\ell)}_{2\t+\ell+4}(s,t) \, .
\label{short}
\ee

To procede we first isolate those contributions to ${\cal C}_0, {\cal B}_0$
and ${\cal A}_0$ in (\ref{defC0}), (\ref{defB0}) and (\ref{defA0}) which
are protected in that there are no corresponding terms involving $\ln s$
which generate anomalous dimensions,
\bea
{\cal C}^{\rm short}_0(s,t) &\ea&\sum_{\t=1,2} \sum_{\ell=0,2,\dots} c_{\t,\ell}\, 
s^{\t} G^{(\ell)}_{2\t+\ell+4}(s,t) \, , \nn \\
{\cal B}^{\rm short}_0(s,t) &\ea&  \sum_{\t=0,2} \sum_{\ell=1,3,\dots} b_{\t,\ell}\, 
s^{\t} G^{(\ell)}_{2\t+\ell+4}(s,t) \, , \nn \\
{\cal A}^{\rm short}_0(s,t)&\ea & \sum_{\t=0,1} \sum_{\ell=0,2,\dots} a_{\t,\ell}\, 
s^{\t} G^{(\ell)}_{2\t+\ell+4}(s,t) \, , 
\eea
where we note that $b_{1,\ell}=c_{2,\ell}=0$. We then define, using (\ref{defFJ}),
\be
A^{\rm short}_{{\cal J}} = F_{{\cal J}}({\A}^{\rm short}_0,{\B}^{\rm short}_0,
{\C}^{\rm short}_0) \, , \qquad {\cal J} = [0,6,0],[1,4,1],[0,4,0] \, .
\ee 
Thus $A_{[0,6,0]}(s,t)=s^2{\cal C}^{\rm short}_0(s,t)$ and it is easy to see that
\be
{d}^{[0,6,0]}_{\t,\ell}  = c_{\t,\ell} \, , \qquad \t=1,2 \, , \
\ell =0,2,\dots \, .
\label{d1}
\ee
which corresponds to an operator product expansion involving protected twist 6 
and 8 operators. The relevant operators may clearly be identified with the
$\C_{[0,4,0]\ell}$ and $\C_{[1,2,1]\ell}$ semi-short supermultiplets. For
$A_{[0,6,0]}$ and $A_{[0,4,0]}$ using
(\ref{recur}) we may show that this requires a non zero
\be
d^{[1,4,1]}_{\t,\ell} \ \tau=1,2,3 \, , \ \ell =1,3,\dots \, , \qquad
d^{[0,4,0]}_{\t,\ell} \ \tau=1,2,3,4 \, , \ \ell =0,2,\dots \, .
\label{d2}
\ee
where the cancellation of possible $\t=-1,0$  terms depends on 
$b_{0,\ell} = - 4 c_{1,\ell-1}$ and $a_{0,\ell{+2}} = -4 c_{2,\ell}$. 
Here the $G^{(\ell)}_\Delta$ appearing in the final result for the OPE correspond 
to just those expected for operators belonging to 
$\C_{[0,4,0]\ell}, \, \C_{[1,2,1]\ell}$ semi-short supermultiplets. Detailed
results for $d^{\cal J}_{\t,\ell}$ are given in Appendix \ref{appF}. It is  
critical that they are positive.

A crucial test is whether all the remaining free field contributions can be
represented as in (\ref{short}). If we subtract off all contributions
corresponding to long multiplets which gain anomalous dimensions we have in
general
\be
A^{\rm short}_{{\cal J}} = H_{\cal J} 
+ F_{{\cal J}}({\A}^{\rm short}_0,{\B}^{\rm short}_0,{\C}^{\rm short}_0) \, , \qquad
H_{\cal J} = A^{\rm free}_{{\cal J}}  - F_{{\cal J}}(\A_0,\B_0,\C_0) \, ,
\label{decomp}
\ee
dropping any $1/N^2$ terms in $A^{\rm free}_{\cal J}$. Results for
$H_{\cal J}(s,t)$ are given in Appendix \ref{appF} which are simple when
expressed in terms of new variables $z,x$. Using this form it is straightforward
to see that $H_{\cal J}$ corresponds to contributions which have only twist 2
or twist 0.

In detail we first consider $A^{\rm short}_{[2,2,2]}$
and $A^{\rm short}_{[3,0,3]}$. In this case only  twist 2 contributions appear
in the expansion of $H_{[2,2,2]}$ and $H_{[3,0,3]}$. Adding on the 
contributions resulting from ${\B}^{\rm short}_0,{\C}^{\rm short}_0$ for this
case as in (\ref{decomp})
there are again non trivial cancellations and we find
non zero expansion coefficients just for
\be
d^{[2,2,2]}_{\t,\ell} \ \tau=1,2,3 \, , \ \ell =0,2,\dots \, , \qquad
d^{[3,0,3]}_{\t,\ell} \ \tau=1,2,3 \, , \ \ell =1,3,\dots \, .
\label{d3}
\ee
For ${\cal J}=[2,0,2],[0,2,0],[1,2,1],[1,0,1]$ $H_{\cal J}$ contains both twist 0
and twist 2 but together with twist 4 such contributions cancel in
$A^{\rm short}_{\cal J}$ leaving
\be
d^{[2,0,2]}_{\t,\ell},d^{[0,2,0]}_{\t,\ell} \ \tau=1,2,3,4 \, ,\ 
\ell =0,2,\dots \, , \qquad
d^{[1,2,1]}_{\t,\ell},d^{[1,0,1]}_{\t,\ell} \ \tau=1,2,3,4 \, , \ 
\ell =1,3,\dots \, ,
\label{d4}
\ee
to be non zero. The singlet case, ${\cal J}=[0,0,0]$ is further restricted in
that there is cancellation of twist 6 terms as well leaving non zero
\be
d^{[0,0,0]}_{\t,\ell} \ \tau=2,3,4 \, , \ \ell =0,2,\dots \, .
\label{d5}
\ee
It is then clear, for all representations ${\cal J}$, by matching (\ref{d1}), 
(\ref{d2}), (\ref{d3}), (\ref{d4}) and (\ref{d5}) with Table 1 that, 
to zeroth order in $1/N$, that only operators belonging at least to the 
$\C_{[0,4,0]\ell}, \, \C_{[1,2,1]\ell}$ semi-short supermultiplets
are necessary in the operator product expansion, although contributions in addition
from $\C_{[2,0,2]\ell}$ are possible.

When $\ell=0,1$ then the results for $d^{\cal J}_{\t,\ell}$ may be modified. First
we note that we must have
\be
d^{[0,0,0]}_{-2,0} = 50 \, ,
\ee
reflecting the contribution of the identity operator. For $\ell=0,1$ all
$d^{\cal J}_{\t,\ell}$ listed above are still present to zeroth order in $1/N$
except for the twist 6 contribution
\be
d^{[0,2,0]}_{1,0} = 0 \, .
\label{absc} 
\ee
From Table 1 we may note that for twist 6 the operators
in the $\C_{[0,4,0]0}, \, \C_{[1,2,1]1}$ semi-short multiplets in the 
$[0,2,0], \, [0,6,0], \, [2,0,2], \, [2,2,2]$ representations have lowest spin 2 while 
for $[1,0,1], \, [3,0,3]$ the lowest spin is 3. Also for $\C_{[1,2,1]1}$ we only
have a twist 8 $[0,6,0]$ operator with spin 2.  From $\C_{[2,0,2]0}$ we may have
operators for $[2,0,2]$ with spin 0 and $[1,0,1], \, [3,0,3]$ with spin 1. The
remaining gaps correspond to operators which are part of short BPS multiplets.
If we denote such multiplets by $\B_{\cal J}$,
where if ${\cal J} = [q,p,q]$ with $q>0$ it is a $\frac{1}{4}$-BPS multiplet and
if $q=0$ it is $\half$-BPS, the relevant multiplets which may occur in the
operator product expansion here with twist 6 or more are listed in Table 2.

\vskip 0.5cm

\begin{tabular}{|c|c|c|c|c|c|c|}\hline
$\B_{[0,6,0]}$ & [0,2,0]& [0,4,0] & [2,2,2]& [0,6,0]& [1,2,1]&[1,4,1]
\\ \hline
6 &&2&&0&&1 \\ \hline
8 &&&0&&1& \\ \hline
10 &0&&&&& \\ \hline
\end{tabular}

\begin{tabular}{|c|c|c|c|c|c|c|c|c|}\hline
$\B_{[2,2,2]}$ & [0,2,0]& [0,4,0] & [2,0,2]& [2,2,2]&[0,6,0]&[1,0,1]&[1,2,1]
&[1,4,1]
\\ \hline 6 &&2&2&0,2&&&1,3&1 \\ \hline
8  &0,2&0,2&0,2&0&0&1&1&1
\\ \hline
10&0&0&0&0&&1&1&
\\ \hline
12&0&&&&&&&
\\ \hline
\end{tabular}

\vskip 0.5cm
\begin{center}
\begin{tabular}{l}
Table 2: Spins of operators for given twist belonging to short BPS multiplets\\
contributing
to the operator product expansion of two $[0,3,0]$ $\half$-BPS multiplets\\
with twist at least 6.
\end{tabular}
\end{center}
\vskip 0.5cm

Both BPS multiplets to account for operators which have been identified has
necessarily present in the operator product expansion. It is evident that (\ref{absc})
is necessary for compatibility with the representation content of short and
semi-short multiplets.

Thus we have shown that the OPE underlying 
the supergravity-induced four-point amplitude 
of $\half$-BPS operators of dimension 3 fulfills the requirements 
of the superconformal symmetry and unitarity. Only those long multiplets
belonging to the representations $[0,2,0]$, $[1,0,1]$ and $[0,0,0]$ may acquire
possible anomalous scaling dimensions.

\section*{Acknowledgements} We are grateful to Sergei Frolov, Paul Heslop
and Paul Howe for many useful discussions. E.S. wishes to thank the Max
Planck Institute f\"ur Gravitationsphysik for the warm hospitality. G.A. was
supported in part by the European Commission RTN programme
HPRN-CT-2000-00131 and by RFBI grant N02-01-00695.

\section*{Appendices}
\appendix
\setcounter{equation}0


\section{The ${\cal N}=2$ reduction formula}\label{-A}
 
Here we present a simplified version of the procedure of Ref. \cite{4pt'} 
for projecting an ${\cal N}=4$ four-point function of $\half$-BPS operators 
onto the ${\cal N}=2$ hypermultiplet (HM) and SYM constituents. The new 
procedure can easily be applied to $\half$-BPS operators of any dimension $k$. 
Further, we briefly recall the use of the ${\cal N}=2$ insertion formula from 
Ref. \cite{EPSS} and apply it to the correlator of weight $k=3$. In this way 
we can reproduce the results of Section \ref{prediction} without reference to
the ${\cal N}=4$ insertion formula.  

The lowest component of the ${\cal N}=4$ field-strength multiplet $W^i$, 
$i=1,\ldots,6$ is a real vector of SO(6). Reducing SO(6) to SU(3), we 
can decompose it into $3+\bar 3$:
\begin{equation}\label{6}
  {\cal W}^i \ \rightarrow \ W^A, \ \bar W_A\;, \quad A=1,2,3\;.
\end{equation}
The further decomposition SU(3) $\rightarrow$ SU(2)$\times$U(1) results 
in
\begin{equation}\label{7}
   W^A \ \rightarrow \ W^a \equiv \phi^a,\ a=1,2; \quad \ W^3 \equiv w\;.
\end{equation}
After projection with SU(2) harmonics $\phi^a$ becomes the lowest 
component of the Grassmann analytic ${\cal N}=2$ HM $q^+ = u^+_a 
\phi^a$; $w$ is the lowest component of the chiral ${\cal N}=2$ field 
strength; their conjugates are $\tilde q^+ = u^{+a} \bar\phi_a$ and the 
antichiral $\bar w$.

The SO(6)-covariant field-strength propagator is $\langle W^i(1) W^j(2) 
\rangle = \langle W^j(1) W^i(2) \rangle \sim \delta^{ij}$. Introducing 
SO(6) harmonics $1_i, 2_i$ and their symmetric contraction $({12})=(21) 
= 1_i\delta^{ij}2_j$, we can write the harmonic-projected SO(6) 
propagator $\langle W(1) W(2) \rangle \sim ({12})$. Next, reducing SO(6) 
to SU(3) we decompose the SO(6) contraction into SU(3) pieces:
\begin{equation}\label{9}
 ({12}) = 1_i\delta^{ij}2_j = 1^A \bar 2_A + \bar 1_A 2^A \equiv [1\bar 2] 
+ [\bar 12]\;.
\end{equation}
In this notation we have ``oriented" propagators for the SU(3)-covariant 
field strengths: $\langle W\bar W \rangle = [1\bar 2]$ and $\langle \bar 
W W \rangle = [\bar 12]$. The further reduction of SU(3) to 
SU(2)$\times$U(1) gives, e.g., $[1\bar 2] = 1^A \bar 2_A = 1^a \bar 2_a 
+ 1^3 \bar 2_3$. This can be split into two independent propagators, one 
for the ${\cal N}=2$ HM:
\begin{equation}\label{10}
 \langle q\tilde q \rangle \sim  1^a \bar 2_a = 1^a \epsilon_{ab}2^b = 
-\bar 1_a 2^a \equiv [12] = -[21]
\end{equation}
and one for the ${\cal N}=2$ field strength, $\langle w\bar w \rangle 
\sim 1^3 \bar 2_3 \equiv 1$ (in the latter there is no need to use 
harmonics, $1^3 \bar 2_3$ is just a ``bookkeeping device").

The $\half$-BPS operator of weight $k$ is ${\rm Tr}({\cal W }^{\{i_1}\cdots 
{\cal W }^{i_k\}})$, where $\{\}$ denotes traceless symmetrisation. 
Projected with SO(6) harmonic, it becomes ${\cal W }^k = {\rm Tr}({\cal 
W }^{i_1}\cdots {\cal W }^{i_k})1_{i_1}\cdots 1_{i_k}$, and the absence 
of traces is guaranteed by the defining properties of the SO(6) 
harmonics. The four-point function for such operators has a harmonic 
structure consisting of all possible pairings of the four sets of 
harmonics. For instance, for $k=2$ we have
\begin{eqnarray}
&&\langle {\cal W }^2 |{\cal W }^2 |{\cal W }^2|{\cal W }^2\rangle \nn \\
&&\quad = A_1\;({12})^2({34})^2 + A_2\;({13})^2({24})^2 + A_3\;({14})^2({23})^2   
\nn \label{0}\\
&&\qquad {}+  B_1\;({13})({14})({23})({24}) + B_2\;({12})({14})({23})({34}) +
B_3\;({12})({13})({24})({34}) \, ,
\end{eqnarray}
(compared to eq.(\ref{befcro2}), we have absorbed the space-time 
propagator factors into the coefficient functions $A,B,C$).

The reduction to either ${\cal N}=2$ HMs or field strengths is 
straightforward. We replace each SO(6) contraction by SU(3) 
contractions: $(pq)= [p\bar q] + [\bar pq]$ and expand each SO(6) 
harmonic structure in (\ref{0}) into products of SU(3) contractions. For 
example,
\begin{eqnarray}
(12)^2({34})^2  &\rightarrow& [\bar 12]^2[\bar 34]^2 + 
2\;[\bar 12]^2[\bar 34][3\bar 4] + 4\;
[1\bar 2][\bar 12][3\bar 4][\bar 34] + \ldots\nn \\
(13)({14})({23})({24}) &\rightarrow& [\bar 13][\bar 14][2\bar 3][2\bar 4] 
+ [\bar 13][\bar 14][2\bar 3][2\bar 4] \nonumber\\ 
&& {}+ [1\bar 3][\bar 14][\bar 23][2\bar 4] + [\bar 13][1\bar 4][2\bar 3][\bar 24] +
\ldots\;, \label{11}
\end{eqnarray}
where we have displayed just the terms relevant for the two HM 
projections considered below. If we want to keep only the HM 
constituents of the composite operators, we need to replace the SU(3) 
contractions by SU(2) ones, taking care of the signs, e.g., $[1\bar 
2]\rightarrow [12]$, $[\bar 12]\rightarrow -[12]$. In this way we 
obtain, for example,
\begin{eqnarray}
\hskip -0.5cm \langle \tilde q^2|q^2 |\tilde q^2|q^2\rangle &\ea & 
A_1\;[12]^2[34]^2  + 0\; [13]^2[24]^2  + A_3\;[14]^2[23]^2   \nn \\
  &&{}+  0\; [13][14][23][24]  - B_2\; [12][14][23][34]  
+ 0\; [12][13][24][34]\;;  \label{1} \\ 
\hskip -0.5cm \langle \tilde q^2|q^2 |q\tilde q|q\tilde q\rangle &\ea& -{2}A_1\; 
[12]^2[34]^2  + 0\; [13]^2[24]^2  + 0\; [14]^2[23]^2   \nn \\
  &&{}+  B_1\; [13][14][23][24]  + B_2\; [12][14][23][34]  
- B_3\; [12][13][24][34]\;.  \label{2}  
 \end{eqnarray}
Finally, with the help of the SU(2) harmonic cyclic identity $[12][34] + 
[13][42] + [14][23] = 0$ we can eliminate, e.g., all factors of the type 
$[13][24]$. Thus, the projection (\ref{2}) becomes
\begin{equation}\label{3'}
\langle \tilde q^2|q^2 |q\tilde q|q\tilde q\rangle =  
(-2A_1-B_3)[12]^2[34]^2 +  B_1[14]^2[23]^2 + (B_1+B_2-B_3) [12][34][14][23] \, .
\end{equation}

Similarly, to obtain the U(1) or chiral-antichiral ${\cal N}=2$ 
field-strength projection we replace every $(pq)$ in eq.(\ref{0}) by 1 
if it corresponds to a Wick contraction of the type $\langle w \bar w 
\rangle$, or by 0 if it corresponds to $\langle w w \rangle$ or to 
$\langle \bar w \bar w \rangle$. In this way we find
\begin{equation}\label{8}
  \langle w^2|\bar w^2|w^2|\bar w^2 \rangle = A_1+A_3+B_2\;.
\end{equation}  

                          
Clearly, this procedure can easily be generalised to any dimension. In 
the case $k=3$ we have the decomposition (\ref{g4pt}) of the ${\cal 
N}=4$ amplitude into ten SU(4) harmonic structures. It has a large 
number of possible ${\cal N}=2$ HM projections, but it turns out that in 
order to derive the consequence of the ${\cal N}=2$ insertion formula it 
is sufficient to consider only one of them, $\langle q^3|\tilde q^3|q^2 
\tilde q|q\tilde q^2 \rangle$. Repeating the steps described above, we 
easily obtain
\begin{eqnarray}
  \langle q^3|\tilde q^3|q^2 \tilde q|q\tilde q^2 \rangle &\ea& 
(-3A_1-B_1)\; [12]^3[34]^3 - B_3\; [14]^3[23]^3 \nn \\
  &&{}+ (-B_1+2B_2+C)\; [12]^2[34]^2[14][23] \nn \\
&&{} + (-B_3-B_4+C)\; [14]^2[23]^2[12][34] \, . \label{12} 
\end{eqnarray}

In order to find the restrictions following from the insertion formula, 
we need not appeal to its ${\cal N}=4$ version, but can rely on the 
safer and well-understood ${\cal N}=2$ one based on the off-shell 
harmonic superspace formulation of the theory. The ${\cal N}=2$ 
insertion formula \cite{hssw,EPSS} predicts that any ${\cal N}=2$ 
amplitude is a product of the dimension 2 polynomial
\begin{equation}\label{4}
  R^{2222}_{{\cal N}=2} = \frac{[12]^2[34]^2}{x_{12}^4\, x_{34}^4}\ s  + 
\frac{[14]^2[23]^2}{x_{14}^4\,x_{23}^4}\ t  +
\frac{[12][34][14][23]}{x_{12}^2\, x_{34}^2\, x_{14}^2\, x_{23}^2}\ (s+t-1) \, ,
\end{equation}
with another factor of dimension $k-2$ which contains arbitrary 
functions of $s,t$. In our case of dimension 3, for the projection in 
eq.(\ref{12}) this gives
\begin{equation}\label{5}
   \langle q^3|\tilde q^3|q^2 \tilde q|q\tilde q^2 \rangle = 
R^{2222}_{{\cal N}=2}\ \left[ F(s,t)\, \frac{[12][34]}{x^2_{12}\,x^2_{34}} +  
G(s,t) \, \frac{[14][23]}{x^2_{14}\, x^2_{23}}  \right]\;.
\end{equation}
Comparing the coefficients of each of the four independent harmonic 
structures in eqs.(\ref{12}) and (\ref{5}) we obtain four equations 
relating the coefficients $A,B,C$ to the newly introduced arbitrary 
functions $F,G$. Further, from the crossing symmetry relations 
(\ref{cr1}), (\ref{cr2'}) and (\ref{cr3}) it follows that only three of 
the coefficients $A,B,C$ are independent. Thus, the four equations 
impose a relation between the functions $F$ and $G$, so that in the end 
everything is expressed in terms of $G$. The independent function 
$\gamma$ (\ref{relabc}) is related to $G$ as follows:
\begin{equation}\label{13}
  \gamma = \frac{1}{3(t-s)}\Big [\hat G - 2\tilde{\hat G} - 
\frac{1}{s} (\tilde G + \hat{\tilde G})\Big ]\;,
\end{equation}
or inversely,
\begin{equation}\label{14}
  G = \frac{t-s+1}{t} \hat\gamma - t\gamma\;.
\end{equation}
Here $\tilde G(s,t) = G(t,s)$ and $\hat G(s,t) = G(s/t, 1/t)$. In 
addition, the crossing symmetry condition $\gamma= \tilde \gamma$ 
is equivalent to the corresponding condition on $G$ in eq.(\ref{13}):
\begin{equation}\label{15}
  \Big [ \hat G + \frac{1}{s}(\tilde G + \hat{\tilde G})\Big ] + 
\Big [ \tilde{\cdots} \Big ] =0  \;.
\end{equation}        

\section{C-algebra}
\label{A}
\subsection{$C$-tensors}
The supergravity fields of the five-dimensional effective action couple 
through SO(6) invariant tensors represented by overlapping integrals of 
spherical harmonics on the five-dimensional sphere: \bea \nonumber 
a_{123}=\int Y^{I_1}Y^{I_2}Y^{I_3}\, ,~~~~~~ t_{123}=\int 
\nabla^{\alpha}Y^{I_1}Y^{I_2}Y^{I_3}_{\alpha}\, ,~~~~~~ p_{123}=\int 
\nabla^{\alpha}Y^{I_1}\nabla^{\beta}Y^{I_2}Y^{I_3}_{(\alpha\beta)}\, . 
\eea These are essentially the Clebsh-Gordon coefficients for the tensor 
product of SO(6) irreps, and it is in terms of these tensors that the 
cubic and quartic couplings of the effective action were expressed in 
Ref. \cite{AF2}.

The irreducible representations of SO(6) which are of interest to us 
here have Dynkin labels $[0,k,0]$ and $[1,k-1,1]$ ($k$ odd), $[2,k-2,2]$ 
($k$ even) and can be described in terms of the canonically normalised 
$C$-tensors with the corresponding Young symmetry. In particular, the 
irrep $[1,k,1]$ is given by a tensor $C^{I}_{m;i_1\ldots i_k}$ which is 
traceless symmetric w.r.t. $i_1,\ldots, i_k$ and has a vanishing 
symmetric part, while $[2,k,2]$ is described by $C^{I}_{mn;i_1\ldots 
i_k}$, traceless and symmetric w.r.t. $i_1,\ldots , i_k$ and $m,n$ separately, 
and obeying the constraint \bea \nonumber C^I_{mn;i_1\ldots 
i_k}+C^I_{mi_1;n\ldots i_k}+\ldots + C^I_{mi_k;i_1\ldots n}=0 \, . \eea 
We assume the following normalisations
$$C^I_{m;i_1...i_k}C^J_{n;i_1...i_k}=\d_{mn}\d^{IJ}\, , ~~~~
C^I_{m_1n_1;i_1...i_k}C^J_{m_2n_2;i_1...i_k}=\d^{IJ}\d_{m_1n_1;m_2n_2}\, .
$$

The relation of the integrals of spherical harmonics to the $C$-tensors is 
as follows 
\bea \label{aviaC} a_{123}&\ea &\frac{\prod_{i=1}^3\frac{k_i! 
z(k_i)}{\alpha_i !}}{\pi^{\frac{3}{2}}(\sigma+2)! \, 2^{\sigma-1}}\
 \langle C^{1}_{[0,k_1,0]}C^{2}_{[0,k_2,0]}C^{3}_{[0,k_3,0]} \rangle \, , \\
\label{tviaC} t_{123}&\ea &\frac{\prod_{i=1}^3 \frac{k_i ! 
z(k_i)}{(\alpha_i-\frac{1}{2})!}}{\pi^{\frac{3}{2}}(k_3+1)(\sigma+\frac{3}{2})! 
\, 2^{\sigma-\frac{3}{2}}}\
 \langle C^{1}_{[0,k_1,0]}C^{2}_{[0,k_2,0]}C^{3}_{[1,k_3-1,1]} \rangle \, , \\
\label{pviaC} p_{123}&\ea & \frac{\alpha_3 \; \prod_{i=1}^3\frac{k_i! 
z(k_i)}{\alpha_i !}}{\pi^{\frac{3}{2}}(\sigma+1)!\, 2^{\sigma}}\
 \langle C^{1}_{[0,k_1,0]}C^{2}_{[0,k_2,0]}C^{3}_{[2,k_3-2,2]} \rangle \, ,
\eea 
where $z(k)=(2^{k-1}(k+1)(k+2))^{1/2}$, $\sigma=\half(k_1+k_2+k_3)$ 
and $\alpha_i=\half(k_j+k_l-k_i)$, $j\neq l\neq i$. We use the notation 
$\langle C^1C^2C^3\rangle $ to denote the unique SO(6) tensor obtained 
by contracting particular subsets of indices of the three $C$-tensors 
$C^I_{[a,b,c]}$ (the subscript denotes the Dynkin labels of the 
corresponding irrep). Explicitly, 
\be \langle 
C^{1}_{[0,k_1,0]}C^{2}_{[0,k_2,0]}C^{3}_{[0,k_3,0]} \rangle = 
C^{I_1}_{i_1\ldots i_{\alpha_2}j_1\ldots j_{\alpha_3} 
}C^{I_2}_{j_1\ldots j_{\alpha_3}l_1\ldots l_{\alpha_1}} 
C^{I_3}_{l_1\ldots l_{\alpha_1}i_1\ldots i_{\alpha_2}}\, , 
\end{equation}
and
\bea 
\hskip-0.5cm
\langle C^{1}_{[0,k_1,0]}C^{2}_{[0,k_2,0]}C^{3}_{[1,k_3-1,1]} \rangle&\ea & 
C^{I_1}_{mi_1\ldots i_{p_2}j_1\ldots j_{p_3} }C^{I_2}_{j_1\ldots 
j_{p_3}l_1\ldots l_{p_3}}
C^{I_3}_{m;l_1\ldots l_{p_1}i_1\ldots i_{p_2}}\nn \\
&&{}- C^{I_1}_{i_1\ldots i_{p_2+1}j_1\ldots j_{p_3} 
}C^{I_2}_{j_1\ldots j_{p_3}l_1\ldots l_{p_1-1}m} C^{I_3}_{m;l_1\ldots 
l_{p_1-1}i_1\ldots i_{p_2+1}} \, , \eea 
where $p_1=\alpha_1+\half$, $p_2=\alpha_2-\half $ and $p_3=\alpha_3-\half$ and, 
\bea \langle 
C^{1}_{[0,k_1,0]}C^{2}_{[0,k_2,0]}C^{3}_{[2,k_3-2,2]} 
\rangle=C^{I_1}_{mi_1\ldots i_{p_2}j_1\ldots j_{p_3}} 
C^{I_2}_{nj_1\ldots j_{p_3}l_1\ldots l_{p_1}} C^{I_3}_{mn;l_1\ldots 
l_{p_1}i_1\ldots i_{p_2}}\, . \eea

Now we specialise these formulae to the case of interest when the legs 1 
and 2 correspond to $k=3$ $\half$-BPS operators, i.e. to the irrep $[0,3,0]$ 
(to simplify the notation, we will not display the Dynkin labels for 
these two legs). It is easy to see that $\langle 
C^{1}C^{2}C^{3}_{[0,k_3,0]} \rangle \neq 0 $ only if $k_3=0,2,4,6$, 
$\langle C^{1}C^{2}C^{3}_{[1,k_3-1,1]} \rangle \neq 0$ only if 
$k_3=1,3,5$ and $\langle C^{1}C^{2}C^{3}_{[2,k_3-2,2]} \rangle\neq 0$ 
only if $k_3=2,4$.

\subsection{Summation formulae}
The four-point function (\ref{g4pt'}) is given in terms of ten 
independent tensor structures $\delta^{12}\delta^{34}$, $C^{1234}$ (and 
permutations) and $S^{1234}$. However, the AdS exchange graphs and the 
quartic couplings are expressed in terms of sums of the type \bea 
\label{sums} \langle C^{1}C^{2}C^{5}\rangle \langle C^{3}C^{4}C^{5} 
\rangle\, , \eea where summation over the representation index at the 
fifth point is assumed. Therefore we need to reexpress these sums in the 
basis of our ten independent tensor structures. This can be achieved by 
using the completeness condition for the C-tensors. Below we obtain the 
corresponding formulae.

In view of possible applications to the computation of correlation 
functions of higher $\half$-BPS operators it is useful to derive a general 
formula for the sum $C_{i_1...i_n}^IC_{j_1...j_n}^I$ expressing the 
completeness condition. We have 
\be \label{complete}  \hskip -0.5cm
C_{i_1...i_n}^IC_{j_1...j_n}^I = \sum_{k=0}^{[\frac{n}{2}]} \, \theta_k \!\!
\sum_{(l_1\ldots l_k)} \d_{i_{l_1}i_{l_2}}\ldots \d_{i_{l_{k-1}}i_{l_k}} 
\d^{(n-k)}{}_{\!\! i_1\ldots \hat{i}_{l_{1}}\ldots\hat{i}_{l_k}\ldots i_{l_n},
(j_{k+1}\ldots j_n}  \d_{j_1j_2}^{\vphantom g}\ldots 
\d_{j_{k-1}j_k)}^{\vphantom g} \, . 
\ee
Here $(\dots )$ stands for total symmetrisation 
of indices, and $\d^{(p)}{}_{\!i_1 \dots i_p,j_1\dots j_p} =
\d^{(p)}{}_{\!(i_1 \dots i_p),(j_1\dots j_p)}$ denotes the 
symmetrised product of $p$ Kronecker deltas $\d_{i_r j_s}$. For every fixed 
$k$ the internal sum runs over all subsets $(l_1\ldots l_k)\in (1\ldots n)$ 
which lead to different products $\d_{i_{l_1}i_{l_2}}\ldots 
\d_{i_{l_{k-1}}i_{l_k}}$, i.e. to expressions that cannot be obtained 
from one another by permuting Kronecker deltas in the product. The 
coefficients $\theta_k$ can be found by requiring the r.h.s. of 
(\ref{complete}) to be traceless w.r.t. any pair of indices from 
$i_1,\ldots ,i_n$ and they are 
\bea 
\theta_0 = 1 \, , \qquad
\theta_{k}=\frac{(-1)^k }{2^k \, (n+1)\ldots (n+2-k)} \, . 
\eea 
As an application of this formula we have, for instance, 
\bea \label{3complete} 
C_{i_1i_2i_3}^IC_{j_1j_2j_3}^I &\ea& 
\frac{1}{6}\Big [\delta_{i_1j_1}\delta_{i_2j_2}\delta_{i_3j_3} 
+ \delta_{i_1j_3}\delta_{i_2j_1}\delta_{i_3 j_2} + 
\delta_{i_1j_2}\delta_{i_2j_3}\delta_{i_3 j_1} \nn \\
&& \qquad{} +
\delta_{i_1j_2}\delta_{i_2j_1}\delta_{i_3 j_3} +
\delta_{i_1j_1}\delta_{i_2j_3}\delta_{i_3 j_2} +
\delta_{i_1j_3}\delta_{i_2j_2}\delta_{i_3 j_1} \Big ] \nn \\
&&{} - 
\frac{1}{8}\Big[ \delta_{i_1i_2}\delta_{i_3(j_1}\delta_{j_2j_3)}
+\delta_{i_1i_3}\delta_{i_2(j_1}\delta_{j_2j_3)}
+\delta_{i_2i_3}\delta_{i_1(j_1}\delta_{j_2j_3)} \Big] 
\, . \eea

Now substituting the completeness condition in the sum (\ref{sums}) with 
$C^5_{[0,k_5,0]}$ and using the definitions (\ref{defC}), we obtain the 
following formulae (the case $k_5=0$ trivially gives $\d^{12}\d^{34}$) 
\bea \nonumber \langle C^{1}C^{2}C^{5}_{[0,2,0]}\rangle\langle 
C^{3}C^{4}C^{5}_{[0,2,0]}\rangle
&\ea &\frac{1}{2}C^{1234}+\frac{1}{2}C^{1243}-\frac{1}{6}\d^{12}\d^{34} \, , \\
\nonumber \langle C^{1}C^{2}C^{5}_{[0,4,0]}\rangle\langle 
C^{3}C^{4}C^{5}_{[0,4,0]}\rangle &=&\frac{2}{3}S^{1234} 
+\frac{1}{6}C^{1324}+\frac{1}{6}C^{1423}-\frac{2}{15}C^{1234}
-\frac{2}{15}C^{1243} \\
\nonumber
&&{}+ \frac{1}{60}\d^{12}\d^{34} \, , \\
\label{sumC} \langle C^{1}C^{2}C^{5}_{[0,6,0]}\rangle\langle 
C^{3}C^{4}C^{5}_{[0,6,0]}\rangle 
&\ea &\frac{1}{20}\left(\d^{13}\d^{24}+\d^{14}\d^{23}+9C^{1342}+
9C^{1432}\right)\nn \\
\nonumber
&&{}- \frac{9}{140}\left(4S^{1234}+C^{1324}+C^{1423} \right)\\ 
&&{}+ \frac{3}{140}\left(C^{1234}+C^{1243}\right)-\frac{1}{700}\d^{12}\d^{34} 
\, . \eea

The sum (\ref{sums}) with $C^5_{[1,k_5-1,1]}$ involves the tensors 
$C_{m;i}^I$, $C_{m;ijk}^I$, $C_{m;ijklp}^I$. The matrix $C_{m;i}$ is 
antisymmetric and the corresponding completeness condition reads 
$C_{i;j}^IC_{k;l}^I=\half(\d_{ik}\d_{jl}-\d_{il}\d_{jk})$. Therefore, for 
$k_5=1$ it gives \be\label{k51} \langle 
C^{1}C^{2}C^{5}_{[1,0,1]}\rangle\langle C^{3}C^{4}C^{5}_{[1,0,1]}\rangle 
= 2(C^{1234}-C^{1243})\;. \end{equation}
For the other two tensors the completeness 
conditions are more involved because of the mixed symmetry of the 
indices. Fortunately, there is another, indirect way to work out the 
corresponding sums. In \cite{AF2} the following reduction relations were 
proved: 
\bea \nonumber 
t_{125}t_{345}&\ea &-\frac{(f_1-f_2)(f_3-f_4)}{4f_5}a_{125}a_{345}
+\frac{1}{4}f_5(a_{145}a_{235}-a_{245}a_{135}) \, , \\
(1-f_5)t_{125}t_{345}&\ea&\frac{1}{4}(f_5^2-
f_5(f_1+f_2+f_3+f_4-4))(a_{145}a_{235}-a_{135}a_{245}) \nn \\
&&{}- \frac{4-f_5}{4f_5}(f_1-f_2)(f_3-f_4)a_{125}a_{345} \label{redrel} \, , 
\eea 
where $f_i=k_i(k_i+4)$. Since we already know the sums (\ref{sumC}) 
and their relation (\ref{aviaC}) to the integrals of the spherical 
harmonics, we can substitute them in the r.h.s. of (\ref{redrel}). 
Further, using (\ref{k51}) and (\ref{tviaC}), we obtain two equations 
for the two remaining sums, whose solution is \bea \langle 
C^{1}C^{2}C^{5}_{[1,2,1]}\rangle\langle C^{3}C^{4}C^{5}_{[1,2,1]}\rangle 
&\ea & \frac{1}{3}
(-C^{1234}+C^{1243}+2C^{1324}-2C^{1423}) \, , \nonumber\\
\langle C^{1}C^{2}C^{5}_{[1,4,1]}\rangle\langle 
C^{3}C^{4}C^{5}_{[1,4,1]}\rangle &\ea &\frac{1}{25}
\Big( C^{1234}-C^{1243}-5C^{1324}+5C^{1423} \nonumber\\
&&{}+ 15C^{1342}-15C^{1432}+5\d^{13}\d^{24}-5\d^{14}\d^{23} \Big) \, . 
\label{sumT} \eea 
Analogously, using formulae (B.10) and (B.11) from 
\cite{AF2} we obtain \bea \nonumber
\langle C^{1}C^{2}C^{5}_{[2,0,2]}\rangle\langle C^{3}C^{4}C^{5}_{[2,0,2]}\rangle 
&\ea &-\frac{2}{15}\Big(20S^{1234}+5C^{1234}+5C^{1243} \nonumber \\
&&\qquad {}-  10C^{1324}-10C^{1423}
- \d^{12}\d^{34} \Big) \, , \nonumber \\
\langle C^{1}C^{2}C^{5}_{[2,2,2]}\rangle\langle 
C^{3}C^{4}C^{5}_{[2,2,2]}\rangle 
&\ea &\frac{8}{15}\Big(\frac{2}{7}C^{1234}+\frac{2}{7}C^{1243}
-C^{1324}-C^{1342}-C^{1423}-C^{1432} \nonumber \\
&&\qquad {}+ \d^{13}\d^{24}+\d^{14}\d^{23}-\frac{1}{35}\d^{12}\d^{34}
  \Big) \, .\label{sumP}
\eea In conclusion, formulae (\ref{sumC}), (\ref{sumT}) and (\ref{sumP}) 
are the necessary tools to express the contribution of the exchange 
graphs and the contact term in terms of the ten independent tensor 
structures.

\subsection{Pairings and normalised projectors}
Given the four-point function (\ref{g4pt}) one can analyze the 
underlying OPE. To this end one needs the (normalised) projectors on 
irreps appearing in the decomposition (\ref{irreps}). This problem has 
already been partially solved since eqs.(\ref{sumC}),(\ref{sumT}) and 
(\ref{sumP}) represent the (non-normalised) projectors on the 
corresponding irreps in (\ref{irreps}). Only the projector on the irrep 
$[3,0,3]$ is missing. To normalize the projectors as well as to find the 
missing one, the pairings among the ten tensor structures of (\ref{g4pt}) 
have to be worked out. This is done by using the completeness relation
(\ref{3complete}) and the results are summarised in Table 3.

\vskip 1cm
\begin{center}
\begin{tabular}{c|ccccccc}
Tensor      & $C^{1234}$ & $C^{1243}$ & $C^{1324}$ & $C^{1342}$ & $C^{1423}$ 
& $C^{1432}$ & $S^{1234}$ \\\hline
&&&&&&&\\
$C^{1234}$  & $\frac{50375}{108}$ & $\frac{1375}{108}$ & $\frac{25525}{108}$
& $\frac{1025}{108}$ & $\frac{1025}{108}$ &$\frac{325}{108}$ & $\frac{575}{24}$\\
&&&&&&&\\
$S^{1234}$  & $\frac{575}{24}$ & $\frac{575}{24}$ & $\frac{575}{24}$& 
$\frac{575}{24}$ & $\frac{575}{24}$ &$\frac{575}{24}$ & $\frac{13225}{144}$   \\
&&&&&&&\\
$\d^{12}\d^{34}$ & $\frac{1250}{3}$ & $\frac{1250}{3}$ & 125 & $\frac{25}{3}$ 
& 125 & $\frac{25}{3}$ & $\frac{125}{3}$   \\
&&&&&&&\\
$\d^{13}\d^{24}$ & 125 & $\frac{25}{3}$ & $\frac{1250}{3}$ & $\frac{1250}{3}$ 
& $\frac{25}{3}$ & 125 & $\frac{125}{3}$ \\
&&&&&&&\\
$\d^{14}\d^{23}$ & $\frac{25}{3}$ & 125 & $\frac{25}{3}$ & 125 & 
$\frac{1250}{3}$ & $\frac{1250}{3}$ & $\frac{125}{3}$
\end{tabular}
\end{center}
\vskip 0.5cm
\begin{center}
\begin{tabular}{l}
Table 3: $C$-algebra. The number appearing at the intersection of a row and a\\
column is the value of the pairing of the corresponding $C$-tensors, 
e.g. the value \\
of $C^{1234}S^{1234}$ is 575/24.
\end{tabular}
\end{center}
\vskip 0.5cm

Using Table 3 we can easily find the expressions for the normalised 
projectors. Their normalisation is fixed to be 
$P^{1234}_{\cal J}P^{1234}_{\cal J}=\nu_{\cal J}$, where $\nu_{\cal J}$ is 
the dimension of the corresponding irrep. We find 
\bea \nonumber
P_{[0,0,0]}^{1234}&\ea &\frac{1}{50}\d^{12}\d^{34}\, , \\
\nonumber
P_{[0,2,0]}^{1234}&\ea &\frac{6}{35}\Big (C^{1234}+C^{1243}-
\frac{1}{3}\d^{12}\d^{34}\Big )\, ,\\
\nonumber P_{[0,4,0]}^{1234}&=&\frac{3}{7}\Big(2S^{1234} 
+\frac{1}{2}C^{1324}+\frac{1}{2}C^{1423}-\frac{2}{5}C^{1234} 
-\frac{2}{5}C^{1243}+\frac{1}{20}\d^{12}\d^{34} \Big)\, , \\
\nonumber 
P_{[0,6,0]}^{1234}&\ea &\frac{1}{20}\Big(\d^{13}\d^{24}+\d^{14}\d^{23}+9C^{1342}+
9C^{1432}\\
\nonumber &&\qquad {}- \frac{9}{7}\left(4S^{1234}+C^{1324}+C^{1423} \right) 
+\frac{3}{7}\left(C^{1234}+C^{1243}\right)-\frac{1}{35}\d^{12}\d^{34}\Big)\, , \\
\nonumber
P_{[1,0,1]}^{1234}&\ea &\frac{9}{70}(C^{1234}-C^{1243})\, , \\
\nonumber
P_{[1,2,1]}^{1234}&\ea &\frac{1}{4}(-C^{1234}+C^{1243}+2C^{1324}-2C^{1423}) \, ,\\
\nonumber
P_{[1,4,1]}^{1234}&\ea &\frac{1}{20}\Big(C^{1234}-C^{1243}-5C^{1324}+5C^{1423} \\
\nonumber &&{}+ 15C^{1342}-15C^{1432}+5\d^{13}\d^{24}-5\d^{14}\d^{23}
\Big) \, , \\
\nonumber 
P_{[3,0,3]}^{1234}&\ea &\frac{1}{28}\Big(2C^{1234}-2C^{1243}-7C^{1324}+7C^{1423}
-21C^{1342}+21C^{1432}\\
\nonumber &&\qquad {}+ 7\d^{13}\d^{24}-7\d^{14}\d^{23}
\Big) \, , \\
\nonumber 
P_{[2,0,2]}^{1234}&\ea &\frac{3}{100}\Big(-20S^{1234}-5C^{1234}-5C^{1243} 
+10C^{1324}+10C^{1423}+\d^{12}\d^{34}
\Big) \, , \\
\nonumber P_{[2,2,2]}^{1234}&\ea &\frac{9}{700}\Big(
-35C^{1324}-35C^{1342}-35C^{1423}-35C^{1432}\\
&&\qquad {}+ 10C^{1234}+10C^{1243}-\d^{12}\d^{34}+35\d^{13}\d^{24}+
35\d^{14}\d^{23} 
\Big) \, . \eea

\section{Contact terms}
\label{B} Here we extract the relevant contact interactions from the 
general quartic effective Lagrangian of \cite{AF2}. This Lagrangian can 
be written in the form \bea {\cal L}_4={\cal L}_4^{(4)}s^1\nabla_\mu s^2 
\nabla\cdot\nabla(s^3\nabla^\mu s^4)+ {\cal L}_4^{(2)}s^1\nabla_\mu s^2 
s^3\nabla^\mu s^4+ {\cal L}_4^{(0)}s^1s^2s^3s^4 \, , \eea where ${\cal 
L}_4^{(4)}$, ${\cal L}_4^{(2)}$ and ${\cal L}_4^{(0)}$ are the 
corresponding couplings considered as functions of the representation 
labels $k_1=x$, $k_2=y$, $k_3=t$, $k_4=w$ of the four scalar fields 
$s^I$ involved ({\it c.f.} Appendix A of \cite{AF2})\footnote{One of the 
couplings in \cite{AF2} was not printed out properly and we therefore 
reproduce it here 
$(S_{p2})^{(0)}_{I_1I_2I_3I_4}=\frac{2}{\d}f_5^2p_{125}
p_{345}(k_1^2+k_2^2+k_3^2+k_4^2-2(k_1+k_2+k_3+k_4)+2 
(k_1k_2+k_3k_4)-4)$. }. For the case of interest we put $x=y=z=t=3$. 
Finally, in comparison with \cite{AF2} we rescale the fields as $s\to 
\frac{\pi^{3/4}}{480^{1/2}}s$ to fit with our normalisations of the 
quadratic and the cubic terms in the actions of Section \ref{amplitude}, 
and we change the overall sign to work with the Euclidean version of the 
AdS space.

\subsection{Four-derivative couplings}
First we sum up the quartic couplings of the four-derivative vertex 
\bea 
\nonumber \!\! {\cal L}_4^{(4)}=-\frac{\pi^3}{480^2} 
\Big[\frac{1}{1024}f_5^3-\frac{7}{32}f_5^2+\frac{369}{32}f_5\Big] 
(a_{145}a_{235}-a_{135}a_{245})+\frac{\pi^3}{2^{18}\! \cdot  \!
75}(f_5-1)^2t_{125}t_{345} \, . \eea 
Next we express all the $a$- and $t$-tensors in terms of $C$-tensors using 
formulae (\ref{aviaC}) and (\ref{tviaC})  and apply the summation formulae 
of Appendix
\ref{A}. As a result we get 
\bea \label{4dir} {\cal L}_4^{(4)}= 
\frac{3}{10240}\left(C^{1234}-C^{1243}+C^{1342}-C^{1432}+C^{1423}-
C^{1324}\right)\, 
. \eea 
On the other hand, for the $AdS_5$ background the four-derivative 
interaction can be written as \bea \nonumber s^1\nabla_\mu s^2 
\nabla\cdot\nabla(s^3\nabla^\mu s^4) =-10 s^1\nabla_\mu s^2 
s^3\nabla^\mu s^4 +2 s^1\nabla_\mu s^2 \nabla_\nu 
s^3\nabla^\mu\nabla^\nu s^4 \, . \eea The first term here is symmetric 
under $2\leftrightarrow 4$ while the second term is symmetric under 
$2\leftrightarrow 3$. These terms are further multiplied by the tensor 
(\ref{4dir}), which is antisymmetric under $2\leftrightarrow 4$ and 
independently under $2\leftrightarrow 3$ and therefore the corresponding 
result vanishes. Hence there is no four-derivative contribution to the 
on-shell value of the gravity action.

\subsection{The remaining couplings}
Proceeding as before, i.e. expressing the total two-derivative coupling 
in terms of the $C$-tensors, we obtain 
\bea \hskip -0.1cm {\cal L}_4^{(2)}&\ea &-\frac{1}{1761607680}\Big[
3125461668\,S^{1234}+323264681\, C^{1234}+123826345\, C^{1243} \nonumber\\
\nonumber &&\qquad\quad{}+ 826347689\, C^{1324}-134041431\,C^{1342}+
733630633\, C^{1423}+369041577\, C^{1432} \\
&& \qquad\quad{}- 206335071\, \d^{12}\d^{34} -
53550175\, \d^{13}\d^{24}+99234721 \,\d^{14}\d^{23} \Big] \, . \label{L42}\eea 
This expression looks rather 
ugly but this is in fact spurious because we have not yet taken into 
account that it is multiplied by $s^1\nabla_\mu s^2 s^3\nabla^\mu s^4$, 
an expression which is symmetric w.r.t. $2\leftrightarrow 4$. Indeed, 
with this in mind we see that eq.(\ref{L42}) is equivalent to 
\bea 
\nonumber {\cal L}_4^{(2)}s^1\nabla_\mu s^2 s^3\nabla^\mu 
s^4 &\ea& -\frac{1}{251658240}\Big[
446494524S+98900894(C^{1234}+C^{1342}+C^{1423})\\
\nonumber &&\qquad\qquad{}-7650025(\d^{12}\d^{34}+\d^{14}\d^{23}+ \d^{13}\d^{24}) 
\Big] s^1\nabla_\mu s^2 s^3\nabla^\mu s^4 \\
&&{} -\frac{3}{32}C^{1243}s^1\nabla_\mu s^2 s^3\nabla^\mu s^4 \, .
\eea 
It is clear that the tensor in the square brackets is totally symmetric under 
permutations of the indices $(2,3,4)$.

Now, consider the expression \bea I=\chi^{1234}s^1\nabla_\mu s^2 
s^3\nabla^\mu s^4 \, , \eea where $\chi^{1234}$ is a totally symmetric 
tensor. Let us show that such an interaction term can be reduced to a 
term without derivatives. Indeed, under integration by parts we obtain 
\bea \nonumber \chi^{1234}s^1\nabla_\mu s^2 s^3\nabla^\mu s^4&\ea & 
\chi^{1234}s^1\nabla_\mu s^3 s^2\nabla^\mu s^4= -\chi^{1234}\nabla_\mu 
s^1 s^2 s^3\nabla^\mu s^4
-\chi^{1234}s^1 s^3 \nabla_\mu  s^2\nabla^\mu s^4\\
\nonumber &&{}- m_4^2\chi^{1234} s^1 s^2 s^3 s^4 \, . \eea Hence we find 
\bea \chi^{1234}s^1\nabla_\mu s^2 s^3\nabla^\mu 
s^4=-\frac{m_4^2}{3}\chi^{1234}s^1s^2s^3s^4 =\chi^{1234}s^1s^2s^3s^4 \, 
, \eea where we have used the fact that $m_4^2=-3$. Therefore we can use 
this trick to reduce the totally symmetric part of the term ${\cal 
L}_4^{(2)}s^1\nabla_\mu s^2 s^3\nabla^\mu s^4$ to a term without 
derivatives.

Next, the coupling without derivatives reads \bea {\cal 
L}_4^{(0)}&=&\frac{68665157}{20971520}\, S^{1234} 
+\frac{100568527}{41943040}\, C^{1234}- 
\frac{4675733}{16777216}\, \d^{12}\d^{34} \, , \eea where we have taken 
into account that this coupling is multiplied by $s^1s^2s^3s^4$ which is 
symmetric under permutations of $2,3,4$.

Finally, summing up ${\cal L}_4^{(0)}$ with the non-derivative term 
obtained from the symmetric part of ${\cal L}_4^{(2)}$, we obtain the 
spectacularly simple final expression in eq.(\ref{cont}).

\section{$D$-functions}
\label{C}
\subsection{${\bf D}$-operators}
Here we collect the necessary facts about the $D$-functions. The 
$D$-functions related to $AdS_{d+1}$ are defined by the formula
\be
\label{defD} 
D_{\D_1\D_2\D_3\D_4}(x_1,x_2,x_3,x_4)= \int \frac{\rmd^{d} 
w~ \rmd w_0}{w_0^{d+1}}\, \prod_{i=1}^4 K_{\D_i}(w,x_i) \, ,
\ee
for
\be K_{\D}(w,x)
= \left(\frac{w_0}{w_0^2+(\w-x)^2}\right)^{\D} \, ,
\label{defK}
\ee 
and where the integral is taken over the space parametrised by 
$w_\mu=(w_0,\w)$, $\w$ being a $d$-dimensional vector, $w_0\ge 0$. From this
definition one can deduce  the following Feynman parameter representation 
\bea \label{FP} 
D_{\D_1\D_2\D_3\D_4}(x_1,x_2,x_3,x_4)=\frac{\pi^{\frac{1}{2}d}\G
\left(\Sigma-\frac{d}{2}\right) \G\left(\Sigma\right) }
{2\prod_i\G(\D_i)}\int \prod_j \rmd \a_j 
\a_j^{\D_j-1} \frac{\d(\sum_j\a_j-1)}{(\sum_{k<l}\a_k\a_l \,
x^2_{kl})^{\Sigma}} \, , \eea 
where $\Sigma=\frac{1}{2}\sum_{i=1}^4\D_i$. From the $D$ functions we may define
corresponding functions of the conformal invariants $s,t$ by
\be
{\prod_{i=1}^4 \Gamma(\Delta_i) \over \Gamma(\Sigma-\half d)} \,
{2\over \pi^{{1\over 2}d}} D_{\D_1\D_2\D_3\D_4}(x_1,x_2,x_3,x_4) =
{(x_{14}^2){}^{\raise 2pt\hbox{$\scriptstyle \! \Sigma-\Delta_1-\Delta_4$}}
(x_{34}^2){}^{\raise 2pt\hbox{$\scriptstyle \! \Sigma-\Delta_3-\Delta_4$}}
\over (x_{13}^2){}^{\raise 2pt\hbox{$\scriptstyle \! \Sigma-\Delta_4$}}
\, (x_{24}^2){}^{\raise 2pt\hbox{$\scriptstyle \! \Delta_2$}}} \,
\oD_{\Delta_1\Delta_2\Delta_3\Delta_4}(s,t) \, .
\end{equation}
For $\D_i=1$ we have
\be
\oD_{1111}(s,t) = \Phi(s,t) \, ,
\label{defP}
\ee
where $\Phi(s,t)$ given in terms of
the standard four-dimensional one-loop (box) integral considered as a 
function of the conformal cross-ratios $s$ and $t$. It has an explicit 
representation in terms of dilogarithms \cite{UD}.

The $D$-functions satisfy the following derivative relation
$$\frac{\pa}{\pa x_{12}^2}D_{\D_1\D_2\D_3\D_4}=
\frac{\D_1\D_2}{\half {d}-\Sigma}D_{\D_1{+1}\,\D_2{+1}\,\D_3\,\D_4}$$ 
and similarly for any other pair of indices
which leads to
\bea
\nonumber
\oD_{\D_1+1\D_2+1\D_3\D_4}&\ea &-\pa_s\oD_{\D_1\D_2\D_3\D_4}\, , \\
\nonumber
\oD_{\D_1\D_2\D_3+1\D_4+1}&\ea &(\D_3+\D_4-\Sigma-s\pa_s)
\oD_{\D_1\D_2\D_3\D_4}\, ,\\
\nonumber
\oD_{\D_1\D_2+1\D_3+1\D_4}&\ea &-\pa_t\oD_{\D_1\D_2\D_3\D_4}\, ,\\
\nonumber
\oD_{\D_1+1\D_2\D_3\D_4+1}&\ea &(\D_1+\D_4-\Sigma-t\pa_t)
\oD_{\D_1\D_2\D_3\D_4}\, ,\\
\nonumber
\oD_{\D_1\D_2+1\D_3\D_4+1}&\ea &(\D_2+s\pa_s+t\pa_t)
\oD_{\D_1\D_2\D_3\D_4} \, , \\
\oD_{\D_1+1\D_2\D_3+1\D_4}&\ea &(\Sigma-\D_4+s\pa_s+t\pa_t)
\oD_{\D_1\D_2\D_3\D_4}\,  .
\label{deriv}
\eea
This allows differential operators
$\overline{\bf D}_{\Delta_1\, \Delta_2\, \Delta_3\, \Delta_4}$ to be 
obtained so that (\ref{dPhi}) is valid whenever $\D_i, \Sigma$ are integers.
The differential operators are not unique, clearly $\oD_{2222}$ can be obtained
by using three separate pairs of equations in (\ref{deriv}) giving the relations
\be
\pr_s s \pr_s \Phi = \pr_t t \pr_t \Phi = (s\pr_s + t \pr_t + 1)^2 \Phi  \, .
\label{Phieq}
\ee
The action of the derivatives  on $\Phi$ is given by \cite{EPSS} 
\bea \label{identity} \pa_s 
\Phi(s,t)&=&\frac{1}{\lambda^2}
\left(\Phi(s,t)(1-s+t)+2\ln s -\frac{s+t-1}{s}\ln t\right) \, , \nn \\
\pa_t \Phi(s,t)&=&\frac{1}{\lambda^2} 
\left(\Phi(s,t)(1-t+s)+2\ln t -\frac{s+t-1}{t}\ln s\right) \, , \eea 
where $\lambda=\sqrt{(1-s-t)^2-4st}$.

Using (\ref{deriv}) we can obtain expressions for all the 
$\oD$-functions we are interested in as differential operators acting on 
$\Phi(s,t)$, up to the arbitrariness following from (\ref{Phieq}).
In particular, the following
$\overline{\bf D}$-operators 
\vskip 0.5cm
\begin{center}
\begin{tabular}{ll}
$\overline{{\bf D}}_{3311}=\pa_s{}^{\!2} \, $  &  
$\overline{{\bf D}}_{3232}=(1+s\pa_s)(2+s\pa_s+t\pa_t)\pa_s\, $ \\
$\overline{{\bf D}}_{3322}=-(2+s\pa_s)\pa_s{}^{\!2} \, $  & 
$\overline{{\bf D}}_{3333}=-(1+s\pa_s)(2+s\pa_s+t\pa_t){}^{\!2}\, $\\
$\overline{{\bf D}}_{2332}=-(1+s\pa_s)\pa_t\pa_s\, $  & 
$\overline{{\bf D}}_{3412}=(3+s\pa_s+t\pa_t)\pa_s{}^{\!2} \, $  \\
$\overline{{\bf D}}_{3421}=-\pa_t\pa_s{}^{\!2}\, $  & 
$\overline{{\bf D}}_{3243}=-(1+s\pa_s)(2+s\pa_s+t\pa_t)s\pa_s{}^{\!2}\, $ \\
$\overline{{\bf D}}_{3423}=-(2+s\pa_s)(3+s\pa_s+t\pa_t)\pa_s{}^{\!2}\, $ & 
$\overline{{\bf D}}_{3342}=-\pa_t(1+s\pa_s)(2+s\pa_s+t\pa_t)\pa_s\, $ \\
$\overline{{\bf D}}_{3432}=\pa_t(2+s\pa_s)(3+s\pa_s+t\pa_t)\pa_s{}^{\!2}\, $  & 
$\overline{{\bf D}}_{4334}=-t\pa_t(1+s\pa_s)(2+s\pa_s+t\pa_t)\pa_s\, $ \\
$\overline{{\bf D}}_{4433}=-\pa_s(1+s\pa_s)(2+s\pa_s+t\pa_t)\pa_s$ &  
\end{tabular}
\end{center}
\vskip 0.3cm 
may be used to determine the coefficient functions of Section \ref{check}. 

\subsection{Relations for $\oD$-functions and simplification of the supergravity 
amplitude}

Here we show how the original coefficient functions (\ref{abcgr}) can be 
further simplified and present an independent proof of the splitting 
property of the supergravity-induced four-point amplitude into ``free'' 
and ``quantum" parts.


By virtue of the obvious permutation symmetries in (\ref{defD}) we have
\bea
\oD_{\Delta_1\, \Delta_2\, \Delta_3\, \Delta_4}(s,t) 
& \ea & t^{-\Delta_2}\oD_{\Delta_1\, \Delta_2\, \Delta_4\, \Delta_3}(s/t,1/t) 
\nn \\
& \ea{}& t^{\Delta_4-\Sigma} \,
\oD_{\Delta_2\, \Delta_1\, \Delta_3\, \Delta_4}(s/t,1/t) \nn  \\
& \ea{}& \oD_{\Delta_3\, \Delta_2\, \Delta_1\, \Delta_4}(t,s) \nn \\
& \ea{}& t^{\Delta_1+\Delta_4-\Sigma} \,
\oD_{\Delta_2\, \Delta_1\, \Delta_4\, \Delta_3}(s,t)  \nn \\
& \ea{}& s^{\Delta_3+\Delta_4-\Sigma} \,
\oD_{\Delta_4\, \Delta_3\, \Delta_2\, \Delta_1}(s,t) \, . 
\label{perm}
\eea
We also have the reflection property
\be
\oD_{\Delta_1\, \Delta_2\, \Delta_3\, \Delta_4}(s,t) = 
\oD_{\Sigma{-\Delta_3}\,\Sigma{-\Delta_4}\,\Sigma{-\Delta_1}\,
\Sigma{-\Delta_2}}(s,t)  \, .
\label{refl}
\end{equation}
In addition there are relations involving $\oD$ functions with differing
$\Sigma$. First
there are what may be referred to as the two up, two down relations
\bea \hskip -0.9cm
(\Delta_2 + \Delta_4 - \Sigma)
\oD_{\Delta_1\, \Delta_2\, \Delta_3\, \Delta_4}(s,t) & \ea &
\oD_{\Delta_1\, \Delta_2{+1}\, \Delta_3\, \Delta_4{+1}}(s,t) -
\oD_{\Delta_1{+1}\, \Delta_2\, \Delta_3{+1}\, \Delta_4}(s,t) \, , \nn \\
\hskip -0.9cm (\Delta_1 + \Delta_4 - \Sigma)
\oD_{\Delta_1\, \Delta_2\, \Delta_3\, \Delta_4}(s,t) & \ea &
\oD_{\Delta_1{+1}\, \Delta_2\, \Delta_3\, \Delta_4{+1}}(s,t) -
t \oD_{\Delta_1\, \Delta_2{+1}\, \Delta_3{+1}\, \Delta_4}(s,t) \, , \nn \\
\hskip -0.9cm (\Delta_3 + \Delta_4 - \Sigma)
\oD_{\Delta_1\, \Delta_2\, \Delta_3\, \Delta_4}(s,t) & \ea &
\oD_{\Delta_1\, \Delta_2\, \Delta_3{+1}\, \Delta_4{+1}}(s,t) - s
\oD_{\Delta_1{+1}\, \Delta_2{+1}\, \Delta_3\, \Delta_4}(s,t) \, .
\label{tup}
\eea
and then the following formula involving  the sum of three $\oD$ functions 
with the same $\Sigma$
\bea
\Delta_4 \oD_{\Delta_1\, \Delta_2\, \Delta_3\, \Delta_4}(s,t) &\ea &
\oD_{\Delta_1\, \Delta_2\, \Delta_3{+1}\, \Delta_4{+1}}(s,t)
+ \oD_{\Delta_1\, \Delta_2{+1}\, \Delta_3\, \Delta_4{+1}}(s,t) \nn \\
& &{}+ \oD_{\Delta_1{+1}\, \Delta_2\, \Delta_3\, \Delta_4{+1}}(s,t) \, .
\label{3rel}
\eea
These results give relations between 
$\oD_{\Delta_1\,\Delta_2\,\Delta_3\,\Delta_4}$ for any fixed $\Sigma$
and are necessary for consistency of (\ref{deriv}). 
If $\Delta_i = n_i$ for integers $n_i$ then for any $\sum_i n_i = 2 \Sigma$ 
even all $\oD$-functions may be related\footnote{Let $(n_1,n_2,n_3,n_4)$ 
be the set of integers corresponding to a $\oD$ function. Let 
$(n_1,n_2,n_3,n_4) \sim (n'_1,n'_2,n'_3,n'_4)$ if the $\oD$ function 
is related up to contributions with lower $\Sigma$. From (\ref{tup}) 
$(n_1,n_2,n_3,n_4) \sim (n_1+1,n_2-1,n_3+1,n_4-1) \sim (n_1+2,n_2-2,n_3,n_4)$ 
and thus we may reduce $(n_1,n_2)$ to one of the form $(m+2,m),(m+1,m),(m,m)$ 
and similarly $(n_3,n_4)$ to
$(n+2,n),(n+1,n),(n,n)$. The relation (\ref{tup}) then allows $m\to m\pm 1$
while $n=n\mp 1$ so that we can take $m=n+1$ or $m=n$. There are then
10 possibilities but these can be reduced to $(n+3,n+1,n+2,n) \sim
(n+2,n+2,n+1,n+1)$, $(n+2,n,n+2,n) \sim (n+1,n+1,n+1,n+1)$, $(n+1,n+1,n+2,n)$
and $(n,n,n+2,n)$.  Using (\ref{3rel}) $(n,n,n+2,n) \sim - (n+1,n,n+1,n)
- (n,n+1,n+1,n)$. Similarly $(n+1,n+1,n+2,n)$ can be reduced to
$(n+1,n+1,n+1,n+1)$.}. 

For one $\Delta_i=0$ the integral (\ref{defD}) reduces to a three point 
function. Using the above this leads to
\bea
&& \lb \oD_{\Delta_1{+1}\, \Delta_2\, \Delta_3{+1}\, \Delta_4} 
+ s\oD_{\Delta_1{+1}\, \Delta_2{+1}\, \Delta_3\, \Delta_4}
+ t\oD_{\Delta_1\, \Delta_2{+1} \, \Delta_3{+1}\, \Delta_4} \rb
\big |_{\Delta_4=\Delta_1+\Delta_2+\Delta_3} \nn \\
&& \qquad = \Gamma(\Delta_1) \Gamma(\Delta_2) \Gamma(\Delta_3) \, . 
\label{D0}
\eea
This leads, for $\D_1=\D_2=\D_3=1$, to the following inhomogeneous differential 
equation for $\Phi$
\be
\Big [ (1-s-t)\pr_s\pr_t - \pr_s{}^{\!2}\, s  - \pr_t{}^{\!2}\, t \Big ] \Phi(s,t)
= \frac{1}{st} \, .
\label{inPhi}
\ee
Both (\ref{Phieq}) and (\ref{inPhi}) follow from (\ref{identity}).

These results may now be used to simplify the results of the supergravity
calculations. For simplicity let
\be
(a_i,b_i,c,\alpha,\beta,\gamma) = \frac{9}{2N^2} 
({\hat a}_i, {\hat b}_i, {\hat c},{\hat\alpha},{\hat \beta},{\hat \gamma}) \, , 
\end{equation}
and then we may write from (\ref{abcgr})
\be
{\hat a}_1 = s^3 \lb (1+t-s) ( \oD_{4433} + \oD_{4422} ) + 5\, \oD_{3333}
- \oD_{3311} \rb \, . 
\label{ah1}
\end{equation}
To simplify this we use (\ref{tup}) and (\ref{3rel}) three times in each
case to obtain
\bea
(1+t-s) \oD_{4433} & \ea & - 2 \, \oD_{4424} + 2 ( \oD_{4323} + \oD_{3423} )
- 3 \, \oD_{3333} \, , \nn \\
(1+t-s) \oD_{4422} & \ea & 2 \, \oD_{4413} + \oD_{4312} + \oD_{3412} 
- 3 \, \oD_{3322} \, .
\eea
Inserting this in (\ref{ah1}) and using (\ref{3rel}) again gives
\be
{\hat a}_1 = - 2 s^3 \lb \oD_{4424} + \oD_{4413} \rb  \, .
\end{equation}

With further use of (\ref{perm}) and (\ref{refl})
we obtain
\bea
{\hat \alpha} &\ea & - 2 \lb s {\oD}_{3335} +  {\oD}_{2235} \rb =
- 2s^2 \lb  {\oD}_{5333} +  {\oD}_{5322 } \rb \, , \nn \\
{\hat \beta} &\ea & - 2 \lb  t {\oD}_{3335} +  {\oD}_{3225} \rb 
= - 2 \lb {\oD}_{4226} -  {\oD}_{3225} \rb \, , \nn \\
{\hat \gamma} &\ea & - 2 \lb  {\oD}_{3335} +  {\oD}_{2325} \rb  =
- 2 s \lb  {\oD}_{5333} +  {\oD}_{5232} \rb \, .
\label{abc}
\eea

For further simplification we first consider $b_1$ where we have from
(\ref{abcgr}),
\bea \hskip -0.5cm
{\hat b}_1 & \ea & s^2 \lb  (1+t-s) {\oD}_{4433} - 4 \, {\oD}_{4334}
- 4 \, {\oD}_{3243} + 4t\, {\oD}_{3342} - {\oD}_{3423} + t\, {\oD}_{3432} \nn \\
&&  \qquad {}+ 8\, {\oD}_{3232} - 4\, {\oD}_{3322} - 2 \, {\oD}_{3412} 
+ 2t \, {\oD}_{3421} + 2 \, {\oD}_{3311} \rb \nn \\
\hskip -0.5cm & \ea & s^2 \lb - 2\, {\oD}_{4424} - 4 \, {\oD}_{4334}
+ 2 ( {\oD}_{4323} + {\oD}_{3423} ) - 3\, {\oD}_{3333}  - 4 \, {\oD}_{3243}
+  4 \, {\oD}_{4233} \nn \\ 
&&  \qquad {} - {\oD}_{3423} + {\oD}_{4323} 
+ 8\, {\oD}_{3232} - 4\, {\oD}_{3322} - 2 \, {\oD}_{3412}
+ 2 \, {\oD}_{4312} + 2 \, {\oD}_{3311} \rb \, .
\label{b1} 
\eea
According to the general results obtained in Section \ref{prediction} this is 
related to
$(t-s-1){\hat \alpha} + s {\hat \gamma}$. From (\ref{abc}) we need
\bea
(t-s-1){\oD}_{5333} & \ea & 2\, {\oD}_{4334} + {\oD}_{5223} - 3
( {\oD}_{4233} + {\oD}_{4323} ) \nn \\
(t-s-1){\oD}_{5322}  & \ea & 2\, {\oD}_{4323} - 3 \, {\oD}_{4222}  
- 2\, {\oD}_{4312}  + \frac{1}{s^2} \, ,
\eea
where the last result depends on using (\ref{D0}). Hence we obtain
\bea
(t-s-1){\hat \alpha} + s {\hat \gamma} &\ea & s^2 \lb  - 4 \, {\oD}_{4334}
- 2 \, {\oD}_{5333} - 2 \, {\oD}_{5223} + 6 ( {\oD}_{4233} + {\oD}_{4323} ) \nn
\\
&&\ {} - 2\, {\oD}_{5232} - 4\, {\oD}_{4323} + 6 \, {\oD}_{4222} + 4\, 
{\oD}_{4312} \rb - 2  \, .
\label{abD}
\eea
Combining (\ref{b1}) and (\ref{abD}) we readily find
\be
{\hat b}_1 = (t-s-1){\hat \alpha} + s {\hat \gamma} + 2 \, . 
\label{bsim}
\end{equation}

The results for $c$ can be similarly simplified. The starting point is
\be
{\hat c} = 32 st \lb - {\ts{3\over 2}} \, {\oD}_{3333} + 
{\oD}_{2332} + {\oD}_{3232} + {\oD}_{3322} \rb \, .
\end{equation}
For comparison we have
\bea
&& (s-t-1) {\hat \alpha} + (t-s-1) {\hat \beta} + (1-s-t) {\hat \gamma} \nn \\
&& \ ={}  - 4s \, {\oD}_{3326} - 4t \, {\oD}_{2336} - 4\, {\oD}_{3236} \nn\\
&& \ \quad {} - 8 (s-t-1) {\oD}_{2235} - 8 (t-s-1) {\oD}_{3225} - 8 (1-s-t)
{\oD}_{2325} \nn \\
&& \ ={} - 4 + 8st \lb - {\oD}_{2253} - {\oD}_{5223} + {\oD}_{3252} +
{\oD}_{5232} - {\oD}_{2523} \rb \nn \\
&& \ \quad {} + 8t \lb {\oD}_{2235}  + {\oD}_{2253} \rb
+ 8s \lb {\oD}_{3225}  + {\oD}_{5223} \rb \, ,
\eea
where we have used (\ref{D0}).
Using (\ref{perm}) and (\ref{tup}) to generate an overall factor of $st$
we may now use (\ref{3rel}) to demonstrate
\be
{\hat c} = (s-t-1){\hat \alpha} + (t-s-1){\hat \beta} + (1-s-t){\hat \gamma}
+ 4 \, .
\label{csim}
\end{equation}
(\ref{bsim}) and (\ref{csim}) are equivalent to (\ref{amaz}).

For determining anomalous dimensions we need to identify the terms in
$\oD_{\D_1 \D_2 \D_3 \D_4}(s,t) $ involving $\ln s $ in an expansion in
terms of $s,1-t$. For $N= \D_1 + \D_2 - \Sigma + 1 = 1,2,\dots$ we have
from \cite{DO1}
\bea
\hskip - 1cm\oD_{\D_1 \D_2 \D_3 \D_4}(s,t) \sim && \!\!\!\!\!\!\!\!\!
\ln s \, \frac{(-1)^N}{(N-1)!} \,
\frac{\Gamma(\D_1) \Gamma(\D_2) \Gamma(\Sigma - \D_3) \Gamma(\Sigma - \D_4)}
{\Gamma(\D_1 + \D_2)} \nn \\
\hskip - 1cm && \!\!\!\!\!\!\!\!\!\!\!\!
{} \times \sum_{m,n=0}^\infty \frac{ (\D_1)_m (\Sigma - \D_3)_m
( \D_2)_{m+n} (\Sigma - \D_4)_{m+n}}{m!n!\,(N)_m (\D_1 + \D_2)_{2m+n}} \, s^m
(1-t)^n \, , 
\label{logexp}
\eea 
where $(x)_m = \Gamma(x+m)/\Gamma(x)$. For $N=2,3,\dots$ there are
additional terms which dominate for $s \sim 0$ and which do not contain $\ln s$,
\bea
&& \hskip -1cm s^{1-N} (N-2)! \,
\frac{\Gamma(\D_1-N+1) \Gamma(\D_2 -N+1) \Gamma(\Sigma - \D_3 -N+1) 
\Gamma(\Sigma - \D_4 -N +1 )}{\Gamma(\D_1 + \D_2- 2N +2 )} \nn \\
&& \hskip -1cm  {} \times \sum_{m=0}^{N-2} \sum_{n=0}^\infty 
\frac{ (\D_1 {-N}{ +1} )_m (\Sigma {- \D_3}{ -N}{ +1} )_m
(\D_2 {-N}{ +1})_{m+n} (\Sigma {-\D_4}{ -N}{ +1} )_{m+n}}
{ m!n!\, (1- N)_m (\D_1 + \D_2 - 2N +2 )_{2m+n}}\, s^m (1-t)^n \, . \nn \\
\label{powexp}
\eea


\section{Exchange graphs of a massive symmetric tensor}\label{D}

Here we show how the method of Ref. \cite{dHFR} can be generalised to 
compute the AdS graphs involving the exchange of a massive symmetric 
tensor.

Recall that an exchange graph is a double integral in the $z$ and $w$ 
variables over the five-dimensional AdS space. One first computes the 
$z$-integral and then recasts the remaining $w$-integral as a sum of 
different $D$-functions.

We start with the equations of motion for the massive symmetric tensor 
$\varphi_{\mu\nu}$. The Lagrangian is given by \cite{AF2} 
\bea 
{\cal 
L}&\ea &-\frac{1}{4}\nabla_{\rho}\varphi_{\mu\nu}\nabla^{\rho}\varphi^{\mu\nu} 
+\frac{1}{2}\nabla_{\mu}\varphi^{\mu\rho}\nabla^{\nu}\varphi_{\nu\rho} 
-\frac{1}{2}\nabla_{\mu}\varphi_{\rho}^{\rho}\nabla_{\nu}\varphi^{\mu\nu}
+\frac{1}{4}\nabla_{\rho}\varphi_{\mu}^{\mu}\nabla^{\rho}\varphi^{\nu}_{\nu}\nn\\
&&{}+ \frac{1}{4}(2-f)\varphi_{\mu\nu}\varphi^{\mu\nu} 
+\frac{1}{4}(2+f)(\phi_{\mu}^{\mu})^2+\alpha T_{\mu\nu}\varphi^{\mu\nu} 
\, . \label{lag} \eea 
Here $\alpha$ is some coupling and the tensor $T_{\mu\nu}$ is 
assumed to be of the form 
\bea \label{masstensor} T_{\mu\nu}= 
\frac{1}{2}\nabla_{\mu} s\nabla_{\nu} s +\frac{1}{2}\nabla_{\nu} 
s\nabla_{\mu} s -\frac{1}{2}g_{\mu\nu}\Big( \nabla^{\rho} s 
\nabla_{\rho} s +\frac{1}{2}(2m^2-f)ss \Big ) \, . \eea The scalar 
field $s$ has mass squared $m^2=\Delta(\Delta-4)$ and $f=k(k+4)$ is the 
mass squared of the symmetric tensor $\varphi_{\mu\nu}$ transforming in 
the irrep $[0,k,0]$ of SO(6). The massless graviton arises as the 
particular case $k=0$. For the sake of clarity we suppress the 
inessential representation index for both the $s^I$ and 
$\varphi_{\mu\nu}^I$ fields.

First of all, the Lagrangian (\ref{lag}) implies the following equation 
\bea \nabla_{\rho}\nabla^{\rho}\phi_{\lambda}^{\lambda}= 
\nabla_{\rho}\nabla_{\lambda}\phi^{\rho\lambda}-\frac{2-f+d(2+f)}{2-d}
\phi_{\lambda}^{\lambda}  -\frac{2\alpha}{2-d}T_{\mu}^{\mu} \, , 
\label{divph}\eea 
where $d$ is the dimension 
of the AdS space, i.e. $d=5$ for $AdS_5$. Using this equation we then find 
\bea 
W_{\mu\nu}^{~\rho\lambda} \phi_{\rho\lambda} & \!\! \equiv \!\! &
-\nabla_{\rho}\nabla^{\rho}\phi_{\mu\nu}+\nabla_{\mu}\nabla^{\rho}
\phi_{\rho\nu}+\nabla_{\nu}\nabla^{\rho} 
\phi_{\rho\mu}-\nabla_{\mu}\nabla_{\nu}\phi_{\rho}^{\rho}
-\Big ( (2-f)\phi_{\mu\nu}+\frac{6+f}{2-d}g_{\mu\nu} \phi_{\lambda}^{\lambda}
\Big )\nn \\
&\ea & \alpha\Big ( 
g_{\mu\rho}g_{\nu\lambda}+g_{\mu\lambda}g_{\nu\rho} 
+\frac{2}{2-d}g_{\mu\nu}g_{\rho\lambda} \Big ) T^{\rho\lambda} \, ,
\eea 
which defines {\it the modified Ricci operator} $W_{\mu\nu}^{~\rho\lambda}$ 
acting on $\phi_{\rho\lambda}$. Applying the derivative $\nabla^{\mu}$ to both 
sides of this equation we find 
\bea \nonumber 
f\left(\nabla^{\mu}\phi_{\mu\nu}-\nabla_{\nu}\phi_{\lambda}^{\lambda}\right)= 
2\alpha \nabla_{\mu}T^{\mu}_{\nu}\, . \eea 
Differentiating this formula, using (\ref{divph}),  
and assuming that $f$ is non-zero we can solve for the trace of 
$\varphi_{\mu\nu}$: \bea \label{gravtrace} \varphi_{\lambda}^{\lambda}= 
-\frac{\alpha}{2f(f+3)}\Big (3 \nabla_{\mu}\nabla_{\nu}T^{\mu\nu} 
+ fT_{\mu}^{\mu}\Big ) \, . \eea

By definition, the $z$-integral describing the exchange of a massive 
symmetric tensor is given by 
\bea \label{zint} 
A_{\mu\nu}(w,x_3,x_4)= \int \frac{\rmd^{d} z~ \rmd z_0}{z_0^{d+1}}\,  
G_{\mu\nu,\lambda\rho}(w,z)T^{\lambda\rho}(z,x_3,x_4) \, . \eea 
Here  the bitensor $G_{\mu\nu,\rho\lambda}(w,z)$ is the bulk-to-bulk propagator 
for the massive symmetric tensor $\varphi_{\mu\nu}$. 
It depends on the AdS invariant distance and obeys the equation 
\bea 
\label{prop} W_{\mu\nu}^{~\rho\lambda}G_{\rho\lambda,\mu'\nu'}(w,z) = 
\Big ( g_{\mu\mu'}g_{\nu\nu'}+g_{\mu\nu'}g_{\nu\mu'} 
+\frac{2}{2-d}g_{\mu\nu}g_{\mu'\nu'} \Big)\d(z,w) \, . \eea 
The tensor $T^{\lambda\rho}(z,x_3,x_4)$, suppressing the argument $z$, is
given by 
\bea \label{masten} T_{\mu\nu}(x_3,x_4)&\ea & 
\frac{1}{2}\nabla_{\mu} K_{\Delta}(x_3)\nabla_{\nu} K_{\Delta}(x_4)
+\frac{1}{2}\nabla_{\nu} K_{\Delta}(x_3)\nabla_{\mu} K_{\Delta}(x_4) \nn \\
&&{}- \frac{1}{2}g_{\mu\nu}\Big ( \nabla^{\rho} K_{\Delta}(x_3) 
\nabla_{\rho} K_{\Delta}(x_4) 
+\frac{1}{2}(2m^2-f)K_{\Delta}(x_3)K_{\Delta}(x_4) \Big ) \, , \eea 
where the functions $K_{\Delta}(z,x)$ were defined in (\ref{defK}).

Following \cite{dHFR}, we use $A_{\mu\nu}(w,x_3,x_4)=
A_{\mu\nu}(w-x_3,0,x_{43})$ and then perform
a conformal inversion on the integral (\ref{zint}) to obtain
\bea 
A_{\mu\nu}(w,0,x) 
=\frac{1}{x^{2\Delta}\,w^4}J_{\mu\lambda}(w)J_{\nu\rho}(w)
I_{\lambda\rho}(w'-x')\, , \qquad w' = \frac{w_{\mu}}{w^2} \, , \quad
x' = \frac{x_{\mu}}{x^2} \, , 
\eea 
where $J_{\mu\nu}(w)=\d_{\mu\nu}-2\frac{w_{\mu}w_{\nu}}{w^2}$.
We then write down the following ansatz for the tensor 
$I_{\mu\nu}$
\bea \label{Ansatz} 
I_{\mu\nu}(w)=g_{\mu\nu}h(t)+P_{\mu}P_{\nu}\phi(t)+
\nabla_{\mu}\nabla_{\nu}X(t)+ 
\nabla_{(\mu}(P_{\nu)}Y(t)) \, . \eea 
Here $t=w_0^2/w^2$, $g_{\mu\nu}= \delta_{\mu\nu}/w_0^2$, 
$P_{\mu}=\d_{0\mu}/w_0$, and $h(t)$, $\phi(t)$, $X(t)$, $Y(t)$ are four 
unknown functions. To find them one has to work out the action of the 
modified Ricci operator on eq.(\ref{Ansatz}). For the individual terms 
in (\ref{Ansatz}) we obtain the following formulae 
\bea \nonumber 
W_{\mu\nu}^{~\rho\lambda}[g_{\rho\lambda}h(t)]&\ea &\Big ( 
4t^2(t-1)h''+4t(t+1)h'+\frac{8}{3}(f+3)h \Big )g_{\mu\nu}
-3\nabla_{\mu}\nabla_{\nu}h \, , \\
\nonumber W_{\mu\nu}^{~\rho\lambda}[P_{\rho}P_{\lambda}\phi(t)]&\ea & 
g_{\mu\nu}\Big (4t(t-1)\phi'+\frac{1}{3}(f+24)\phi\Big )
-\nabla_{\mu}\nabla_{\nu}\phi\, \\
\nonumber &&\! {}+ P_{\mu}P_{\nu}\Big (4t^2(1-t)\phi''-8t^2\phi'+f\phi\Big ) 
+\frac{\d_{0(\mu} 
w_{\nu)}w_0}{(w^2)^2}\Big (4t(t-1)\phi''+8t\phi'\Big ) \, . \eea 
and 
\bea \nonumber 
W_{\mu\nu}^{~\rho\lambda}[\nabla_{\rho}\nabla_{\lambda}X(t)] 
&\ea &g_{\mu\nu}\Big (-\frac{4}{3}t^2(t-1)X''-\frac{4}{3}t(t+1)X'\Big )f
+f\nabla_{\mu}\nabla_{\nu}X  \, , \\
\nonumber 
W_{\mu\nu}^{~\rho\lambda}[\nabla_{\rho}(P_{\lambda}Y)+\nabla_{\lambda}(P_{\rho}Y)] 
&\ea &
g_{\mu\nu}\Big (-\frac{14}{3}Y-\frac{4}{3}t(t-1)Y'\Big )f \\
\nonumber &&{}+  P_{\mu}P_{\nu}(2Y+4tY')f+ \frac{\d_{0(\mu} 
w_{\nu)}w_0}{(w^2)^2}(-2Y'f) \, . \eea

The basic equation allowing us to determine the unknown functions in 
(\ref{Ansatz}) is derived by using the fundamental equation 
(\ref{prop}): \bea 
W_{\mu\nu}^{~\rho\lambda}[I_{\rho\lambda}(w)]&\ea &
\frac{1}{3}g_{\mu\nu}(2m^2-f)t^{\Delta}+ 
P_{\mu}P_{\nu}(2\Delta^2 t^{\Delta})-2\Delta^2t^{\Delta-1} \frac{\d_{0\{ 
\mu} w_{\nu\}}w_0}{(w^2)^2} \, . \eea
To solve this equation we first equate the terms involving 
$\nabla_{\mu}\nabla_{\nu}$ on both sides: 
\bea \nonumber 
\nabla_{\mu}\nabla_{\nu}(-3h-\phi+fX)=0 \eea 
and pick up the trivial solution 
\bea \label{h} h(t)=-\frac{1}{3}\phi+\frac{f}{3}X\, . \eea  
Now we equate the coefficients of the expression 
$\frac{\d_{0(\mu} w_{\nu)}w_0}{(w^2)^2}$ and find 
\bea \nonumber 
4t(t-1)\phi''+8t\phi'-2fY'=-2\Delta^2t^{\Delta-1} \, . \eea This 
equation is trivially integrated to give \bea \label{Y} 
Y(t)=a+\frac{1}{2f}\left(4t(t-1)\phi'+4\phi+2\Delta t^{\Delta}\right) \, 
, \eea 
where $a$ is an integration constant. Equating the coefficients 
of $P_{\mu}P_{\nu}$ we get 
\bea 
4t^2(1-t)\phi''-8t^2\phi'+f\phi+2Yf+4tfY'=2\Delta^2t^{\Delta} \, . \eea 
Substituting here (\ref{Y}) we find a closed equation for $\phi$: \bea 
\label{phi} 
4t^2(t-1)\phi''+4t(3t-1)\phi'+(f+4)\phi+2af+2\Delta(\Delta+1) 
t^{\Delta}=0 \, . \eea 
Finally we equate the coefficients of $g_{\mu\nu}$ and substitute their 
(\ref{h}) and (\ref{Y}). The resulting equation is used to find $X$: 
\bea \nonumber 
X(t)&\ea &\frac{1}{8f(f+3)}\Big[
12t^2(2t-1)(t-1)\phi''+12t(4t^2+t-3)\phi'+(36+5f)\phi\\
\label{X} &&\qquad\qquad\qquad {}+ 
42af-3t^{\Delta}\left(f+2\Delta(\Delta-2\Delta t-3)\right) 
\Big] \, . \eea Thus, solving eqs.(\ref{h}),(\ref{Y}),(\ref{phi}) 
and (\ref{X}) we can establish the $z$-integral corresponding to a 
massive spin 2 tensor exchange.
 
Now we solve equation (\ref{phi}) for $\Delta=3$, $f=12$. The solution 
regular at both $t=1$ and $t=0$ has the form \bea \nonumber 
\phi(t)=-\frac{3}{2}a-\frac{3}{4}t^2 \, , ~~~ 
h(t)=\frac{3}{2}a+\frac{3}{20}t^2\, ,~~~~ 
X(t)=\frac{1}{4}a-\frac{1}{40}t^2\, , ~~~~ 
Y(t)=\frac{3}{4}a+\frac{1}{8}t^2\, . \eea

It is easy to see that upon substituting the coefficient functions found 
into the ansatz (\ref{Ansatz}), all the terms proportional to the 
arbitrary integration constant $a$ cancel out. Combining all the 
pieces together we obtain 
\be \label{Ansatz1} 
I_{\mu\nu}(w)=\frac{3}{20}t^2g_{\mu\nu} 
-\frac{3}{4}t^2~P_{\mu}P_{\nu}-\frac{1}{40}\nabla_{\mu}\nabla_{\nu}t^2+ 
\frac{1}{8}\nabla_{(\mu}\left(P_{\nu )}t^2\right) \, . \ee  
Working out the action of the derivatives this formula can be written in 
the form 
\be \label{Ansatz2} 
I_{\mu\nu}(w)=-\frac{3}{5}\, t^2~\frac{w_{\mu}w_{\nu}}{(w^2)^2}\, . \ee
This is an amazingly simple expression, even simpler than the 
corresponding result for the graviton exchange.

To restore the $z$-integral from $I_{\mu\nu}$ one has to let  
\bea t = \frac{w'^2_0}{(w'-x')^2}\to q=x_{34}^2 
~\frac{w_0}{w_0^2+(\vec{w}-x_3)^2}\frac{w_0}{w_0^2+(\vec{w}-x_4)^2}\, ,
\eea
and also
\be
\frac{1}{w^2 (w'-x')^2} J_{\mu\nu}(w) (w'-x')_\nu \to
Q_{\mu}=\frac{(x_3-w)_{\mu}}{(x_3-w)^2}-\frac{(x_4-w)_{\mu}}{(x_4-w)^2}\, .
\ee
In this way we find the following result for the $z$-integral 
describing the exchange of the spin 2 tensor of $m^2=12$: 
\be
\label{zintf} A_{\mu\nu}(w,x_3,x_4) = -\frac{3}{5 
x_{34}^2}Q_{\mu}Q_{\nu}K_2(w,x_3)K_2(w,x_4) \, . \ee 
The result (\ref{zintf}) is symmetric under the exchange 
$x_3\leftrightarrow x_4$, as it should be. Owing to the conformal 
property of $Q_{\mu}$ one can also see that $A_{\mu\nu}$ indeed 
transforms as a tensor under inversions. 

Finally, taking the trace of (\ref{zintf}) and using
$Q_{\mu}Q^{\mu}=\frac{x_{34}^2w_0^2}{(w-x_3)^2(w-x_4)^2}$  (the contraction
involves the AdS metric), we get 
\be \label{trcheck} \phi_{\lambda}^{\lambda}= 
A_{\lambda}^{\lambda}(w,x_3,x_4)=-\frac{3}{5}\, K_3(w,x_3)K_3(w,x_4) \, . 
\ee 
On the other hand, this formula should coincide with 
(\ref{gravtrace}) for $\alpha=1$. Using the formula (\ref{masstensor}) 
one can show that eq.(\ref{gravtrace}) indeed coincides with 
eq.(\ref{gravtrace}). This provides a consistency check on our 
calculation.

The computation of the remaining $w$-integral does not present any 
difficulty and is based on the general formula \bea \label{id} 
\nabla^{\mu}_w K_{\Delta}(w,x_1)Q_{\mu}=\Delta 
K_{\Delta+1}(w,x_1)\Biggl[x_{14}^2\frac{w_0}{(w-x_4)^2} 
-x_{13}^2\frac{w_0}{(w-x_3)^2}\Biggr] \, . \eea

\section{Results for Short Multiplet Expansions}\label{appF}

Using the variables $z,x$ defined by
\be
s=zx \, , \qquad t=(1-z)(1-x) \, ,
\ee
the functions $H_{\cal J}$, defined in (\ref{decomp}), can be expressed as
\be
H_{\cal J}(s,t) = \frac{1}{z-x} \Big (z\, g_{\cal J}(z) - \half s\, g_{\cal J}(z)
+ s \, f_{\cal J}(z) - (z \leftrightarrow x) \Big ) \, ,
\ee
where, from the results in \cite{DO}, $g_{\cal J}$ represents contributions 
corresponding to just  twist 0 and $f_{\cal J}$ to twist 2. We find
\bea
g_{[2,2,2]}(z) &\ea & 0 \, , \qquad\qquad\qquad\qquad\qquad\quad  \ \,
f_{[2,2,2]}(z) = \frac{10}{9} \Big ( z^3 + \frac{z^3}{(1-z)^3} \Big ) \, , \nn \\
g_{[3,0,3]}(z) &\ea & 0 \, , \qquad\qquad\qquad\qquad\qquad\quad  \ \,
f_{[3,0,3]}(z) = -2 
\Big ( \frac{2}{z}-1 \Big ) \Big ( z^3 + \frac{z^3}{(1-z)^3} \Big ) \, , \nn \\ 
g_{[1,2,1]}(z) &\ea & -\Big (z^3 + \frac{z^3}{(1-z)^3}\Big ) \, , 
\qquad\qquad  \
f_{[1,2,1]}(z) = \frac{3}{2} \Big ( z^3 - \frac{z^3}{(1-z)^3} \Big ) \, , \nn \\ 
\hskip -1cm g_{[2,0,2]}(z) &\ea & \frac{5}{3} \Big ( \frac{2}{z} -1 \Big ) 
\Big (z^3 + \frac{z^3}{(1-z)^3} \Big ) \, , \quad 
f_{[2,0,2]}(z) = - \frac{5}{2} \Big ( \frac{2}{z} -1 \Big )
 \Big ( z^3 - \frac{z^3}{(1-z)^3} \Big ) \, , \nn \\ \hskip -1cm
g_{[0,2,0]}(z) &\ea & - \frac{35}{8}
\Big (z^3 - \frac{z^3}{(1-z)^3} \Big ) \, , \qquad \quad \
f_{[0,2,0]}(z) = \frac{45}{16} \Big ( z^3 + \frac{z^3}{(1-z)^3} \Big ) \, ,\nn \\ 
\hskip -1cm g_{[1,0,1]}(z) &\ea & \frac{35}{6}\Big ( \frac{2}{z} -1 \Big ) 
\Big (z^3 - \frac{z^3}{(1-z)^3} \Big ) \, , \ \,
f_{[1,0,1]}(z) = -\frac{15}{4}\Big ( \frac{2}{z} -1 \Big )
\Big ( z^3 + \frac{z^3}{(1-z)^3} \Big)   \, , \nn \\
\hskip -1cm g_{[0,0,0]}(z) &\ea & -50 + \frac{25}{2} 
\Big ( z^3 - \frac{z^3}{(1-z)^3} \Big) +\frac{45}{2}
\Big ( \frac{2}{z} -1 \Big )\Big (z^3 + \frac{z^3}{(1-z)^3} \Big ) \, , \nn \\
\hskip-1cm && f_{[0,0,0]}(z) = -\frac{15}{4}
\Big ( z^3 + \frac{z^3}{(1-z)^3} \Big)  -\frac{35}{4}
\Big ( \frac{2}{z} -1 \Big ) \Big ( z^3 - \frac{z^3}{(1-z)^3} \Big ) \, .
\eea
The expansion coefficients for the individual contributions of short 
representations are given by
\bea\hskip-0.6cm 
d_{1,\ell}^{[1,4,1]}&\! \ea & 2^{\ell-2}\frac{(\ell+2)!(\ell+4)!}{(2\ell+3)!}
(\ell+1) \, , \qquad 
d_{2,\ell}^{[1,4,1]} = 2^{\ell-2}\frac{3(\ell+3)!(\ell+5)!}{5(2\ell+5)!}
(\ell+2) \, , \nn \\ \hskip-0.6cm
d_{3,\ell}^{[1,4,1]} &\!\ea & 2^{\ell-3}
\frac{3(\ell+4)!(\ell+5)!}{35(2\ell+7)!}(\ell+2)(\ell+7) \, , \nn \\ \hskip-0.6cm
d_{1,\ell}^{[0,4,0]}&\! \ea & 2^{\ell-1}
\frac{(\ell+2)!(\ell+3)!}{15(2\ell+3)!}(4\ell^2+20\ell+51) \, , \quad  
d_{2,\ell}^{[0,4,0]} = 2^{\ell-4}\frac{7(\ell+3)!(\ell+4)!}{5(2\ell+5)!}
(\ell^2+7\ell+18 ) \, , \nn \\ \hskip-0.6cm
d_{3,\ell}^{[0,4,0]} &\!\ea & 2^{\ell-5}
\frac{9(\ell+4)!(\ell+6)!}{25(2\ell+7)!}(\ell+3) \, , \quad \,
d_{4,\ell}^{[0,4,0]} = 2^{\ell-2}
\frac{(\ell+5)!(\ell+6)!}{315(2\ell+9)!}(\ell+3)(\ell+8) \, , \nn \\ \hskip-0.6cm
d_{1,\ell}^{[2,2,2]}&\! \ea & 2^{\ell-2}\frac{(\ell+2)!(\ell+4)!}{(2\ell+3)!}
(\ell+1) \, , \qquad
d_{2,\ell}^{[2,2,2]} = 2^{\ell-3}\frac{7(\ell+3)!(\ell+5)!}{9(2\ell+5)!}
(\ell+2) \, , \nn \\ \hskip-0.6cm
d_{3,\ell}^{[2,2,2]} &\!\ea & 2^{\ell-4}
\frac{5(\ell+4)!(\ell+6)!}{21(2\ell+7)!}(\ell+3) \, , \nn \\ \hskip-0.6cm
d_{1,\ell}^{[3,0,3]}&\! \ea & 2^{\ell-2}\frac{(\ell+2)!(\ell+4)!}{(2\ell+3)!}
(\ell+1) \, , \qquad
d_{2,\ell}^{[3,0,3]} = 2^{\ell-3}\frac{3(\ell+3)!(\ell+5)!}{5(2\ell+5)!}
(\ell+2) \, , \nn \\ \hskip-0.6cm
d_{3,\ell}^{[3,0,3]} &\!\ea & 2^{\ell-4}
\frac{9(\ell+4)!(\ell+6)!}{35(2\ell+7)!}(\ell+3) \, , \nn \\ \hskip-0.6cm
d_{1,\ell}^{[1,2,1]}&\! \ea & 2^{\ell-4}
\frac{9(\ell+2)!(\ell+3)!}{5(2\ell+3)!}(\ell^2+5\ell+12) \, , \quad \!\!
d_{2,\ell}^{[1,2,1]} = 2^{\ell-2}\frac{(\ell+3)!(\ell+4)!}{35(2\ell+5)!}
(8\ell^2+56\ell+129) \, , \nn \\ \hskip-0.6cm
d_{3,\ell}^{[1,2,1]} &\!\ea & 2^{\ell-5}
\frac{9(\ell+4)!(\ell+5)!}{175(2\ell+7)!}(7\ell^2+63\ell+162) \, , \quad
d_{4,\ell}^{[1,2,1]} = 2^{\ell-3}
\frac{(\ell+5)!(\ell+7)!}{147(2\ell+9)!}(\ell+4) \, , \nn \\ \hskip-0.6cm
d_{1,\ell}^{[2,0,2]}&\! \ea & 2^{\ell-4}
\frac{5(\ell+2)!(\ell+3)!}{3(2\ell+3)!}(\ell^2+5\ell+12) \, , \quad
d_{2,\ell}^{[2,0,2]} = 2^{\ell-4}\frac{(\ell+3)!(\ell+4)!}{7(2\ell+5)!}
(5\ell^2+35\ell+78 ) \, , \nn \\ \hskip-0.6cm
d_{3,\ell}^{[2,0,2]} &\!\ea & 2^{\ell-3}
\frac{3(\ell+4)!(\ell+5)!}{35(2\ell+7)!}(\ell^2+9\ell+23) \, , \quad
d_{4,\ell}^{[2,0,2]} = 2^{\ell-3}
\frac{(\ell+5)!(\ell+7)!}{147(2\ell+9)!}(\ell+4) \, , \nn \\
\hskip-0.6cm
d_{1,\ell}^{[0,2,0]}&\! \ea & 2^{\ell-7}
\frac{27(\ell+2)!(\ell+4)!}{5(2\ell+3)!}(\ell+1) \, , \quad
d_{2,\ell}^{[0,2,0]} = 2^{\ell-7}\frac{3(\ell+3)!(\ell+4)!}{5(2\ell+5)!}
(7\ell^2+49\ell+106 ) \, , \nn \\ \hskip-0.6cm
d_{3,\ell}^{[0,2,0]} &\!\ea & 2^{\ell-7}
\frac{3(\ell+4)!(\ell+5)!}{175(2\ell+7)!}(59\ell^2+531\ell+1552) \, , \
d_{4,\ell}^{[0,2,0]} = 2^{\ell-6}
\frac{(\ell+5)!(\ell+6)!}{105(2\ell+9)!}(7\ell^2+77\ell+268) \, , \nn \\
\hskip-0.6cm
d_{1,\ell}^{[1,0,1]}&\! \ea & 2^{\ell-5}
\frac{(\ell+2)!(\ell+4)!}{(2\ell+3)!}(\ell+1) \, , \qquad
d_{2,\ell}^{[1,0,1]} = 2^{\ell-3}\frac{(\ell+3)!(\ell+4)!}{5(2\ell+5)!}
(\ell^2+7\ell+15) \, , \nn \\ \hskip-0.6cm
d_{3,\ell}^{[1,0,1]} &\!\ea & 2^{\ell-6}
\frac{3(\ell+4)!(\ell+5)!}{35(2\ell+7)!}(5\ell^2+45\ell+118) \, ,
\quad
d_{4,\ell}^{[1,0,1]} = 2^{\ell-4}
\frac{(\ell+5)!(\ell+6)!}{63(2\ell+9)!}(\ell^2+11\ell+36) \, , \nn \\
\hskip-0.6cm
d_{2,\ell}^{[0,0,0]}&\! \ea &
2^{\ell-5}\frac{3(\ell+3)!(\ell+5)!}{7(2\ell+5)!}
(\ell+2) \, , \qquad
d_{3,\ell}^{[0,0,0]} = 2^{\ell-3}\frac{9(\ell+4)!(\ell+6)!}{35(2\ell+7)!}
(\ell+3) \, , \nn \\ \hskip-0.6cm
d_{4,\ell}^{[0,0,0]} &\!\ea & 2^{\ell-3}
\frac{(\ell+5)!(\ell+7)!}{147(2\ell+9)!}(\ell+4) \, . 
\label{genl}
\eea

For $\ell=0,1$ the general formulae are no longer correct in all cases, for
those cases which are different from what would be given from (\ref{genl})
we have
\bea
d_{1,0}^{[0,4,0]} & \ea & 2 \, , \qquad d_{2,0}^{[0,4,0]} = {\ts \frac{9}{5}} 
\, , \qquad d_{3,0}^{[0,4,0]} = {\ts \frac{11}{100}} \, , \qquad 
d_{4,0}^{[0,4,0]} = {\ts \frac{2}{441}}  \, , \nn \\
d_{1,1}^{[1,2,1]} & \ea & {\ts \frac{162}{35}} \, , \qquad  
d_{2,1}^{[1,2,1]} = {\ts \frac{1912}{1225}}
\, , \qquad d_{1,1}^{[1,0,1]} = {\ts \frac{3}{5}} \, , \qquad 
d_{2,1}^{[1,0,1]} = {\ts \frac{64}{105}}  \, , \nn \\
d_{1,0}^{[2,0,2]} & \ea & 2 \, , \qquad \ \ \, d_{2,0}^{[2,0,2]} = {\ts \frac{9}{14}}
\, , \qquad d_{3,0}^{[2,0,2]} = {\ts \frac{32}{245}} \, , \qquad 
d_{2,0}^{[2,2,2]} = {\ts \frac{109}{105}}  \, , \nn \\
d_{1,0}^{[0,2,0]} & \ea & 0 \, , \qquad \ \ \, d_{1,2}^{[0,2,0]} = {\ts \frac{33}{20}}
\, , \qquad d_{2,0}^{[0,2,0]} = {\ts \frac{45}{112}} \, , \qquad
d_{3,0}^{[0,2,0]} = {\ts \frac{1829}{19600}}  \, , \nn \\
d_{2,0}^{[0,0,0]} & \ea & {\ts \frac{9}{70}} \, , \qquad \
d_{3,0}^{[0,0,0]} = {\ts \frac{61}{1960}} \, .
\eea

\renewcommand{\baselinestretch}{0.6}

\end{document}